\pdfoutput=1

\documentclass[article]{aa}
\newcommand{\refcom}[1]{#1}
\usepackage{lineno}

\usepackage{txfonts}    
\usepackage{pdflscape}
\usepackage{multirow}
\usepackage[english]{babel}
\usepackage{longtable}

\usepackage{xspace}

\newcommand\Tstrut{\rule{0pt}{2.6ex}}         

\newcommand{\ang}{\AA\xspace}
\newcommand{\sne}{SNe~Ia\xspace}
\newcommand{\sn}{SN\ Ia\xspace}
\newcommand{\cahk}{\ion{Ca}{H\&K}~$\lambda~3945$\xspace}
\newcommand{\cahkfeat}{\ion{Ca}{H\&K}+\ion{Si}{ii}\xspace}
\newcommand{\cahkfull}{\ion{Ca}{H\&K}~$\lambda \lambda~3934, 3969$\xspace}
\newcommand{\ewcahk}{EW(\ion{Ca}{H\&K} $ \lambda 3945$)\xspace}
\newcommand{\cair}{\ion{Ca}{IR}~$\lambda 8579$\xspace}
\newcommand{\sisix}{\ion{Si}{ii}~$\lambda 6355$\xspace}

\newcommand{\sifour}{\ion{Si}{ii}~$\lambda 4138$\xspace}
\newcommand{\sithree}{\ion{Si}{ii}~$\lambda 3858$\xspace}
\newcommand{\ewsifour}{EW(\ion{Si}{ii}~$\lambda 4138$)\xspace}
\newcommand{\ewsisix}{EW(\ion{Si}{ii}~$\lambda 6355$)\xspace}
\newcommand{\vsisix}{v(\ion{Si}{ii}~$\lambda 6355$)\xspace}

\newcommand{\mej}{$M_{ej}$\xspace}
\newcommand{\mni}{$M_{Ni}$\xspace}
\newcommand{\nifive}{${}^{56}$Ni\xspace}

\newcommand{\salt}{\texttt{SALT2.4}\xspace}
\newcommand{\rsi}{$\mathcal{R}($Si$)$}
\newcommand{\rca}{$\mathcal{R}($Ca$)$}

\newcommand{\todo}[1]{}
\newcommand{\rfc}[1]{#1}
\newcommand{\rfctwo}[1]{#1}
\newcommand{\rtr}[1]{#1}
\newcommand{\rtrt}[1]{#1}

\newcommand{\uige}{\uni}

\newcommand{\uniw}{$\lambda($uNi$)$\xspace}
\newcommand{\utiw}{$\lambda($uTi$)$\xspace}
\newcommand{\ucaw}{$\lambda($uCa$)$\xspace}
\newcommand{\usiw}{$\lambda($uSi$)$\xspace}
\newcommand{\uni}{uNi\xspace}
\newcommand{\uti}{uTi\xspace}
\newcommand{\uca}{uCa\xspace}
\newcommand{\usi}{uSi\xspace}

\newcommand{\rbias}{$0.163 \pm 0.029$}
\newcommand{\rsigma}{$5.7 \sigma$}
\newcommand{\rbiasm}{$0.119 \pm 0.032$}

\newcommand{\cutlc}{five\xspace}
\newcommand{\cutbg}{four\xspace}

\newcommand{\samplesize}{$92$\xspace}
\newcommand{\samplehubblefitsize}{$73$\xspace}
\newcommand{\rmsti}{$0.116 \pm 0.011$\xspace}
\newcommand{\rmstica}{$0.086 \pm 0.010$\xspace}

\newcommand{\saltrms}{$0.135$\xspace}

\newcommand{\rmssalt}{$0.135 \pm 0.011$\xspace}
\newcommand{\nbrti}{$57$\xspace}
\newcommand{\deltchitisalt}{$20.5$\xspace}
\newcommand{\cahubblersi}{$0.43$\xspace}
\newcommand{\rmssaltca}{$0.122 \pm 0.012$\xspace}

\newcommand{\dispti}{$0.070 \pm 0.009$\xspace}
\newcommand{\dispsalt}{$0.090 \pm 0.008$\xspace}
\newcommand{\randomtica}{two\xspace}   
\newcommand{\randomsaltca}{zero\xspace}
\newcommand{\mstepsalt}{$0.098 \pm 0.031$\xspace}
\newcommand{\lstepsalt}{$-0.151 \pm 0.028$\xspace}
\newcommand{\mstepti}{$0.042 \pm 0.031$\xspace}
\newcommand{\lstepti}{$-0.075 \pm 0.031$\xspace}
\newcommand{\msteptica}{$0.022 \pm 0.030$\xspace}
\newcommand{\lsteptica}{$-0.065 \pm 0.030$\xspace}

\begin{document}

\title{Understanding Type Ia supernovae through their U-band spectra }

\author{J.~Nordin \inst{\ref{berlin}}
  \and   G.~Aldering \inst{\ref{lbnl}}
  \and   P.~Antilogus \inst{\ref{lpnhe}}
  \and   C.~Aragon \inst{\ref{lbnl}}
  \and   S.~Bailey \inst{\ref{lbnl}}
  \and   C.~Baltay \inst{\ref{yale}}
  \and   K.~Barbary \inst{\ref{lbnl},\ref{bccp}}
  \and   S.~Bongard \inst{\ref{lpnhe}}
  \and   K.~Boone \inst{\ref{lbnl},\ref{ucb}}
  \and   V.~Brinnel \inst{\ref{berlin}}
  \and   C.~Buton \inst{\ref{ipnl}}
  \and   M.~Childress \inst{\ref{southampton}}
  \and   N.~Chotard \inst{\ref{ipnl}}
  \and   Y.~Copin \inst{\ref{ipnl}}
  \and   S.~Dixon \inst{\ref{lbnl}}
  \and   P.~Fagrelius \inst{\ref{lbnl},\ref{ucb}}
  \and   U.~Feindt \inst{\ref{okc}}
  \and   D.~Fouchez \inst{\ref{cppm}}
  \and   E.~Gangler \inst{\ref{lpc}}
  \and   B.~Hayden \inst{\ref{lbnl}}
  \and   W.~Hillebrandt \inst{\ref{garching}}
  \and   A.~Kim \inst{\ref{lbnl}}
  \and   M.~Kowalski \inst{\ref{berlin},\ref{desy}}
  \and   D.~Kuesters \inst{\ref{berlin}}
  \and   P.-F.~Leget, \inst{\ref{lpc}}
  \and   S.~Lombardo \inst{\ref{berlin}}
  \and   Q.~Lin \inst{\ref{china}}
  \and   R.~Pain \inst{\ref{lpnhe}}
  \and   E.~Pecontal \inst{\ref{cral}}
  \and   R.~Pereira \inst{\ref{ipnl}}
  \and   S.~Perlmutter \inst{\ref{lbnl},\ref{ucb}}
  \and   D.~Rabinowitz \inst{\ref{yale}}
  \and   M.~Rigault \inst{\ref{berlin}}
  \and   K.~Runge \inst{\ref{lbnl}}
  \and   D.~Rubin \inst{\ref{lbnl},\ref{stsci}}
  \and   C.~Saunders \inst{\ref{lbnl},\ref{lpnhe}}
  \and   G.~Smadja \inst{\ref{ipnl}}
  \and   C.~Sofiatti \inst{\ref{lbnl},\ref{ucb}}
  \and   N.~Suzuki \inst{\ref{lbnl},\ref{ipmu}}
  \and   S.~Taubenberger \inst{\ref{garching},\ref{eso}}
  \and   C.~Tao \inst{\ref{cppm},\ref{china}}
  \and   R.~C.~Thomas \inst{\ref{nersc}}\\
  \textsc{The Nearby Supernova Factory}
}

\institute{
  Institut fur Physik, Humboldt-Universitat zu Berlin,
  Newtonstr. 15, 12489 Berlin \label{berlin}
  \and
  Physics Division, Lawrence Berkeley National Laboratory,
  1 Cyclotron Road, Berkeley, CA, 94720 \label{lbnl}
  \and
  Laboratoire de Physique Nucl\'eaire et des Hautes \'Energies,
  Universit\'e Pierre et Marie Curie Paris 6, Universit\'e Paris Diderot Paris 7, CNRS-IN2P3,
  4 place Jussieu, 75252 Paris Cedex 05, France \label{lpnhe}
  \and
  Department of Physics, Yale University,
  New Haven, CT, 06250-8121 \label{yale}
  \and
  Berkeley Center for Cosmological Physics, University of California Berkeley,
  Berkeley, CA, 94720 \label{bccp}
  \and
  Department of Physics, University of California Berkeley,
  366 LeConte Hall MC 7300, Berkeley, CA, 94720-7300 \label{ucb}
  \and
  Universit\'e de Lyon, F-69622, Lyon, France ; Universit\'e de Lyon 1, Villeurbanne ;
  CNRS/IN2P3, Institut de Physique Nucl\'eaire de Lyon. \label{ipnl}
  \and
  Department of Physics and Astronomy, University of Southampton,
  Southampton, Hampshire, SO17 1BJ, UK \label{southampton} 
  \and
  The Oskar Klein Centre, Department of Physics, AlbaNova, Stockholm
  University, SE-106 91 Stockholm, Sweden \label{okc}
  \and
  Aix Marseille Universit\'e, CNRS/IN2P3, CPPM UMR 7346, 13288,
  Marseille, France \label{cppm}
  \and
  Clermont Universit\'e, Universit\'e Blaise Pascal, CNRS/IN2P3, Laboratoire de Physique Corpusculaire,
  BP 10448, F-63000 Clermont-Ferrand, France \label{lpc}
  \and
  Max-Planck Institut f\"ur Astrophysik, Karl-Schwarzschild-Str. 1,
  85748 Garching, Germany \label{garching}
  \and
  Deutsches Elektronen-Synchrotron, D-15735 Zeuthen, Germany \label{desy}
  \and
  Tsinghua Center for Astrophysics, Tsinghua University, Beijing
  100084, China \label{china}
  \and
  Centre de Recherche Astronomique de Lyon, Universit\'e Lyon 1,
  9 Avenue Charles Andr\'e, 69561 Saint Genis Laval Cedex,
  France \label{cral}
  \and
  Space Telescope Science Institute, 3700 San Martin Drive,
  Baltimore, MD 21218 \label{stsci}
  \and
  European Southern Observatory, Karl-Schwarzschild-Str. 2, 85748
  Garching, Germany \label{eso}
  \and
  Computational Cosmology Center, Computational Research Division, Lawrence Berkeley National Laboratory,
  1 Cyclotron Road MS 50B-4206, Berkeley, CA, 94720 \label{nersc}
  \and
  Kavli Institute for the Physics and Mathematics of the Universe,
      University of Tokyo, 5-1-5 Kashiwanoha, Kashiwa, Chiba, 277-8583, Japan \label{ipmu}
}

\date{Received ; accepted}

\abstract
    {Observations of Type Ia supernovae (\sne) can be used to derive accurate cosmological distances through empirical standardization techniques. 
      Despite this success neither the progenitors of \sne nor the explosion process are fully understood. The U-band \refcom{region} has been less well observed for nearby SNe, due to technical challenges, but is the most readily accessible band for high-redshift SNe.
    }
    {Using \refcom{spectrophotometry}
      from the Nearby Supernova Factory, we study the origin and extent of U-band \refcom{spectroscopic} variations in \sne and explore consequences for their standardization and \rfctwo{the potential for providing new insights into the explosion process. }}
    {\rfc{We divide the U-band \refcom{spectrum} into four wavelength regions  \uniw, \utiw, \usiw and \ucaw. Two of these span the \cahkfull complex. We employ spectral synthesis using \texttt{SYNAPPS} to associate the two bluer regions with Ni/Co and Ti.}}
    {(1) The \refcom{flux of the} uTi feature is an extremely sensitive temperature/luminosity indicator, standardizing the SN peak luminosity to \rmsti mag RMS. \rfctwo{A traditional \salt fit on the same sample yields a \saltrms mag RMS. Standardization using uTi also reduces the difference in corrected magnitude between SNe originating from different host galaxy environments.} (2) Early U-band spectra can be used to probe the Ni+Co distribution in the ejecta, thus offering a rare window into the source of lightcurve power. (3) The uCa flux further improves standardization, yielding a \rmstica mag RMS \rfctwo{without the need to include an additional intrinsic dispersion to reach $\chi^2/\mathrm{dof}\sim 1$}. \rtrt{This reduction in RMS is partially driven by an improved standardization of Shallow Silicon and 91T-like SNe.} 
     }
    {}
    

\maketitle

\section{Introduction}

Type Ia supernovae (\sne) are standardizable candles, and measuring their relative distances first led to the discovery of dark energy \citep{riess98,perlmutter99}. Their origin as thermonuclear explosions of CO WDs is well accepted, as is their importance as producers of heavy elements in the Universe \citep[e.g.][]{2014ARA&A..52..107M}. \rfctwo{It is further accepted that the light curve is powered by the decay of \nifive, which has a half-life of $\sim 6$ days for the dominant decay channel to Co.}
However, the progenitor system configuration and the path to detonation is still not understood. While a small sample of nearby \sne now have tight limits on companions set by the lack of interaction or non-detection of hydrogen \citep{2007ApJ...670.1275L,2012ApJ...744L..17B,2016MNRAS.457.3254M}, some SNe do show signs of H interaction \citep{2003Natur.424..651H,2006ApJ...650..510A, 2012Sci...337..942D} and theoretical explanations of the observed diversity prefer multiple explosion channels \citep{2013MNRAS.436..333S,2016IJMPD..2530024M}.

Current lightcurve-based empirical \sn standardization methods yield a $\sim 0.1$ mag intrinsic dispersion (after accounting for measurement uncertainties), interpreted as a random scatter in magnitude and/or color \citep{2014AA...568A..22B,2014ApJ...780...37S,2015ApJ...813..137R}. 
At least part of this observed scatter is due to differences in the explosion process, seen for example in a significantly reduced intrinsic dispersion when comparing spectroscopic twins \citep[][]{2015ApJ...815...58F}. These differences can, in turn, be expected to evolve differently with cosmic time and thus propagate into systematic limits on cosmological constraints from \sne if left unresolved.
\rfctwo{Simultaneously, the origin of the peak-brightness correction for reddening is not fully understood, with lingering differences between empirical reddening curves and dust-like extinction in the U-band and between individual well-measured SNe \citep{2010A&A...523A...7G,2011A&A...529L...4C,2014ApJ...789...32B, 2015MNRAS.453.3300A,2016arXiv160904470M,2017ApJ...836..157H}}.
Evidence for an incomplete \sn standardization is also implied by the correlations between standardized magnitude and host-galaxy environment \citep{2010MNRAS.406..782S,2013ApJ...770..108C,2013AA...560A..66R}.\todo{add kelly ref} Most recently, \citet[][R17]{rigualtsub} have used the specific star formation rate measured at the projected SN location (local specific Star Formation Rate, LsSFR) to statistically classify individual \sne as younger or older. Younger \sne are found to be $0.16\pm0.03$ mag fainter than older (after \salt standardization). The fraction of young  SNe is expected to greatly increase as a function of redshift, an era where cosmology aims to be sensitive to slight deviations from the $\Lambda CDM$ model requires full understanding of all such effects.

\rtrt{Spectral features in the rest-frame U-band region ($\sim 3200$ to $4000$~\ang) are less well explored compared with those at redder wavelengths.}
Empirical studies of U-band \refcom{spectroscopic} variability are few, and due to the atmosphere cutoff, usually dependent on high-$z$ or space-based data. Comparisons of samples at different redshifts have suggested an evolution of the mean U/UV properties \citep{2008ApJ...674...51E,2009ApJS..185...32K,2012AJ....143..113F,2015ApJ...803...20M}.
\citet{2016MNRAS.461.1308F} examined UV spectra of ten nearby \sne obtained within five days of peak light, finding variations connected to optical lightcurve shape, and an increasing dispersion bluewards of $4000$~\ang~\rtr{($\lambda>4000$~\AA)}.
\rfctwo{The \cahk ``feature'', dominated by \sithree and the \cahkfull doublet, is more frequently observed, but with conflicting interpretations.
\citet{2012MNRAS.426.2359M} and \citet{2013MNRAS.435..273F} find differences in \ewcahk between samples divided by lightcurve width, but differ as to whether the \cahk velocity correlates with lightcurve width. 
\cite{2011ApJ...742...89F} found high \ion{Ca}{ii} velocity SNe to be intrinsically redder.} 
High Velocity Features (HVFs) -- absorption features that are detached from a ``photospheric'' component -- have long been observed in early \sne spectra \citep{2005ApJ...623L..37M,2006ApJ...645..470T}, and potentially yield insights into the outer ejecta density or ionization \citep{2013MNRAS.429.2127B}.
Several studies have tried to systematically map the presence of detached HVFs in the Ca IR and H\&K regions. 
Fitting coupled Gaussian functions to the Ca features, \citet{2014MNRAS.437..338C} found that neither rapidly declining SNe nor SNe with high photospheric velocity show HVFs at peak light.
\citet{2015MNRAS.451.1973S} reached similar conclusions based on a larger sample.

\rfctwo{
  Individual SNe with early U-band spectroscopy include PTF13asv, showing an initially suppressed U-band flux at day $-14$ that brightened significantly during the following five days \citep{2016ApJ...823..147C}, possibly due to an outer, thin region of radioactive material.
PS1-10afx was first reported as a new transient type by \citet{2013ApJ...767..162C}, but later found to be a gravitationally magnified high-z \sn \citep{2013ApJ...768L..20Q,2014Sci...344..396Q}. The spectral comparison of \citet{2013ApJ...767..162C} (their Fig.\xspace 7) shows a brighter flux in the bluest part of the U-band, differing from comparison \sne (SN2011fe, SN2011iv). 
The UV/U-band \refcom{spectrum} of SN2004dt is discussed in \citet{2012ApJ...749..126W}, where low-resolution HST-ACS spectra show excess U-band flux close to peak. 
\refcom{Further studies of individual SNe with U-band spectroscopic coverage have been presented for example by
  \citet[][]{1996MNRAS.278..111P,2004AJ....128..387G,2007A&A...475..585A,2007A&A...469..645S,2007A&A...471..527G,2008AJ....135.1598M,2009ApJ...697..380W,2009ApJ...700.1456B,2012AJ....143..126B,2012MNRAS.425.1789S,2012ApJ...749..126W,2015ApJ...813...30S}}. \rtr{Significant variation in flux for different phases of observation}  and between objects can be found, but due to uneven candidate and cadence selection this variability has proven hard to quantify.}

The $3000$-$4000$~\ang region is situated at the opacity transition region, going from fully dominated by Iron Group Element (IGE) line absorption in the UV to electron-scattering in the B-band, with the SN Spectral Energy Distribution (SED) shaped by a mix of IGE and Intermediate Mass Element (IME) features. Theoretical predictions of the U-band \refcom{spectrum} thus have to take both effects correctly into account \citep{pintoeastman}, making it hard to gauge the level of systematic uncertainties on the U-band SED predictions from current models.
\rfctwo{Simulations have found progenitor metallicity to cause significant variations bluewards of the U-band \refcom{wavelength region}, but with few clear predictions within this region} \citep{2003ApJ...590L..83T, 2012MNRAS.427..103W}.

Here we attempt to gain a deeper understanding of the U-band region, both to provide data for comparison with predictions from explosion scenarios and to improve the use of \sne as standardizable candles. This will have particular implications for SNe at high-$z$, where restframe U-band \refcom{observations are} naturally obtained by ground-based imaging surveys and important for constraining  the transient type (either at initial detection to trigger follow-up, or during a final photometry-only analysis). 

\rfctwo{The initial motivation for this study and the approach of dividing} the U-band \refcom{spectrum} into four \refcom{wavelength} subregions arose from the comparison of individual supernovae having similar BVR spectra and lightcurve properties but exhibiting spectroscopic differences in the U-band.
  We present the sample and introduce this subdivision in Sec.~\ref{sec:data}, and study U-band \refcom{spectroscopic} variability in Sec.~\ref{sec:uvary}. Sec.~\ref{sec:prog} contains a  discussion of how the \sn explosion can be examined using U-band \refcom{absorption} features.
  In Sec.~\ref{sec:cosmo} we explore the consequences for \sn standardization. Finally, in Sec.~\ref{sec:disc} we discuss the presence of \sn subclasses based on U-band \refcom{flux measurements}, return to the question of HVFs, and provide a brief outlook. 
  We conclude in Sec.~\ref{sec:conc}.

\section{Data and U-band measurables}\label{sec:data}

\subsection{Nearby Supernova Factory}

The Nearby Supernova Factory (SNfactory) has obtained \refcom{time-series spectrophotometry} of a large number of \sne in the Hubble flow. Observations have been performed
with our SuperNova Integral Field Spectrograph \citep[SNIFS,][]{Aldering2002, 2004SPIE.5249..146L}. SNIFS is a fully integrated instrument optimized for automated observation of point sources on a structured background over the full ground-based optical window at moderate spectral resolution ($R\sim500$). It consists of a high throughput wide-band lenslet integral field spectrograph, a multi-filter photometric channel to image the field in the vicinity of the IFS for atmospheric transmission monitoring simultaneously with spectroscopy, and an acquisition/guiding channel. The IFS possesses a fully-filled $6.4\arcsec\times6.4\arcsec$ spectroscopic field of view subdivided into a grid of $15 \times 15$ spatial elements, a dual-channel spectrograph covering $3200-5200$~\ang and $5100-10000$~\ang simultaneously, and an internal calibration unit (continuum and arc lamps). SNIFS is mounted on the south bent Cassegrain port of the University of Hawaii $2.2$ m telescope on Mauna Kea, and is operated remotely. Observations are reduced using a dedicated data reduction pipeline, similar to that presented in \S4 of \citet{2001MNRAS.326...23B}. Discussion of the software pipeline is given in \citet{2006ApJ...650..510A} and updated in \citet{2010ApJ...713.1073S}. Of particular importance for this analysis is the flux calibration and Mauna Kea atmosphere model presented in \citet{2013A&A...549A...8B}. 
This provides accurate flux calibration down to $\sim 3300$~\ang with a residual $\sim 2\%$ grey scatter. The extension to bluer wavelengths compared with most ground-based observations is a key prerequisite for this study.
Host-galaxy subtraction is performed as described in \citet{2011MNRAS.418..258B}, \rfctwo{a methodology subsequently improved and made more flexible\footnote{https://github.com/snfactory/cubefit}.} Each spectrum is corrected for Milky Way dust extinction \citep{1998ApJ...500..525S}, blue-shifted to rest-frame based on the heliocentric host-galaxy redshift \citep[][R17]{2013ApJ...770..107C} and the fluxes are converted to luminosity assuming distances expected for the supernova redshifts in the CMB frame and assuming a dark energy equation of state $w=-1$. 
For the purpose of fitting LCs with \salt and calculating Hubble residuals, magnitudes are generated through integration over the following top-hat profiles: $\mathrm{U}_{\mathrm{SNf}}$ ($3300-4102$~\ang), $\mathrm{B}_{\mathrm{SNf}}$ ($4102-5100$~\ang),  $\mathrm{V}_{\mathrm{SNf}}$ ($5200-6289$~\ang),  $\mathrm{R}_{\mathrm{SNf}}$ ($6289-7607$~\ang) and $\mathrm{I}_{\mathrm{SNf}}$ ($7607-9200$~\ang).

\subsection{Sample}

The sample is based on a slightly enlarged version of the data presented in previous SNfactory publications \citep{2011A&A...529L...4C,2013AA...560A..66R,2015A&A...578C...1F,2015ApJ...815...58F}.\footnote{\rfc{Note that \citet{2011A&A...529L...4C} used slightly different filter definitions.}} \rfc{We require all SNe to have a first high signal-to-noise spectrum prior to $-2$ days. We have updated SALT lightcurve fits to the latest version \citep[\salt ,][]{2014AA...568A..22B}, and require these to provide a good fit to \rtr{synthetic broadband photometry generated from} the data (\cutlc failed this in a blinded inspection). $\mathrm{U}_{\mathrm{SNf}}$ photometry was not included in these fits \rtr{since \citet{2015ApJ...800...57S} demonstrated that the UV is not well described by the \salt model}. Finally, \cutbg 91bg-like SNe were removed. The final sample consists of \samplesize \sne.} \rfctwo{Out of this set, \samplehubblefitsize SNe are found in the $0.03<z<0.1$ redshift range. Spectra shown below are cut redward of $\sim 6500$ \AA\xspace for display purposes only.}

Host-galaxy properties were presented in \citet{2013ApJ...770..108C} and R17, where $z$, \salt $x_1$, $c$ and Hubble residuals have also been tabulated.\todo{either say this is a subset of data from R17, or point to C17 for all data?}
Besides the frequently-used global stellar mass \citep{2010MNRAS.406..782S, 2013ApJ...770..108C}, this includes the local specific star formation rate (LsSFR).  The LsSFR combines the local $H\alpha$ flux (driven by UV-emission from young stars) with the estimated local stellar mass, thus producing an estimate for the fraction of young stars at the SN location. SNe with large LsSFR values are more likely to originate from a young progenitor, while SNe in low LsSFR environments likely originate from older systems. This provides refined information compared with earlier global measurements as star forming galaxies can have locally passive (old) regions.
The lightcurve width, color and host-galaxy mass distributions of this sample closely match that of the combined SNLS and SDSS data in the JLA sample \citep{2014AA...568A..22B}. 

\rfctwo{Measurements of the equivalent width (EW) and velocity of the \sisix absorption feature were made on spectra within $\pm 2.5$ days from (B-band) maximum. These spectroscopic-indicator measurements are further described in \citet{nicothesis}, and a spectroscopic-indicator analysis based on the full SNfactory sample will be presented in Chotard et al. (in prep.).}

We focus on spectra in three restframe phase bins, \emph{pre-peak} ($-8$ to $-4$ days with respect to B-band peak as determined by \salt), \emph{peak} ($-2$ to $2$ days) and \emph{post-peak} ($4$ to $8$ days). This selection strikes a balance between capturing how quickly SNe~Ia vary \rtrt{while limiting the number of new parameters to inspect.} 
Spectra are dereddened based on the optical colors so that, to first order, intrinsic spectroscopic variations can be distinguished from extinction by dust. The correct color curve to apply, which could vary from object to object, \rfctwo{is not fully understood.} We further do not want to assume that intrinsic U-band features do not correlate with reddening, as these potentially could be indirectly caused by relations between intrinsic SN-features and the progenitor environment.
In order to minimize the potential impact of systematic differences due to reddening, we take two conservative steps: (1)
Remove reddened SNe with \salt color $c>0.2$ from further analysis and (2) make an initial dereddening correction assuming the extinction curve of \citet[][F99]{1999PASP..111...63F}  with $R = 3.1$.
Three SNe\footnote{These are SN2007le, SNF20080720-001 and SN2006X} have $c>0.2$ and thus will not be included in the main analysis, unless otherwise noted.
The $E(B-V)$ values used when dereddening are based on \salt $c$ measurements, derived from the $\mathrm{B}_{\mathrm{SNf}}\mathrm{V}_{\mathrm{SNf}}\mathrm{R}_{\mathrm{SNf}}$ bands, and converted to E$($B$-$V$)$ according to Eq.~6 of \citet{2010A&A...523A...7G}.
In Sec.~\ref{sec:reddening} we will return to how this choice of method of accounting for reddening by dust influences measurements.

\subsection{Definition of U-band indices}\label{sec:meas}

\begin{figure*}
\includegraphics[angle=0,width=0.49 \textwidth]{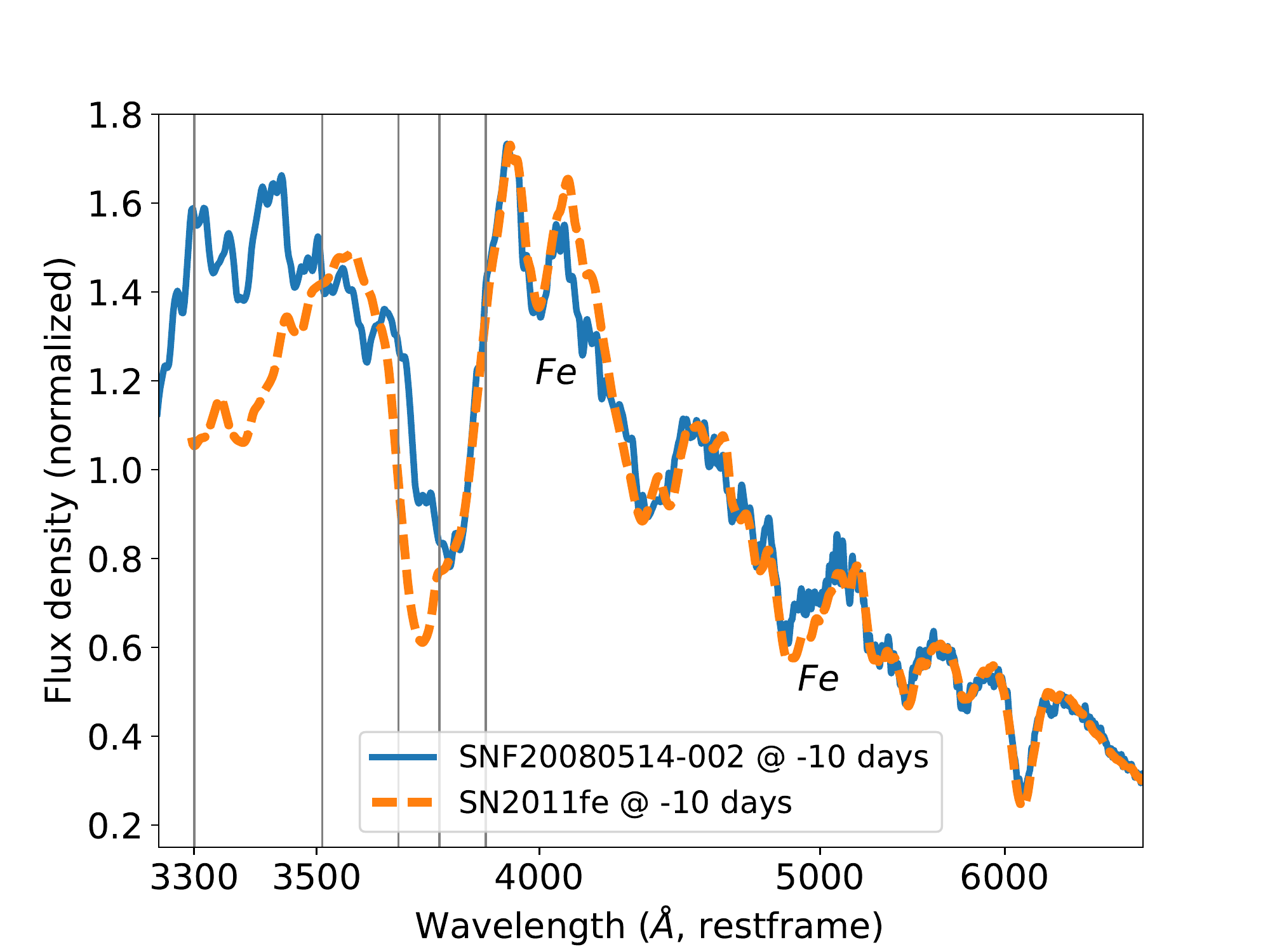}
\includegraphics[angle=0,width=0.49 \textwidth]{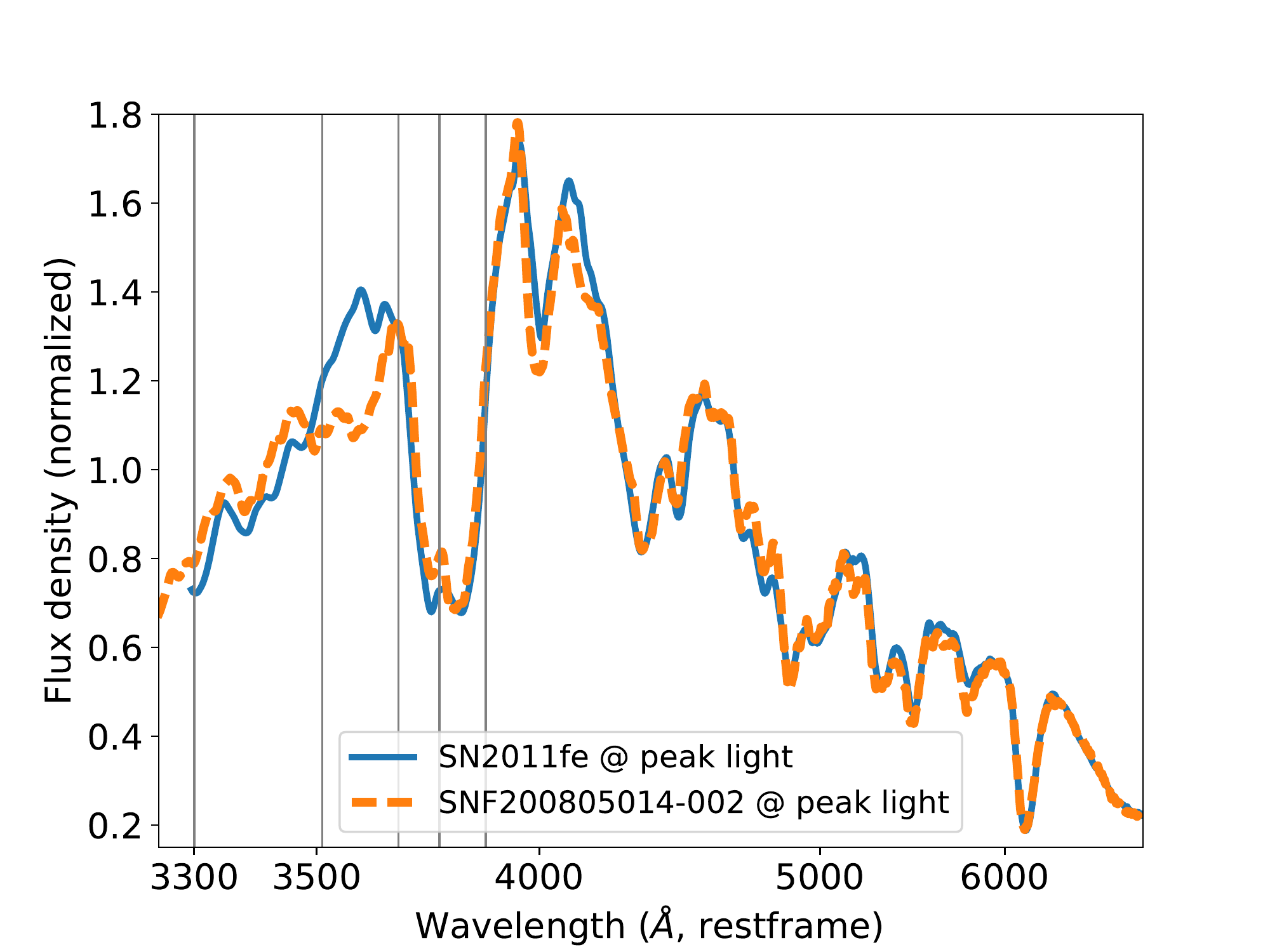}
\caption{ Comparison of the spectra of SN2011fe (\salt $x_1 = -0.4$ and $c=-0.06$) and SNF200805014-002 (\salt $x_1 = -1.5$ and $c=-0.12$) at early (\emph{left}) and peak (\emph{right}) phases. Small spectroscopic differences are found redwards of $4000$~\ang (stable \ion{Fe}{} marked in left panel). Much larger deviation is found bluewards of this limit. The comparison \refcom{at early phases} differs most strongly around $3400$~\AA\xspace and the high velocity edge of the \cahk feature, the comparison \refcom{at later phases} mainly around $3550$~\AA. These differences led to the subdivision of the U-band into four indices, as described in the text. Vertical lines show the feature limits thus defined. 
}
\label{fig:syn_comp}
\end{figure*}

\rfc{Here we examine SN2011fe and SNF20080514-002 as a sample pair of SNe with (relatively) similar $B_{\mathrm{SNf}}V_{\mathrm{SNf}}R_{\mathrm{SNf}}$ spectra and lightcurve properties, but large absorption feature differences in the U-band.}
Their early and peak spectra are compared in Fig.~\ref{fig:syn_comp}.
Spectral differences between these objects can be localized to different behavior in four subdivisions of the U-band \refcom{wavelength region}: \uniw ($3300$--$3510$~\ang), \utiw ($3510$--$3660$~\ang), \usiw ($3660$--$3750$~\ang) and \ucaw ($3750$--$3860$~\ang). 
The two redder regions, roughly covering the \cahk feature, are dominated by \sithree \& \cahkfull absorption. These wavelength regions are thus labeled \usiw ($3660$--$3750$~\AA) and \ucaw ($3750$--$3860$~\AA).
The blue half of the U-band is also divided into two parts. The left (early) panel of Fig.~\ref{fig:syn_comp} shows differences up to $\sim 3500$~\ang, a region labeled \uniw.
The peak spectra (right panel of Fig.~\ref{fig:syn_comp} agree in this region but differ in the subsequent $\sim 3500$ to $\sim 3700$~\ang section, here denoted \utiw. 
The ions most frequently found to dominate these regions are used as labels, but it is clear that absorbing elements will vary with phase and between SNe (see Sec.~\ref{sec:lines} for further discussions). In particular, High Velocity Features dominate variations at early phases in the \utiw and \usiw regions, but have more limited effects at other times and regions (see Sec.~\ref{sec:subset}.)

\begin{figure*}
\includegraphics[angle=0,width=0.99\textwidth,clip, trim=0 0 0.cm 0]{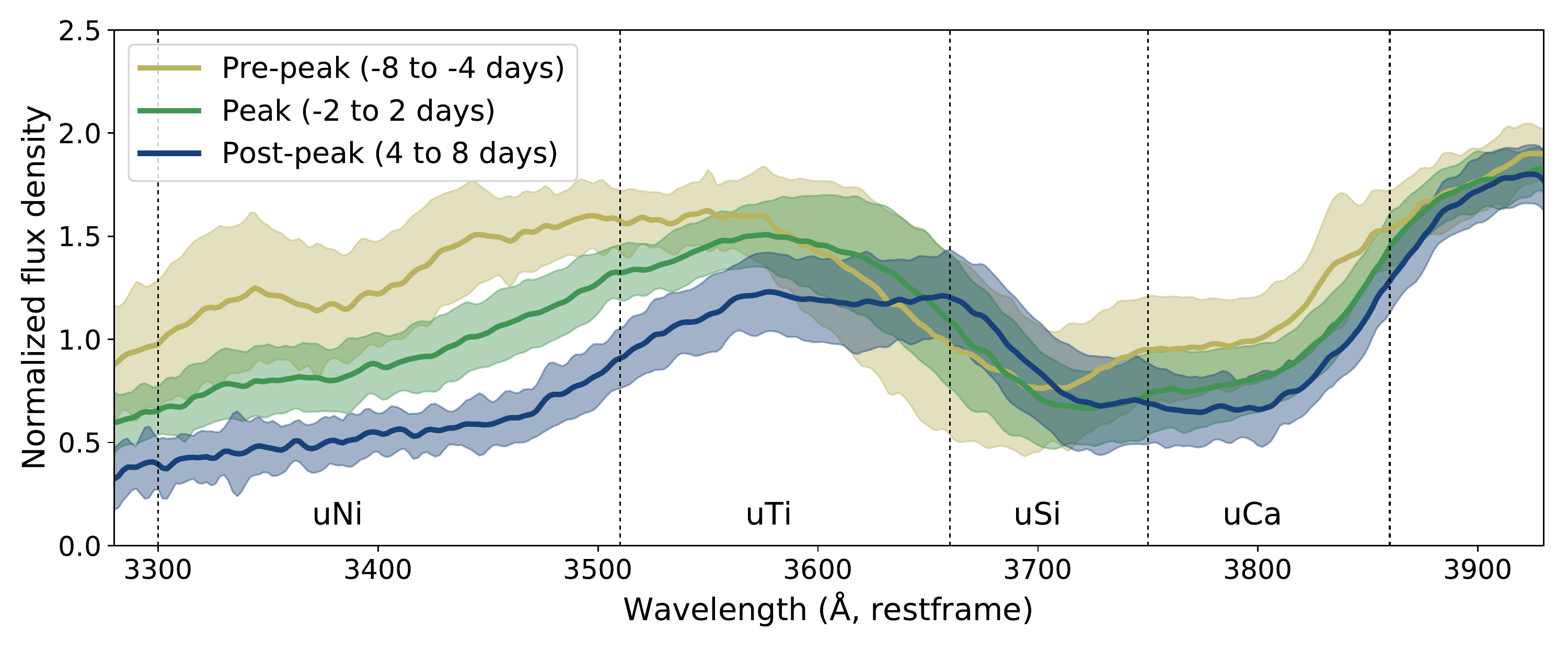}
\caption{Mean spectra and $\pm1$ standard deviation in sample at three representative phases.
\rfctwo{The four wavelength regions considered are separated by dashed lines.}
The interpretation of these indices will be phase dependent, as e.g. the Ti region at early phases is clearly a part of the \cahk + \sithree feature complex and Cr/Fe dominates \uni at late phases.
}
\label{fig:ufeatures}
\end{figure*}

We use the most straightforward and simple method to quantify U-feature variations --
integration of flux within each wavelength region defined above, normalized by the $B_{\mathrm{SNf}}$-band flux at the same phase. We thus determine the spectral index \uni as
  
\begin{equation}\label{eq}
  \mathrm{uNi} = -2.5 \log \left(  \frac{  \int_{3300}^{3510} L_{dered}^{rest}(\lambda) d\lambda }{ \int_{4102}^{5100} L_{dered}^{rest}(\lambda) d\lambda }  \right).
\end{equation}

\rfctwo{The spectral indices} \uti, \usi, and \uca are calculated in the same manner by simply changing the wavelength limits of the numerator. Each \rfctwo{spectral index} is thus a restframe, phase-dependent, dereddened color.
For SNe with multiple spectra within a phase bin, measurements from these spectra are averaged, using the inverse variance from photon counting as weights.
Within each phase bin we find weak dependencies with phase. For each feature/phase combination we fit a linear slope using the full sample and use this to correct individual measurements to the central bin phase ($-6$, $0$ or $6$ days). 

\begin{table*}[!htbp] 
   \centering 
   \caption{SN U-band properties and Ca High Velocity Feature classification. 
   } 
   \resizebox{\textwidth}{!}{ 
   \begin{tabular}{|l|c|c|c|c|c|c|c|c|c|c|c|c|c|}
      \hline 
Name  & \multicolumn{4}{c|}{Phase $-8$ to $-4$}& \multicolumn{4}{c|}{Phase $-2$ to $2$}& \multicolumn{4}{c|}{Phase $4$ to $8$}& Ca HVF\\ [0.1em] 
 & \multicolumn{1}{c}{\uni}& \multicolumn{1}{c}{\usi}& \multicolumn{1}{c}{\uca}& \multicolumn{1}{c|}{\uti}& \multicolumn{1}{c}{\uni}& \multicolumn{1}{c}{\usi}& \multicolumn{1}{c}{\uca}& \multicolumn{1}{c|}{\uti}& \multicolumn{1}{c}{\uni}& \multicolumn{1}{c}{\usi}& \multicolumn{1}{c}{\uca}& \multicolumn{1}{c|}{\uti}& \\ [0.1em] 
\hline  
\hline 
\Tstrut SNF20050728-006 & $1.76 \pm 0.04$  & $3.41 \pm 0.04$  & $2.53 \pm 0.03$  & $1.97 \pm 0.04$  & $2.04 \pm 0.05$  & $3.38 \pm 0.06$  & $2.77 \pm 0.04$  & $1.92 \pm 0.04$  & $2.62 \pm 0.04$  & $3.16 \pm 0.03$  & $3.00 \pm 0.03$  & $2.12 \pm 0.03$  & Y\\ [0.1em] 
SNF20050821-007 & $2.00 \pm 0.04$  & $3.51 \pm 0.04$  & $2.29 \pm 0.03$  & $2.28 \pm 0.04$  & $2.32 \pm 0.04$  & $3.73 \pm 0.04$  & $2.50 \pm 0.03$  & $2.24 \pm 0.04$  & $2.61 \pm 0.04$  & $3.96 \pm 0.04$  & $2.70 \pm 0.03$  & $2.47 \pm 0.04$  & Y\\ [0.1em] 
SNF20051003-004 & $1.57 \pm 0.04$  & $2.57 \pm 0.03$  & $2.44 \pm 0.02$  & $1.96 \pm 0.03$  & --  & --  & --  & --  & $2.65 \pm 0.04$  & $2.79 \pm 0.03$  & $2.90 \pm 0.02$  & $2.15 \pm 0.03$  & --\\ [0.1em] 
SNF20060512-001 & $1.94 \pm 0.04$  & $2.71 \pm 0.03$  & $2.23 \pm 0.02$  & $1.76 \pm 0.03$  & $2.25 \pm 0.04$  & $2.76 \pm 0.03$  & $2.42 \pm 0.02$  & $1.78 \pm 0.03$  & $2.66 \pm 0.04$  & $2.97 \pm 0.03$  & $2.67 \pm 0.02$  & $2.01 \pm 0.03$  & Y\\ [0.1em] 
SNF20060526-003 & $1.75 \pm 0.04$  & $3.03 \pm 0.04$  & $2.46 \pm 0.03$  & $2.09 \pm 0.04$  & $2.14 \pm 0.04$  & $3.04 \pm 0.03$  & $2.71 \pm 0.03$  & $2.01 \pm 0.03$  & --  & --  & --  & --  & Y\\ [0.1em] 
SNF20060609-002 & --  & --  & --  & --  & $1.77 \pm 0.04$  & $2.69 \pm 0.03$  & $2.72 \pm 0.03$  & $1.89 \pm 0.03$  & $2.35 \pm 0.04$  & $2.73 \pm 0.03$  & $2.90 \pm 0.03$  & $2.09 \pm 0.03$  & --\\ [0.1em] 
SNF20060618-023 & $1.32 \pm 0.05$  & $2.41 \pm 0.05$  & $2.15 \pm 0.04$  & $1.45 \pm 0.04$  & --  & --  & --  & --  & $2.48 \pm 0.05$  & $2.76 \pm 0.04$  & $2.76 \pm 0.04$  & $1.99 \pm 0.04$  & --\\ [0.1em] 
SNF20060621-015 & $1.88 \pm 0.04$  & $3.02 \pm 0.03$  & $2.57 \pm 0.02$  & $1.91 \pm 0.03$  & $2.21 \pm 0.04$  & $3.00 \pm 0.03$  & $2.82 \pm 0.02$  & $1.95 \pm 0.03$  & $2.82 \pm 0.04$  & $2.96 \pm 0.03$  & $3.05 \pm 0.03$  & $2.24 \pm 0.03$  & N\\ [0.1em] 
SNF20060907-000 & --  & --  & --  & --  & $2.24 \pm 0.04$  & $3.00 \pm 0.03$  & $2.88 \pm 0.02$  & $2.00 \pm 0.03$  & $2.98 \pm 0.04$  & $3.07 \pm 0.03$  & $3.13 \pm 0.03$  & $2.38 \pm 0.03$  & --\\ [0.1em] 
SNF20060916-002 & $1.59 \pm 0.04$  & $2.87 \pm 0.03$  & $2.46 \pm 0.03$  & $2.00 \pm 0.03$  & --  & --  & --  & --  & $2.54 \pm 0.04$  & $2.89 \pm 0.03$  & $2.90 \pm 0.03$  & $2.17 \pm 0.03$  & N\\ [0.1em] 
SNF20061011-005 & --  & --  & --  & --  & --  & --  & --  & --  & --  & --  & --  & --  & Y\\ [0.1em] 
SNF20061021-003 & $1.67 \pm 0.04$  & $3.19 \pm 0.03$  & $2.66 \pm 0.03$  & $1.90 \pm 0.03$  & $2.05 \pm 0.04$  & $3.20 \pm 0.03$  & $2.94 \pm 0.03$  & $1.92 \pm 0.03$  & --  & --  & --  & --  & N\\ [0.1em] 
SNF20070330-024 & $1.93 \pm 0.04$  & $3.20 \pm 0.04$  & $2.35 \pm 0.03$  & $2.36 \pm 0.04$  & $2.26 \pm 0.04$  & $3.34 \pm 0.03$  & $2.77 \pm 0.03$  & $2.16 \pm 0.03$  & $2.67 \pm 0.04$  & $3.15 \pm 0.03$  & $2.79 \pm 0.02$  & $2.21 \pm 0.03$  & --\\ [0.1em] 
SNF20070420-001 & $1.67 \pm 0.05$  & $3.30 \pm 0.06$  & $2.42 \pm 0.04$  & $2.33 \pm 0.05$  & --  & --  & --  & --  & $2.82 \pm 0.06$  & $3.21 \pm 0.05$  & $2.90 \pm 0.04$  & $2.24 \pm 0.04$  & --\\ [0.1em] 
SNF20070424-003 & $1.86 \pm 0.04$  & $3.42 \pm 0.04$  & $2.52 \pm 0.03$  & $2.09 \pm 0.03$  & $2.23 \pm 0.05$  & $3.23 \pm 0.05$  & $2.80 \pm 0.04$  & $2.06 \pm 0.04$  & $2.84 \pm 0.06$  & $3.04 \pm 0.04$  & $2.95 \pm 0.04$  & $2.34 \pm 0.04$  & Y\\ [0.1em] 
SNF20070506-006 & $1.92 \pm 0.04$  & $2.79 \pm 0.03$  & $2.26 \pm 0.02$  & $1.82 \pm 0.03$  & $2.11 \pm 0.04$  & $2.88 \pm 0.03$  & $2.47 \pm 0.02$  & $1.85 \pm 0.03$  & $2.62 \pm 0.04$  & $3.15 \pm 0.03$  & $2.74 \pm 0.02$  & $2.10 \pm 0.03$  & --\\ [0.1em] 
SNF20070630-006 & --  & --  & --  & --  & $2.07 \pm 0.04$  & $3.28 \pm 0.04$  & $2.73 \pm 0.03$  & $1.97 \pm 0.03$  & $2.66 \pm 0.06$  & $3.20 \pm 0.05$  & $2.94 \pm 0.05$  & $2.23 \pm 0.04$  & Y\\ [0.1em] 
SNF20070712-003 & $1.75 \pm 0.04$  & $2.78 \pm 0.03$  & $2.63 \pm 0.03$  & $1.82 \pm 0.03$  & $2.16 \pm 0.04$  & $2.87 \pm 0.03$  & $2.87 \pm 0.03$  & $1.90 \pm 0.03$  & $2.72 \pm 0.05$  & $2.76 \pm 0.03$  & $3.08 \pm 0.03$  & $2.21 \pm 0.03$  & N\\ [0.1em] 
SNF20070717-003 & $2.03 \pm 0.05$  & $3.50 \pm 0.05$  & $2.71 \pm 0.03$  & $2.01 \pm 0.04$  & $2.48 \pm 0.07$  & $3.20 \pm 0.06$  & $3.03 \pm 0.06$  & $2.17 \pm 0.04$  & $2.98 \pm 0.07$  & $3.10 \pm 0.05$  & $3.17 \pm 0.05$  & $2.38 \pm 0.04$  & N\\ [0.1em] 
SNF20070802-000 & $1.68 \pm 0.04$  & $3.65 \pm 0.03$  & $2.40 \pm 0.02$  & $2.29 \pm 0.03$  & $2.09 \pm 0.04$  & $3.84 \pm 0.04$  & $2.72 \pm 0.03$  & $2.14 \pm 0.03$  & $2.60 \pm 0.08$  & $3.48 \pm 0.08$  & $2.87 \pm 0.05$  & $2.32 \pm 0.05$  & Y\\ [0.1em] 
SNF20070803-005 & $1.53 \pm 0.04$  & $2.29 \pm 0.03$  & $2.10 \pm 0.02$  & $1.73 \pm 0.03$  & $2.02 \pm 0.04$  & $2.39 \pm 0.03$  & $2.27 \pm 0.02$  & $1.84 \pm 0.03$  & $2.64 \pm 0.04$  & $2.39 \pm 0.03$  & $2.45 \pm 0.02$  & $2.06 \pm 0.03$  & --\\ [0.1em] 
SNF20070810-004 & $1.90 \pm 0.04$  & $3.75 \pm 0.05$  & $2.66 \pm 0.03$  & $2.01 \pm 0.03$  & --  & --  & --  & --  & $2.77 \pm 0.05$  & $3.37 \pm 0.04$  & $3.07 \pm 0.04$  & $2.28 \pm 0.04$  & N\\ [0.1em] 
SNF20070817-003 & --  & --  & --  & --  & $2.29 \pm 0.04$  & $3.09 \pm 0.03$  & $2.79 \pm 0.03$  & $2.17 \pm 0.03$  & $3.17 \pm 0.08$  & $3.11 \pm 0.05$  & $3.04 \pm 0.05$  & $2.51 \pm 0.04$  & N\\ [0.1em] 
SNF20070818-001 & --  & --  & --  & --  & $2.24 \pm 0.06$  & $3.59 \pm 0.08$  & $2.71 \pm 0.04$  & $2.30 \pm 0.05$  & $3.13 \pm 0.15$  & $3.42 \pm 0.13$  & $3.00 \pm 0.10$  & $2.47 \pm 0.07$  & --\\ [0.1em] 
SNF20070831-015 & --  & --  & --  & --  & $2.17 \pm 0.04$  & $3.22 \pm 0.03$  & $2.48 \pm 0.03$  & $1.86 \pm 0.03$  & $2.58 \pm 0.04$  & $3.32 \pm 0.04$  & $2.64 \pm 0.03$  & $2.04 \pm 0.03$  & Y\\ [0.1em] 
SNF20070902-018 & --  & --  & --  & --  & $2.18 \pm 0.04$  & $2.84 \pm 0.03$  & $2.82 \pm 0.03$  & $1.98 \pm 0.03$  & --  & --  & --  & --  & --\\ [0.1em] 
SNF20071003-004 & --  & --  & --  & --  & --  & --  & --  & --  & $2.91 \pm 0.07$  & $3.28 \pm 0.05$  & $3.05 \pm 0.04$  & $2.27 \pm 0.04$  & --\\ [0.1em] 
SNF20071021-000 & $1.97 \pm 0.04$  & $3.93 \pm 0.03$  & $2.52 \pm 0.02$  & $2.59 \pm 0.03$  & $2.35 \pm 0.04$  & $3.92 \pm 0.03$  & $2.82 \pm 0.02$  & $2.36 \pm 0.03$  & $2.91 \pm 0.04$  & $3.52 \pm 0.03$  & $2.98 \pm 0.02$  & $2.55 \pm 0.03$  & --\\ [0.1em] 
SNF20080507-000 & $1.78 \pm 0.04$  & $2.79 \pm 0.03$  & $2.25 \pm 0.03$  & $1.69 \pm 0.03$  & $2.24 \pm 0.04$  & $2.96 \pm 0.03$  & $2.50 \pm 0.02$  & $1.84 \pm 0.03$  & $2.75 \pm 0.05$  & $3.22 \pm 0.04$  & $2.69 \pm 0.03$  & $2.10 \pm 0.03$  & --\\ [0.1em] 
SNF20080510-001 & $1.86 \pm 0.04$  & $3.19 \pm 0.04$  & $2.66 \pm 0.03$  & $1.88 \pm 0.03$  & $2.33 \pm 0.04$  & $3.22 \pm 0.04$  & $3.06 \pm 0.04$  & $2.03 \pm 0.03$  & $2.81 \pm 0.04$  & $3.06 \pm 0.03$  & $3.10 \pm 0.03$  & $2.25 \pm 0.03$  & --\\ [0.1em] 
SNF20080512-010 & --  & --  & --  & --  & $2.07 \pm 0.04$  & $3.01 \pm 0.03$  & $2.65 \pm 0.03$  & $2.08 \pm 0.03$  & $2.72 \pm 0.04$  & $2.94 \pm 0.03$  & $2.77 \pm 0.03$  & $2.38 \pm 0.03$  & N\\ [0.1em] 
SNF20080514-002 & $1.75 \pm 0.04$  & $2.78 \pm 0.03$  & $2.63 \pm 0.02$  & $2.12 \pm 0.03$  & $2.19 \pm 0.04$  & $2.85 \pm 0.03$  & $2.80 \pm 0.02$  & $2.25 \pm 0.03$  & $2.87 \pm 0.04$  & $2.91 \pm 0.03$  & $2.95 \pm 0.02$  & $2.60 \pm 0.03$  & --\\ [0.1em] 
SNF20080522-000 & $1.39 \pm 0.04$  & $2.34 \pm 0.03$  & $2.17 \pm 0.02$  & $1.76 \pm 0.03$  & $1.87 \pm 0.04$  & $2.49 \pm 0.03$  & $2.48 \pm 0.02$  & $1.87 \pm 0.03$  & $2.48 \pm 0.04$  & $2.56 \pm 0.03$  & $2.63 \pm 0.02$  & $2.04 \pm 0.03$  & --\\ [0.1em] 
SNF20080522-011 & $1.81 \pm 0.04$  & $3.19 \pm 0.03$  & $2.60 \pm 0.02$  & $2.00 \pm 0.03$  & $2.12 \pm 0.04$  & $3.18 \pm 0.03$  & $2.87 \pm 0.02$  & $1.99 \pm 0.03$  & $2.72 \pm 0.04$  & $3.10 \pm 0.03$  & $3.05 \pm 0.02$  & $2.20 \pm 0.03$  & Y\\ [0.1em] 
SNF20080531-000 & $1.90 \pm 0.04$  & $3.70 \pm 0.03$  & $2.66 \pm 0.02$  & $2.23 \pm 0.03$  & $2.28 \pm 0.04$  & $3.54 \pm 0.03$  & $2.90 \pm 0.02$  & $2.15 \pm 0.03$  & $2.90 \pm 0.04$  & $3.26 \pm 0.03$  & $3.06 \pm 0.03$  & $2.44 \pm 0.03$  & Y\\ [0.1em] 
SNF20080610-000 & --  & --  & --  & --  & $2.35 \pm 0.04$  & $3.13 \pm 0.04$  & $3.03 \pm 0.04$  & $1.99 \pm 0.03$  & $2.94 \pm 0.05$  & $3.11 \pm 0.04$  & $3.18 \pm 0.04$  & $2.26 \pm 0.04$  & --\\ [0.1em] 
SNF20080614-010 & $1.47 \pm 0.05$  & $3.12 \pm 0.06$  & $2.69 \pm 0.05$  & $1.96 \pm 0.04$  & $2.09 \pm 0.05$  & $3.26 \pm 0.05$  & $2.96 \pm 0.04$  & $2.20 \pm 0.04$  & $2.82 \pm 0.06$  & $3.11 \pm 0.05$  & $3.00 \pm 0.05$  & $2.53 \pm 0.05$  & N\\ [0.1em] 
SNF20080620-000 & --  & --  & --  & --  & $2.32 \pm 0.04$  & $3.10 \pm 0.03$  & $2.97 \pm 0.02$  & $2.07 \pm 0.03$  & --  & --  & --  & --  & --\\ [0.1em] 
SNF20080623-001 & $2.05 \pm 0.04$  & $3.64 \pm 0.03$  & $2.60 \pm 0.02$  & $2.30 \pm 0.03$  & $2.29 \pm 0.04$  & $3.52 \pm 0.03$  & $2.93 \pm 0.02$  & $2.10 \pm 0.03$  & $2.94 \pm 0.04$  & $3.37 \pm 0.03$  & $3.13 \pm 0.02$  & $2.40 \pm 0.03$  & Y\\ [0.1em] 
SNF20080626-002 & $1.69 \pm 0.04$  & $3.36 \pm 0.04$  & $2.47 \pm 0.03$  & $2.34 \pm 0.04$  & $2.11 \pm 0.04$  & $3.43 \pm 0.03$  & $2.79 \pm 0.02$  & $2.15 \pm 0.03$  & $2.66 \pm 0.04$  & $3.26 \pm 0.03$  & $2.95 \pm 0.02$  & $2.24 \pm 0.03$  & Y\\ [0.1em] 
SNF20080720-001 & --  & --  & --  & --  & $1.96 \pm 0.04$  & $3.60 \pm 0.03$  & $2.69 \pm 0.02$  & $2.12 \pm 0.03$  & $2.57 \pm 0.04$  & $3.31 \pm 0.03$  & $2.89 \pm 0.02$  & $2.25 \pm 0.03$  & Y\\ [0.1em] 
SNF20080725-004 & $1.88 \pm 0.04$  & $3.48 \pm 0.03$  & $2.42 \pm 0.03$  & $2.14 \pm 0.03$  & $2.19 \pm 0.04$  & $3.56 \pm 0.03$  & $2.66 \pm 0.03$  & $2.08 \pm 0.03$  & $2.78 \pm 0.06$  & $3.53 \pm 0.06$  & $2.97 \pm 0.04$  & $2.25 \pm 0.04$  & Y\\ [0.1em] 
SNF20080803-000 & $1.74 \pm 0.07$  & $3.42 \pm 0.11$  & $2.51 \pm 0.06$  & $2.00 \pm 0.06$  & $1.93 \pm 0.04$  & $3.18 \pm 0.04$  & $2.63 \pm 0.03$  & $1.88 \pm 0.03$  & $2.84 \pm 0.11$  & $3.05 \pm 0.07$  & $2.84 \pm 0.07$  & $2.13 \pm 0.05$  & --\\ [0.1em] 
SNF20080810-001 & $1.61 \pm 0.04$  & $2.69 \pm 0.03$  & $2.62 \pm 0.03$  & $1.84 \pm 0.03$  & $2.10 \pm 0.04$  & $2.74 \pm 0.03$  & $2.82 \pm 0.02$  & $1.98 \pm 0.03$  & $2.88 \pm 0.05$  & $2.80 \pm 0.03$  & $2.94 \pm 0.03$  & $2.44 \pm 0.03$  & N\\ [0.1em] 
SNF20080821-000 & $1.54 \pm 0.04$  & $3.09 \pm 0.04$  & $2.42 \pm 0.03$  & $1.92 \pm 0.03$  & $1.94 \pm 0.04$  & $3.05 \pm 0.03$  & $2.70 \pm 0.03$  & $1.95 \pm 0.03$  & $2.55 \pm 0.05$  & $3.01 \pm 0.04$  & $2.90 \pm 0.03$  & $2.09 \pm 0.03$  & N\\ [0.1em] 
SNF20080825-010 & $1.65 \pm 0.04$  & $3.00 \pm 0.03$  & $2.58 \pm 0.02$  & $1.83 \pm 0.03$  & $2.06 \pm 0.04$  & $3.09 \pm 0.03$  & $2.78 \pm 0.02$  & $1.95 \pm 0.03$  & $2.74 \pm 0.04$  & $3.00 \pm 0.03$  & $2.90 \pm 0.03$  & $2.32 \pm 0.03$  & --\\ [0.1em] 
SNF20080908-000 & $2.04 \pm 0.04$  & $3.41 \pm 0.03$  & $2.61 \pm 0.02$  & $2.10 \pm 0.03$  & --  & --  & --  & --  & $2.96 \pm 0.04$  & $3.10 \pm 0.03$  & $3.07 \pm 0.03$  & $2.31 \pm 0.03$  & Y\\ [0.1em] 
SNF20080909-030 & $2.05 \pm 0.04$  & $2.76 \pm 0.03$  & $2.33 \pm 0.02$  & $1.80 \pm 0.03$  & $2.27 \pm 0.04$  & $2.80 \pm 0.03$  & $2.47 \pm 0.02$  & $1.83 \pm 0.03$  & --  & --  & --  & --  & Y\\ [0.1em] 
SNF20080913-031 & --  & --  & --  & --  & $2.58 \pm 0.04$  & $3.32 \pm 0.03$  & $2.67 \pm 0.03$  & $1.96 \pm 0.03$  & --  & --  & --  & --  & Y\\ [0.1em] 
SNF20080914-001 & $1.82 \pm 0.04$  & $3.49 \pm 0.03$  & $2.60 \pm 0.02$  & $1.82 \pm 0.03$  & $2.06 \pm 0.04$  & $3.27 \pm 0.03$  & $2.77 \pm 0.02$  & $1.95 \pm 0.03$  & --  & --  & --  & --  & --\\ [0.1em] 
SNF20080918-000 & --  & --  & --  & --  & $1.83 \pm 0.04$  & $3.60 \pm 0.04$  & $2.53 \pm 0.03$  & $2.07 \pm 0.03$  & --  & --  & --  & --  & Y\\ [0.1em] 
SNF20080918-004 & $1.97 \pm 0.04$  & $3.22 \pm 0.03$  & $2.71 \pm 0.03$  & $2.17 \pm 0.03$  & $2.48 \pm 0.04$  & $2.97 \pm 0.03$  & $2.92 \pm 0.03$  & $2.20 \pm 0.03$  & --  & --  & --  & --  & N\\ [0.1em] 
SNF20080919-000 & $1.48 \pm 0.04$  & $3.15 \pm 0.04$  & $2.40 \pm 0.03$  & $1.94 \pm 0.04$  & $1.80 \pm 0.04$  & $3.14 \pm 0.03$  & $2.63 \pm 0.03$  & $1.93 \pm 0.03$  & --  & --  & --  & --  & Y\\ [0.1em] 
SNF20080919-001 & $1.85 \pm 0.04$  & $2.62 \pm 0.03$  & $2.24 \pm 0.02$  & $1.65 \pm 0.03$  & $2.27 \pm 0.04$  & $2.71 \pm 0.03$  & $2.50 \pm 0.02$  & $1.75 \pm 0.03$  & $2.66 \pm 0.04$  & $2.90 \pm 0.03$  & $2.77 \pm 0.02$  & $2.05 \pm 0.03$  & N\\ [0.1em] 
SNF20080919-002 & $1.58 \pm 0.04$  & $2.93 \pm 0.04$  & $2.83 \pm 0.04$  & $1.88 \pm 0.04$  & --  & --  & --  & --  & --  & --  & --  & --  & N\\ [0.1em] 
SNF20080920-000 & --  & --  & --  & --  & $1.98 \pm 0.04$  & $3.51 \pm 0.03$  & $2.59 \pm 0.02$  & $2.16 \pm 0.03$  & --  & --  & --  & --  & Y\\ [0.1em] 
SNF20080926-009 & $2.00 \pm 0.04$  & $3.34 \pm 0.04$  & $2.60 \pm 0.03$  & $1.85 \pm 0.03$  & --  & --  & --  & --  & $2.64 \pm 0.05$  & $3.33 \pm 0.04$  & $2.95 \pm 0.03$  & $2.17 \pm 0.03$  & --\\ [0.1em] 
SN2004ef & $1.53 \pm 0.04$  & $3.48 \pm 0.03$  & $2.24 \pm 0.02$  & $2.48 \pm 0.03$  & $2.12 \pm 0.04$  & $3.73 \pm 0.03$  & $2.61 \pm 0.02$  & $2.29 \pm 0.03$  & $2.73 \pm 0.04$  & $3.52 \pm 0.03$  & $2.81 \pm 0.02$  & $2.57 \pm 0.03$  & Y\\ [0.1em] 
SN2005cf & $1.98 \pm 0.04$  & $3.59 \pm 0.03$  & $2.41 \pm 0.02$  & $2.20 \pm 0.03$  & --  & --  & --  & --  & --  & --  & --  & --  & Y\\ [0.1em] 
SN2005el & $1.75 \pm 0.04$  & $3.20 \pm 0.03$  & $2.61 \pm 0.02$  & $2.10 \pm 0.03$  & $2.27 \pm 0.04$  & $3.26 \pm 0.03$  & $2.86 \pm 0.02$  & $2.28 \pm 0.03$  & $2.87 \pm 0.04$  & $3.12 \pm 0.03$  & $2.96 \pm 0.02$  & $2.47 \pm 0.03$  & N\\ [0.1em] 
SN2005hc & --  & --  & --  & --  & $2.09 \pm 0.04$  & $3.36 \pm 0.03$  & $2.51 \pm 0.02$  & $2.08 \pm 0.03$  & $2.63 \pm 0.04$  & $3.30 \pm 0.03$  & $2.72 \pm 0.03$  & $2.30 \pm 0.03$  & --\\ [0.1em] 
SN2005hj & --  & --  & --  & --  & $1.73 \pm 0.04$  & $2.62 \pm 0.03$  & $2.50 \pm 0.03$  & $1.81 \pm 0.03$  & $2.30 \pm 0.04$  & $2.67 \pm 0.03$  & $2.74 \pm 0.03$  & $1.91 \pm 0.03$  & --\\ [0.1em] 
SN2005ir & --  & --  & --  & --  & $1.96 \pm 0.05$  & $3.35 \pm 0.05$  & $2.64 \pm 0.04$  & $2.05 \pm 0.04$  & --  & --  & --  & --  & N\\ [0.1em] 
SN2006X & --  & $3.60 \pm 0.03$  & $2.26 \pm 0.03$  & $2.90 \pm 0.04$  & --  & --  & --  & --  & --  & $3.70 \pm 0.03$  & $2.58 \pm 0.03$  & $2.43 \pm 0.04$  & --\\ [0.1em] 
SN2006cj & $1.60 \pm 0.04$  & $2.83 \pm 0.03$  & $2.48 \pm 0.03$  & $1.85 \pm 0.03$  & $1.97 \pm 0.04$  & $2.90 \pm 0.03$  & $2.79 \pm 0.02$  & $1.93 \pm 0.03$  & $2.59 \pm 0.04$  & $2.91 \pm 0.03$  & $2.95 \pm 0.03$  & $2.13 \pm 0.03$  & N\\ [0.1em] 
SN2006dm & $1.84 \pm 0.04$  & $3.16 \pm 0.03$  & $2.72 \pm 0.03$  & $2.15 \pm 0.04$  & $2.33 \pm 0.04$  & $3.09 \pm 0.03$  & $2.92 \pm 0.03$  & $2.23 \pm 0.03$  & $2.93 \pm 0.04$  & $3.09 \pm 0.03$  & $3.01 \pm 0.03$  & $2.75 \pm 0.04$  & N\\ [0.1em] 
SN2006do & --  & --  & --  & --  & $2.26 \pm 0.04$  & $3.24 \pm 0.03$  & $2.62 \pm 0.03$  & $2.32 \pm 0.04$  & $2.97 \pm 0.04$  & $3.23 \pm 0.03$  & $2.85 \pm 0.03$  & $2.64 \pm 0.04$  & N\\ [0.1em] 
SN2006ob & --  & --  & --  & --  & $2.18 \pm 0.05$  & $2.97 \pm 0.04$  & $2.75 \pm 0.03$  & $2.04 \pm 0.04$  & --  & --  & --  & --  & N\\ [0.1em] 
SN2007bd & --  & --  & --  & --  & $2.12 \pm 0.04$  & $3.39 \pm 0.03$  & $2.80 \pm 0.02$  & $2.15 \pm 0.03$  & $2.75 \pm 0.04$  & $3.21 \pm 0.03$  & $2.93 \pm 0.02$  & $2.32 \pm 0.03$  & N\\ [0.1em] 
SN2007le & $1.47 \pm 0.04$  & $3.47 \pm 0.03$  & $2.35 \pm 0.02$  & $2.41 \pm 0.03$  & --  & --  & --  & --  & $2.55 \pm 0.04$  & $3.30 \pm 0.03$  & $2.80 \pm 0.03$  & $2.25 \pm 0.03$  & --\\ [0.1em] 
SN2008ec & $1.48 \pm 0.04$  & $3.04 \pm 0.03$  & $2.70 \pm 0.02$  & $1.84 \pm 0.03$  & $1.96 \pm 0.04$  & $3.06 \pm 0.03$  & $2.92 \pm 0.02$  & $1.98 \pm 0.03$  & $2.77 \pm 0.04$  & $2.97 \pm 0.03$  & $3.01 \pm 0.02$  & $2.37 \pm 0.03$  & N\\ [0.1em] 
SN2010dt & $2.09 \pm 0.04$  & $3.22 \pm 0.03$  & $2.73 \pm 0.02$  & $1.92 \pm 0.03$  & $2.46 \pm 0.04$  & $3.13 \pm 0.03$  & $3.02 \pm 0.02$  & $2.00 \pm 0.03$  & $3.01 \pm 0.04$  & $3.02 \pm 0.03$  & $3.20 \pm 0.02$  & $2.37 \pm 0.03$  & --\\ [0.1em] 
SN2011bc & $1.97 \pm 0.04$  & $3.26 \pm 0.03$  & $2.31 \pm 0.02$  & $1.90 \pm 0.03$  & $2.19 \pm 0.04$  & $3.42 \pm 0.03$  & $2.52 \pm 0.02$  & $1.92 \pm 0.03$  & $2.72 \pm 0.04$  & $3.58 \pm 0.03$  & $2.81 \pm 0.02$  & $2.16 \pm 0.03$  & Y\\ [0.1em] 
SN2011be & $1.65 \pm 0.04$  & $2.74 \pm 0.03$  & $2.51 \pm 0.02$  & $1.90 \pm 0.03$  & $1.99 \pm 0.04$  & $2.81 \pm 0.03$  & $2.71 \pm 0.02$  & $1.96 \pm 0.03$  & --  & --  & --  & --  & N\\ [0.1em]
PTF09dlc & $1.86 \pm 0.04$  & $3.52 \pm 0.03$  & $2.38 \pm 0.02$  & $2.03 \pm 0.03$  & $2.19 \pm 0.04$  & $3.60 \pm 0.03$  & $2.64 \pm 0.02$  & $2.08 \pm 0.03$  & $2.74 \pm 0.04$  & $3.38 \pm 0.03$  & $2.87 \pm 0.02$  & $2.35 \pm 0.03$  & Y\\ [0.1em] 
PTF09dnl & $1.57 \pm 0.04$  & $3.17 \pm 0.03$  & $2.19 \pm 0.02$  & $2.37 \pm 0.03$  & $2.00 \pm 0.04$  & $3.27 \pm 0.02$  & $2.46 \pm 0.02$  & $2.22 \pm 0.03$  & $2.58 \pm 0.04$  & $3.19 \pm 0.03$  & $2.62 \pm 0.02$  & $2.38 \pm 0.03$  & Y\\ [0.1em] 
PTF09fox & $1.71 \pm 0.04$  & $3.21 \pm 0.03$  & $2.54 \pm 0.03$  & $2.07 \pm 0.03$  & $2.10 \pm 0.04$  & $3.20 \pm 0.03$  & $2.82 \pm 0.03$  & $1.98 \pm 0.03$  & $2.76 \pm 0.04$  & $3.11 \pm 0.03$  & $3.00 \pm 0.03$  & $2.16 \pm 0.03$  & N\\ [0.1em] 
PTF09foz & $1.92 \pm 0.04$  & $3.28 \pm 0.03$  & $2.54 \pm 0.03$  & $1.97 \pm 0.03$  & $2.32 \pm 0.04$  & $3.25 \pm 0.03$  & $2.77 \pm 0.03$  & $2.00 \pm 0.03$  & $3.15 \pm 0.04$  & $3.26 \pm 0.03$  & $3.03 \pm 0.03$  & $2.50 \pm 0.03$  & --\\ [0.1em] 
PTF10hmv & --  & --  & --  & --  & $2.03 \pm 0.04$  & $3.02 \pm 0.03$  & $2.59 \pm 0.02$  & $1.81 \pm 0.03$  & $2.49 \pm 0.04$  & $3.21 \pm 0.03$  & $2.84 \pm 0.02$  & $2.06 \pm 0.03$  & Y\\ [0.1em] 
PTF10icb & $1.62 \pm 0.04$  & $2.80 \pm 0.03$  & $2.55 \pm 0.02$  & $1.81 \pm 0.03$  & $2.01 \pm 0.04$  & $2.79 \pm 0.03$  & $2.81 \pm 0.02$  & $1.89 \pm 0.03$  & $2.71 \pm 0.04$  & $2.82 \pm 0.03$  & $3.03 \pm 0.02$  & $2.17 \pm 0.03$  & N\\ [0.1em] 
PTF10mwb & $1.85 \pm 0.04$  & $3.02 \pm 0.03$  & $2.76 \pm 0.02$  & $1.91 \pm 0.03$  & $2.31 \pm 0.04$  & $3.05 \pm 0.03$  & $2.96 \pm 0.02$  & $2.06 \pm 0.03$  & $3.02 \pm 0.04$  & $3.00 \pm 0.03$  & $3.15 \pm 0.02$  & $2.49 \pm 0.03$  & N\\ [0.1em] 
PTF10ndc & --  & --  & --  & --  & $1.96 \pm 0.04$  & $3.25 \pm 0.03$  & $2.69 \pm 0.02$  & $2.04 \pm 0.03$  & $2.56 \pm 0.04$  & $3.12 \pm 0.03$  & $2.87 \pm 0.03$  & $2.12 \pm 0.03$  & N\\ [0.1em] 
PTF10qjl & $1.94 \pm 0.04$  & $3.22 \pm 0.03$  & $2.41 \pm 0.02$  & $2.06 \pm 0.03$  & --  & --  & --  & --  & $2.67 \pm 0.04$  & $3.30 \pm 0.03$  & $2.89 \pm 0.03$  & $2.23 \pm 0.03$  & Y\\ [0.1em] 
PTF10qjq & $1.46 \pm 0.04$  & $2.52 \pm 0.03$  & $2.45 \pm 0.02$  & $1.81 \pm 0.03$  & $1.94 \pm 0.04$  & $2.66 \pm 0.03$  & $2.64 \pm 0.02$  & $1.97 \pm 0.03$  & $2.64 \pm 0.04$  & $2.70 \pm 0.03$  & $2.81 \pm 0.02$  & $2.23 \pm 0.03$  & --\\ [0.1em] 
PTF10qyz & $1.84 \pm 0.05$  & $3.37 \pm 0.07$  & $2.59 \pm 0.04$  & $2.12 \pm 0.04$  & $2.42 \pm 0.04$  & $3.13 \pm 0.03$  & $2.84 \pm 0.02$  & $2.25 \pm 0.03$  & $3.26 \pm 0.04$  & $3.06 \pm 0.03$  & $3.01 \pm 0.03$  & $2.59 \pm 0.03$  & --\\ [0.1em] 
PTF10tce & $1.79 \pm 0.04$  & $3.37 \pm 0.03$  & $2.26 \pm 0.02$  & $2.19 \pm 0.03$  & $2.20 \pm 0.04$  & $3.52 \pm 0.03$  & $2.57 \pm 0.02$  & $2.10 \pm 0.03$  & $2.71 \pm 0.04$  & $3.36 \pm 0.03$  & $2.75 \pm 0.02$  & $2.33 \pm 0.03$  & Y\\ [0.1em] 
PTF10ufj & $1.93 \pm 0.04$  & $3.46 \pm 0.03$  & $2.52 \pm 0.03$  & $2.39 \pm 0.03$  & $2.32 \pm 0.04$  & $3.38 \pm 0.04$  & $2.84 \pm 0.03$  & $2.22 \pm 0.03$  & $2.92 \pm 0.05$  & $3.29 \pm 0.04$  & $3.07 \pm 0.04$  & $2.38 \pm 0.03$  & Y\\ [0.1em] 
PTF10wnm & $1.56 \pm 0.04$  & $2.86 \pm 0.03$  & $2.50 \pm 0.02$  & $1.91 \pm 0.03$  & $1.91 \pm 0.04$  & $2.93 \pm 0.03$  & $2.73 \pm 0.03$  & $1.92 \pm 0.03$  & $2.62 \pm 0.04$  & $2.91 \pm 0.03$  & $2.97 \pm 0.03$  & $2.11 \pm 0.03$  & N\\ [0.1em] 
PTF10wof & $2.05 \pm 0.04$  & $3.57 \pm 0.03$  & $2.64 \pm 0.03$  & $2.12 \pm 0.03$  & $2.37 \pm 0.04$  & $3.48 \pm 0.03$  & $2.86 \pm 0.02$  & $2.08 \pm 0.03$  & $2.94 \pm 0.04$  & $3.32 \pm 0.03$  & $3.04 \pm 0.03$  & $2.39 \pm 0.03$  & N\\ [0.1em] 
PTF10xyt & $1.76 \pm 0.05$  & $3.26 \pm 0.06$  & $2.50 \pm 0.04$  & $2.09 \pm 0.04$  & $2.47 \pm 0.06$  & $3.32 \pm 0.06$  & $2.81 \pm 0.05$  & $2.07 \pm 0.04$  & --  & --  & --  & --  & Y\\ [0.1em] 
PTF10zdk & $2.45 \pm 0.04$  & $3.71 \pm 0.03$  & $2.49 \pm 0.02$  & $2.32 \pm 0.03$  & $2.39 \pm 0.04$  & $3.81 \pm 0.03$  & $2.70 \pm 0.02$  & $2.09 \pm 0.03$  & --  & --  & --  & --  & Y\\ [0.1em]
      \hline 
   \end{tabular} }
   \tablefoot{See \ref{sec:meas}\xspace for parameter definitions and \ref{sec:subset} for HVF classification.}
   \label{tab:snprops} 
\end{table*} 


These measurements of \uni, \uti, \usi and \uca can be found in Table~\ref{tab:snprops}, \rfctwo{which also identifies whether any \cahk HVF is detected based on visual inspection for each SN (see Sec.\ref{sec:subset}).}

\section{U-band variation}\label{sec:uvary}

\subsection{Mean and variation of the U-band spectrum}

\begin{figure*}
  \includegraphics[angle=0,width=0.99\textwidth,clip, trim=1.cm 0 3.cm 0]{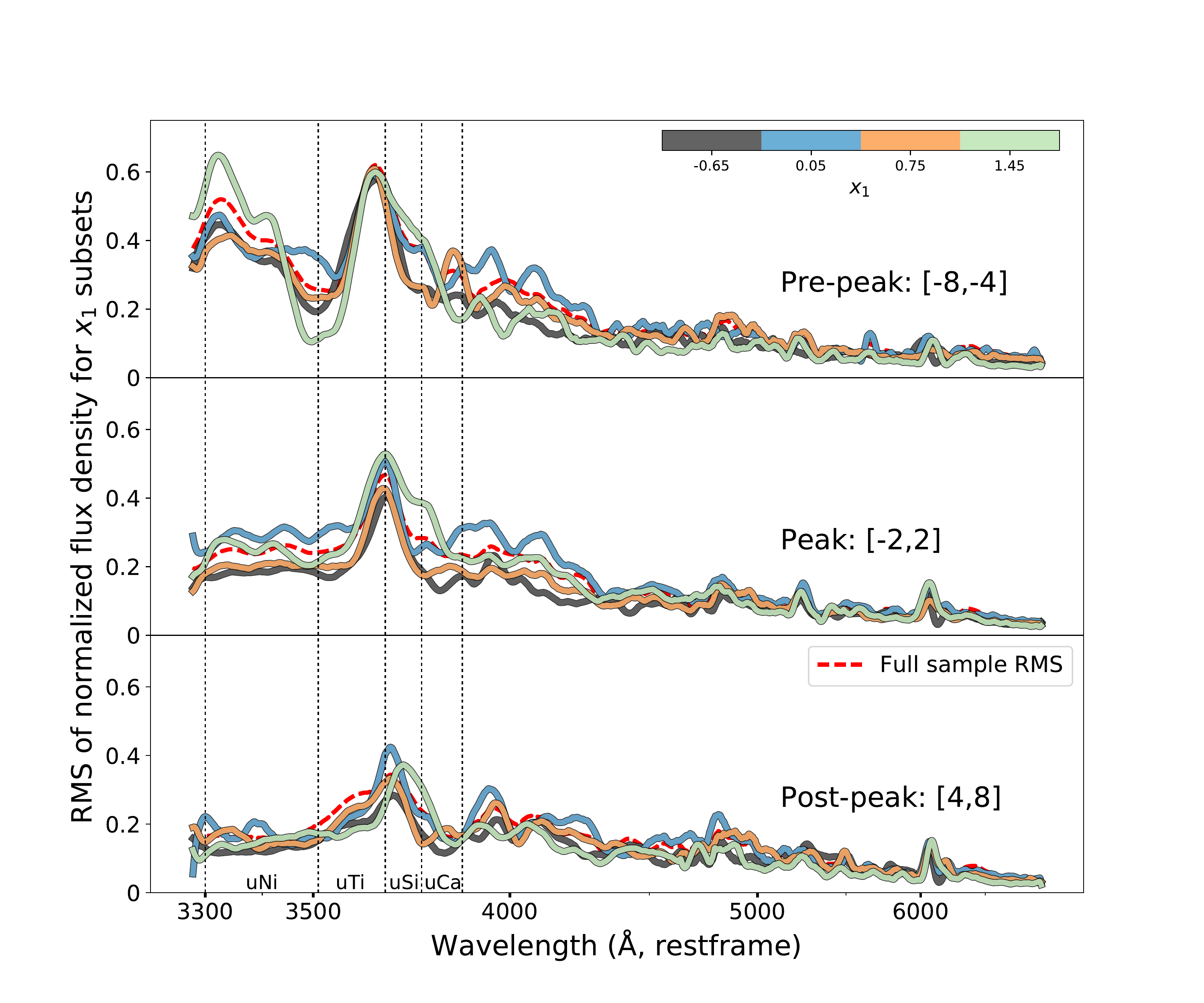}
  \caption{Intrinsic flux RMS (vs. wavelength) for SNe after division into four bins according to increasing lightcurve width (dark to light lines). Spectra were initially normalized to have median flux of unity over the $[3300,6900]$~\ang wavelength region.
 Panels from top to bottom show the three sample phase regions (pre-peak, peak and post-peak). Vertical (dotted) lines mark boundaries of the four U-band regions. The red dashed red line shows the RMS for the full sample.}
\label{fig:urms_x1bin}
\end{figure*}

The mean dereddened spectrum, and the $1\sigma$ sample variation, for each phase bin after dereddening and normalizing to a common median $B_{\mathrm{SNf}}$-band flux, can be seen in Fig.~\ref{fig:ufeatures}. The large variation in the \cahkfeat feature is obvious at pre-peak phases. Also, the bluest region (\uniw) shows large early variation. At roughly one week after peak, the dispersion has significantly decreased at all wavelengths. 
\rfctwo{This RMS vs wavelength is directly displayed as the dashed red line in Fig.~\ref{fig:urms_x1bin} \rtr{with normalization to a wider wavelength range.} Here, we also show the dispersion of subsets of the sample, divided according to the \salt $x_1$ parameter \rtr{and with the mean recalculated for each subset.} All subsets show scatter similar to that of the full sample, with the possible exception of the post-peak \utiw region.
The variability in the post-peak phase bin can be described as a smooth component, only slowly declining with wavelength, with Si line variability superimposed. The \uniw RMS here is $0.16$~mag, comparable to that of the $B_{\mathrm{SNf}}$-band ($0.14$~mag).
}

\subsection{Spectroscopic variations with common \sn properties}\label{sec:composite}

\begin{figure*}
  \includegraphics[angle=0,width=0.32 \textwidth]{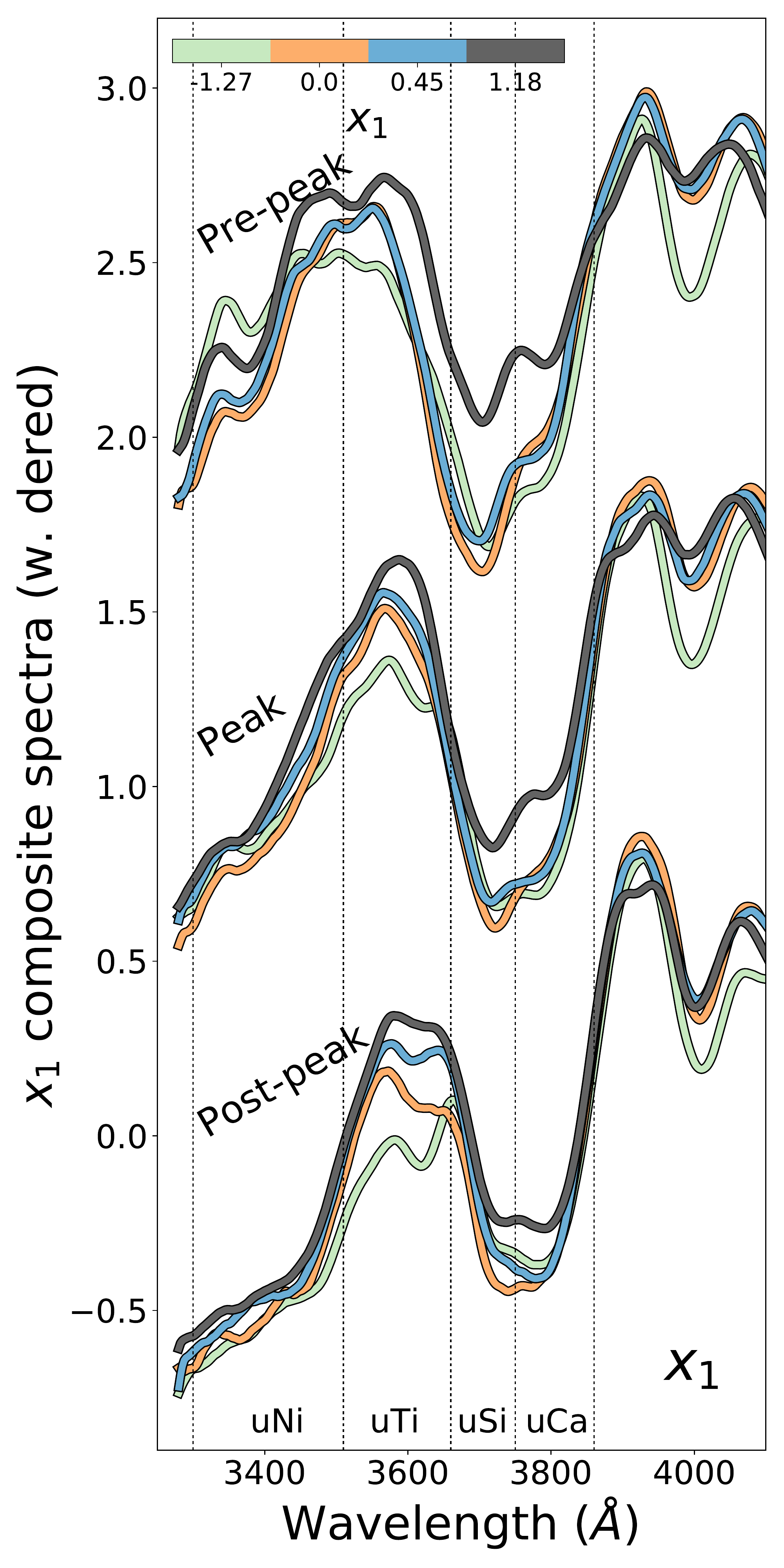}
  \includegraphics[angle=0,width=0.32 \textwidth]{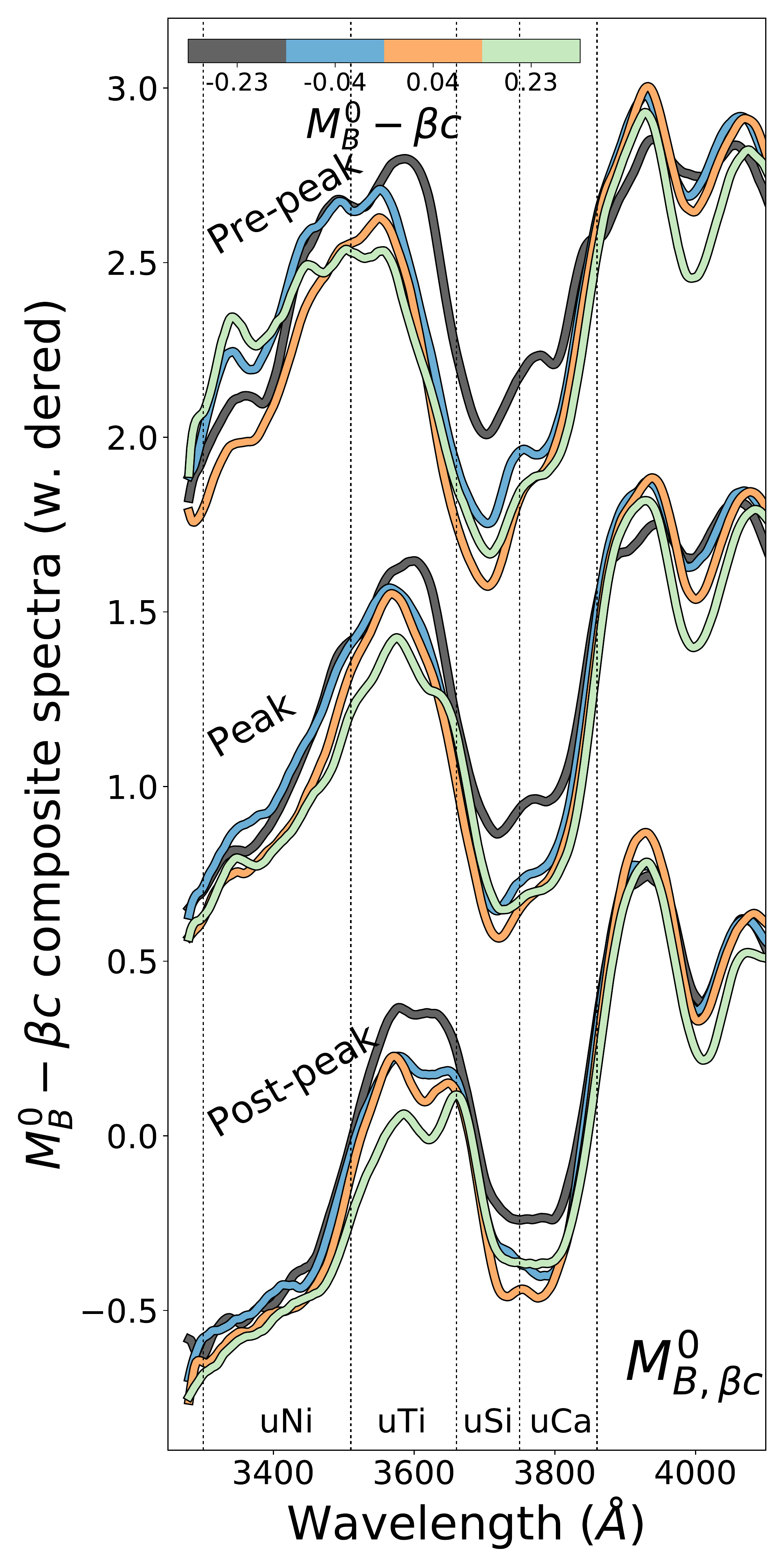}
  \includegraphics[angle=0,width=0.32 \textwidth]{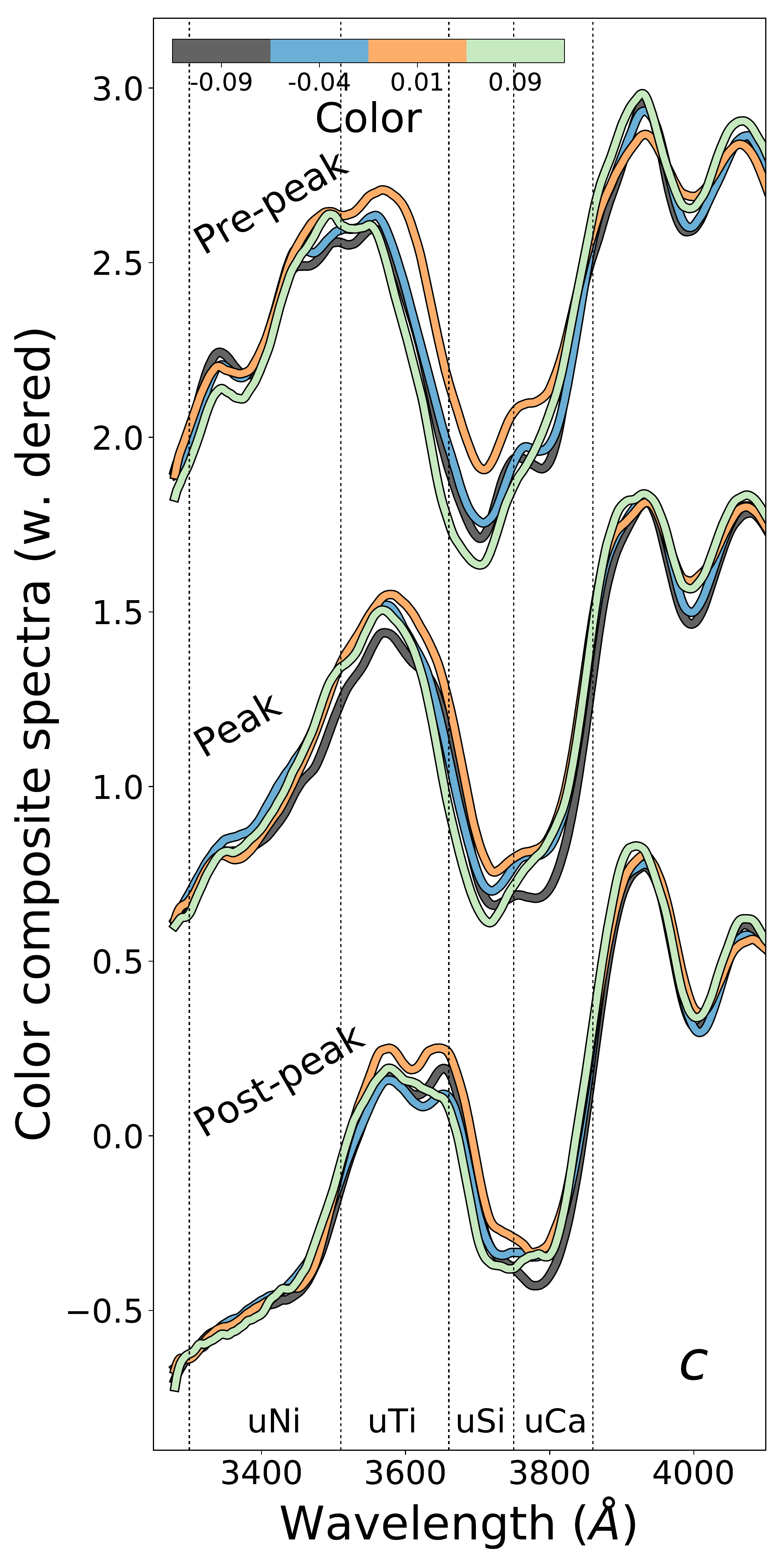}
  \caption{U-band spectroscopic variation spanning the range of common SN properties: $x_1$, dereddened $M_B$, $\mathrm{Color}$ (left to right).
    \rfctwo{Composite spectra were calculated after dividing the sample into four quartiles based on each property, which was repeated at three representative phase ranges ($-6$, $0$, $6$) for each subset.} Subset composites were drawn such that the line color becomes darker as the parameter value decreases. Dotted lines indicate U-band spectral-index sub division boundaries.}
\label{fig:specvary0}
\end{figure*}

\begin{figure*}
  \includegraphics[angle=0,width=0.32 \textwidth]{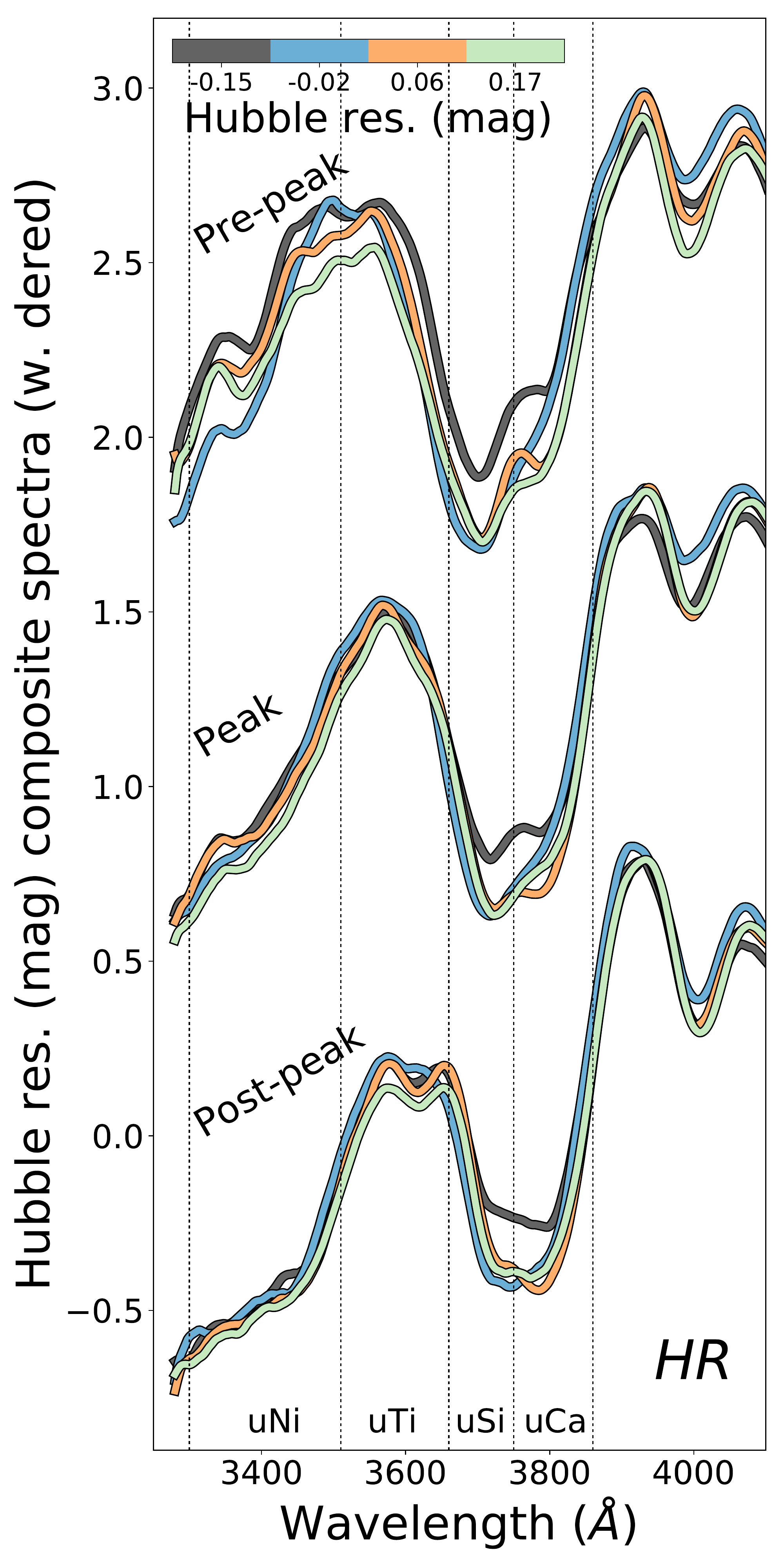}
  \includegraphics[angle=0,width=0.32 \textwidth]{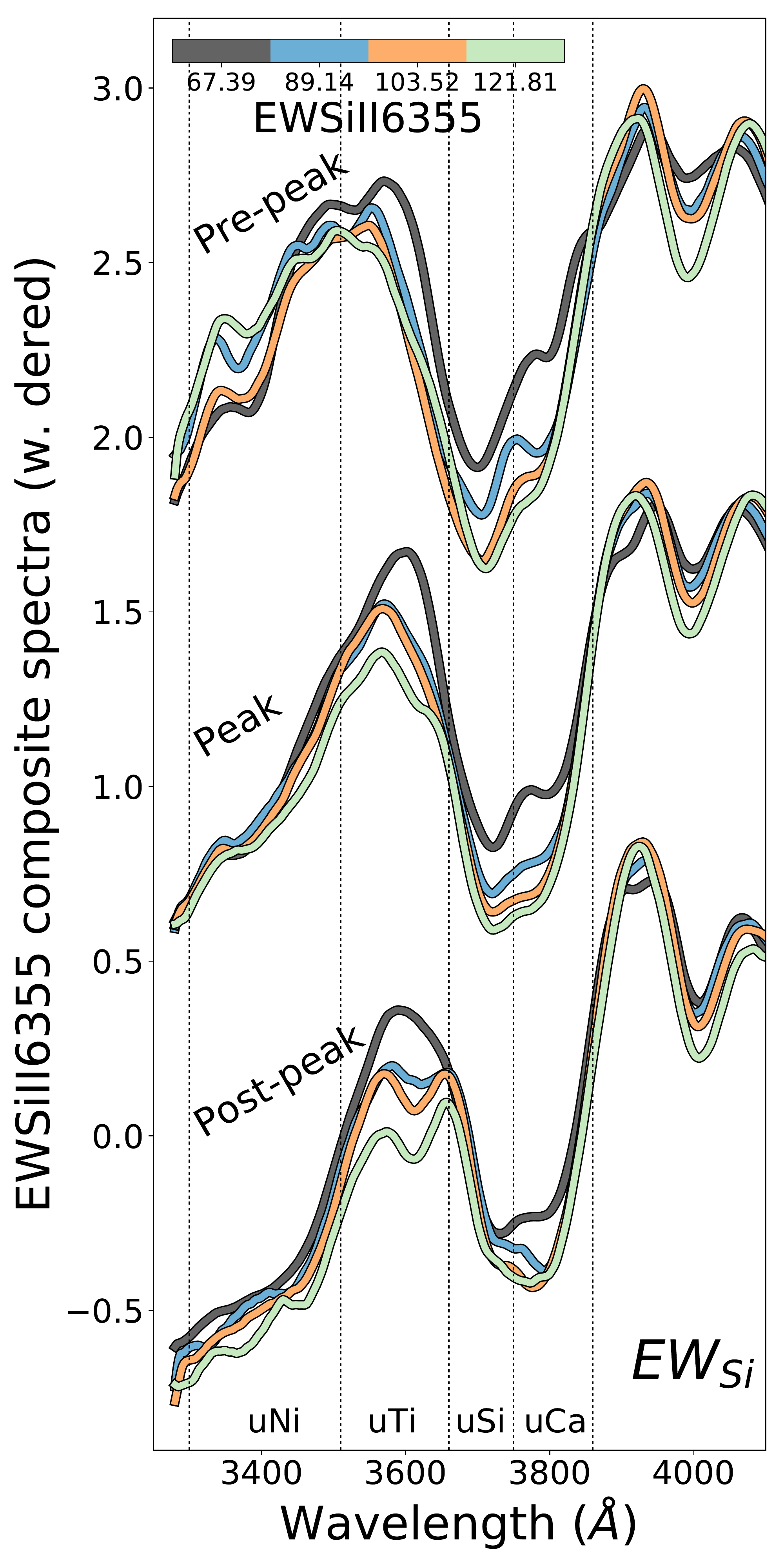}  
  \includegraphics[angle=0,width=0.32 \textwidth]{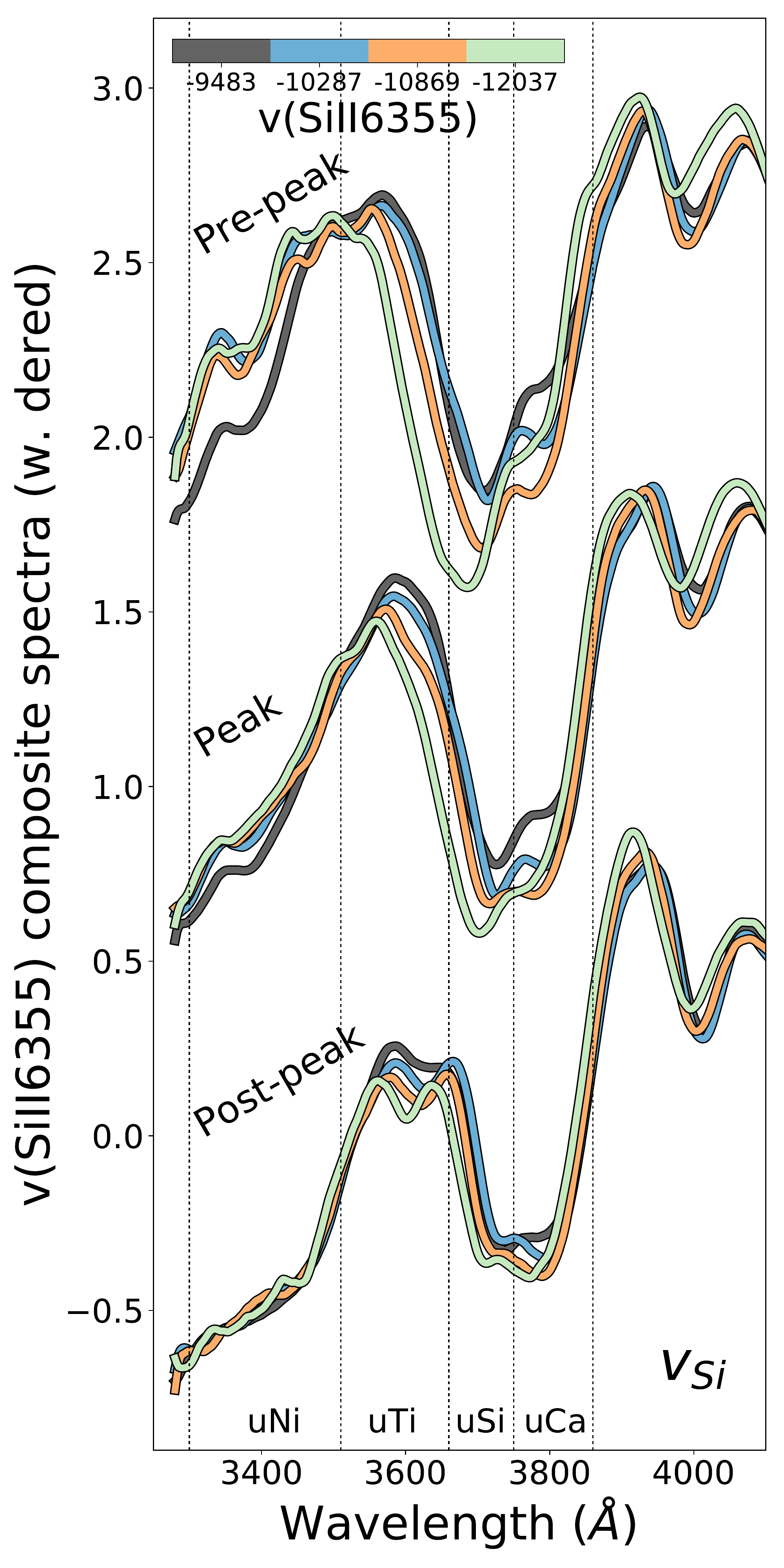}
  \caption{U-band spectroscopic variation spanning the range of common SN properties: Hubble residuals standardized by \salt lightcurve parameters, \ewsisix and \vsisix (left to right).
    \rfctwo{Composite spectra were calculated after dividing the sample into four quartiles based on each property, which was repeated at three representative phase ranges ($-6$, $0$, $6$) for each subset.} Subset composites were drawn such that the line color becomes darker as the parameter value decreases. Dotted lines indicate U-band spectral-index sub division boundaries.}
\label{fig:specvary1}
\end{figure*}

Here we examine how the U-band varies with commonly used \sn properties by comparing composite spectra generated from subsets of the data. These are shown in Figs.~\ref{fig:specvary0} and ~\ref{fig:specvary1}.
Subsets are constructed as follows: We retrieve all (restframe) spectra within a phase bin and normalize using the median flux of the $B_{\mathrm{SNf}}$ band. If a SN has two spectra from different nights within this range, the closest to the center of the bin is used. If the mean phase of two (sequential) spectra is closer to the center phase, their mean spectrum is used instead. 
At each phase we show four composite spectra constructed by dividing the full sample according to quartiles made from a secondary property, i.e. the $25\%$ SNe with lowest value form one subset, the next $25\%$ another and so forth. Quartile \rfc{mean} composites  are shown, from low to high, with increasingly lighter colors \rfctwo{(inverted for $x_1$)}.
A spectral feature correlation with the secondary property, spanning the full sample and not driven by outliers,  will lead to all subset composites being arranged in "color" order.
\rfc{Through examination of these panels we observe the following:}

\begin{itemize}
\item \emph{$x_1$ and $M^0_{B,\beta c}$ (lighcurve width/luminosity)}: The shallow Intermediate Mass Element (IME) features of slow-declining supernovae are seen in the \cahk feature at all times (lightly shaded line). Besides this,  the dominant feature is the strong correlation of the uTi-feature flux level with luminosity/lightcurve width. This dependence grows stronger after peak.
 \item \emph{Color}:  We find no signs of persistent feature variations with \salt color after dereddening, thus visually confirming that our dereddening procedure worked as intended.
 \item \emph{\salt standardization residuals}:
   We identify the pre-peak \uca and \uti indices as potentially interesting for use in standardization.
 \item \emph{\ewsisix}: \refcom{ The width of several features vary with \ewsisix, including \cahkfeat, \ewsifour, and \uti at late phases. This suggests that the \citet{2006PASP..118..560B} subdivision into \sne with broader/narrower spectral features is present, to some extent, in the U-band. }   
 \item \emph{ \vsisix }: As expected, \sne with large \ion{Si}{} velocities at peak also demonstrate blueshifting in the \sithree and \sifour features.
\end{itemize}

\section{Results: Understanding the explosion }\label{sec:prog}

\subsection{Origin of U-band index variations}\label{sec:lines}

\rfc{We use} synthetic \verb SYNAPPS \xspace fits \rfc{ to determine which element changes are needed to remove the dissimilarities between the SN2011fe and SN20080514-002 spectra.}
\verb SYNAPPS \xspace is a C implementation of the original \verb Synow \ code (\verb syn++ ) with an added optimizer, to find the best ion temperature, velocity and optical depth combinations to fit input spectra under the Sobolev approximation for e$^-$ scattering \citep{synapps}. 
Fits presented here include \ion{Mg}{ii}, \ion{Si}{ii}, \ion{Si}{iii}, \ion{S}{ii}, \ion{Ca}{ii}, \ion{Ti}{ii}, \ion{Fe}{ii}, \ion{Fe}{iii}, \ion{Cr}{ii}, \ion{Co}{ii}, \ion{Co}{iii} , \ion{Ni}{ii}.
\rfctwo{We focus on the region bluewards of the \cahkfeat region and thus do not attempt to reconstruct \cahkfeat HVFs. No detached ions were included.}
\rfc{The full fits including these ions capture the observed spectra well.}
Fits were also remade while iteratively deactivating one (or a combination) of the ions.

\subsubsection{\uniw : $3300$ -- $3510$ \ang}

\begin{figure*}
\includegraphics[clip,trim=1.5cm 0.5cm 1.5cm 1.5cm, angle=0,width=0.99 \textwidth]{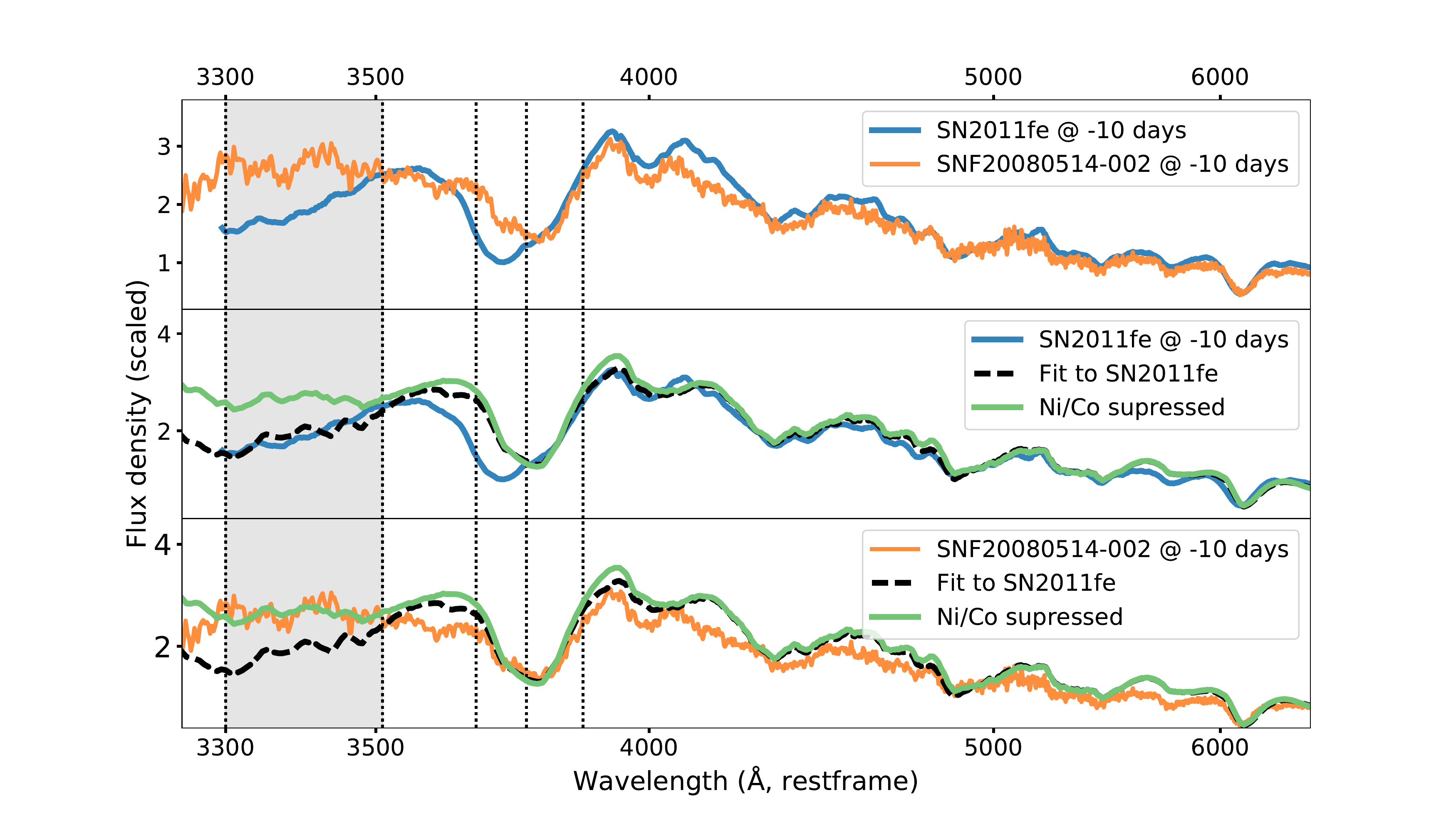}
\caption{
Probing the origin of variation \refcom{in the \uniw region} through SYNAPPS model comparisons. The top panel compares SN2011fe and \refcom{SNF20080514-002} at an early phase ($-10$ days). The second panel repeats the early SN2011fe spectrum (blue line) together with the best SYNAPPS fit (black dashed line). The green line shows the same fit, but with the optical depth ($\tau$) of all \ion{Ni}{} and \ion{Co}{ii} decreased by one dex, \rfc{effectively suppressing these ions}. The third panel compares the early \refcom{SNF20080514-002} spectrum with the same SN2011fe SYNAPPS fits. The SN2011fe fit with suppressed \ion{Ni}{}, \ion{Co}{ii} optical depth matches the \refcom{SNF20080514-002} \uniw  region well. 
Vertical dotted lines indicate the U-band spectral index boundaries, with \uniw lightly shaded grey.
}
\label{fig:syn_ni}
\end{figure*}

\rfctwo{SYNAPPS fits and their implication for the \uniw \refcom{spectroscopic region} are shown in Fig.~\ref{fig:syn_ni}.}
We find that the \uniw window can be tied to the presence of \ion{Ni}{} and \ion{Co}{}, the decay product of ${}^{56}$Ni at these phases. Other elements, investigated using \verb SYNAPPS \ runs without Ni/Co, cannot replicate the observed spectra without distorting other parts of the spectrum.
The effects of \ion{Ni}{} and \ion{Co}{} can be directly seen by manually decreasing the optical depths of these ions relative to the SN2011fe fit: a change of order $1$ dex creates a \verb syn++ \ spectrum that matches \refcom{SNF20080514-002} well in the \uniw region (lower panels of Fig.~\ref{fig:syn_ni}).
\rfc{We show contributions of all included ions to the full fit in the Appendix (Figure~\ref{fig:fesyn}).}
\citet{2008ApJ...677..448T}, using abundance modifications to the W7 density profile to fit SN2002bo, also found this wavelength region to be sensitive to the amount of outer Ni (see their Fig.~2). Similar results can be seen in a number of studies using different techniques: \citet{2013MNRAS.429.2127B} presented one-dimensional delayed-detonation models that show \ion{Co}{ii} to dominate here; \citet{2013MNRAS.429.2228H} performed a spectral tomography analysis of SN2010jn where an early spectrum was dominated by \ion{Fe}{} absorption at $3000$~\ang and the \ion{Ni}{}/\ion{Co}{} around $3200$~\ang, and a SYNAPPS study by \citet{2015ApJ...813...30S} also shows strong \ion{Ni}{}/\ion{Co}{} absorption at $\sim 3300$~\ang (although no \ion{Cr}{} or \ion{Ti}{} was included in these fits).
Like \citet{2013MNRAS.429.2228H}, we find that measurements of this spectral region are necessary for determining the spatial distribution of IGE elements, a key characteristic distinguishing theoretical explosion models. %
\citet{2017MNRAS.464.4476C} found that \ion{V}{II}{} provided improved fits to very early spectra of SN2015F, which would also impact the \uniw region.

\subsubsection{uTi : $3510$ -- $3660$ \ang}

\begin{figure*}
\includegraphics[clip,trim=1.5cm 0.5cm 1.5cm 1.5cm, angle=0,width=0.99 \textwidth]{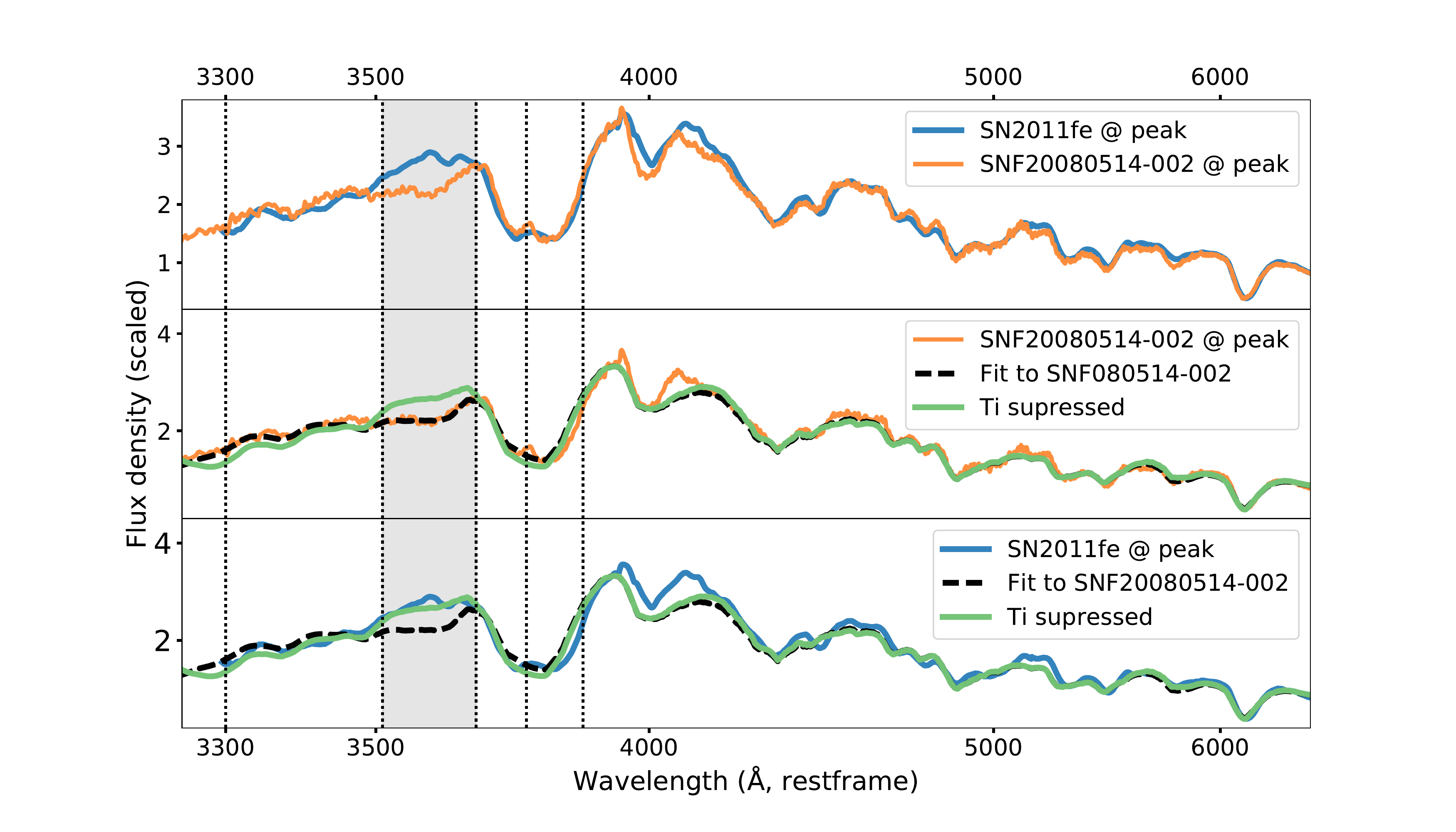}
\caption{
Probing the origin of uTi variation through SYNAPPS model comparisons. The top panel compares SN2011fe and \refcom{SNF20080514-002} at peak light. The second panel shows \refcom{SNF20080514-002} (orange line) together with the best SYNAPPS fit of this spectrum (black dashed line). The green line shows the same fit, but with the optical depth ($\tau$) of \ion{Ti}{ii}  decreased by one dex, effectively suppressing these ions. The third panel compares the SN2011fe spectrum with the same \refcom{SNF20080514-002} SYNAPPS fits. The \refcom{SNF200805014-002} fit with suppressed \ion{Ti}{ii} optical depth matches the SN2011fe \utiw region well. 
Vertical dotted lines indicate the U-band spectral index boundaries, with \utiw shaded light grey.
}
\label{fig:syn_ti}
\end{figure*}

\rfctwo{SYNAPPS fits and their implication for the \utiw region are shown in Fig.~\ref{fig:syn_ti}.}
Among the \verb SYNAPPS \ ions included, only \ion{Ti}{ii} produces significant absorption in the $3510$ to $3660$~\ang region without distorting other parts of the spectrum.  We display this association by manually decreasing the \ion{Ti}{ii} optical depth. We do this for \refcom{SNF20080514-002} since it shows the larger absorption, and again find that a change of order $-1$ dex produces a spectrum that matches the SN2011fe uTi region well.
\rfc{More complete radiative transfer models} are needed to determine whether uTi variations are fully explained by \ion{Ti}{ii} absorption, but we note that the \ion{Ti}{} lines at  $3685$, $3759$ and $3761$~\ang would land in the uTi wavelength region for typical \sn velocities. 
The post-peak \utiw region is, as can be seen in Figs.~\ref{fig:specvary0} and~\ref{fig:specvary1}, strongly correlated with \ewsisix. As no strong \ion{Si}{} lines are expected in this region this \rfctwo{reflects} a general connection between \rtr{widths of spectroscopic features in \sn spectra}.
\rfc{We show contributions of all included ions to the full fit in the Appendix (Fig.~A\ref{fig:snfsyn}).}

\subsection{Explosion models and progenitor scenarios}\label{sec:models}

The single degenerate scenario -- mass transfer from a red giant or main sequence star onto a Carbon-Oxygen White Dwarf -- \refcom{is no longer considered as likely to explain all (or most) \sne}. Challenges come from the lack of companion stars close to nearby SNe \citep[e.g. SN2011fe][]{2011Natur.480..348L,2012Natur.481..164S,2012ApJ...747L..19E}, the statistical absence of early lightcurve variations due to ejecta interaction with the companion \citep{2010ApJ...722.1691H}, an insufficient number of such systems formed \citep{2011MNRAS.417..408R}, and  a large range of ejecta masses \citep{2014MNRAS.440.1498S}.
A number of scenarios, possibly existing in parallel, are currently being investigated. \rfc{These predict similar spectral energy distributions in the $4000$ to $7000$~\ang region around lightcurve peak and have thus proven hard to rule out using such data} \citep{2012ApJ...750L..19R}. 
Bluer wavelengths, on the other hand, show significant differences between current theoretical models.
We have compared output spectra from delayed detonation \citep[model N110,][]{2014MNRAS.444..350S,2013MNRAS.436..333S}, violent merger \citep[model 11+09, ][]{2012ApJ...747L..10P}, sub-Chandra double detonation \citep[model 3m,][]{2010ApJ...719.1067K} and sub-Chandra WD detonation \citep{2010ApJ...714L..52S} models. We find that none of these describe the observed U-band variations.

\rfc{\citet{2016ApJ...824...59M} found constant \ion{Si}{}{} but varying \ion{Ca}{}{} abundances to be a robust consequence of changing the progenitor model metallicity. However, they do not find this to cause strong changes in the observed spectra. We have nonetheless searched for \uca changes, not visible in \usi, for indications of a relationship to metallicity, but did not observe anything significant. The region identified by \citet{2016ApJ...824...59M}  as strongly and consistently affected by metallicity was the \ion{Ti}{}{} absorption at $\sim 4300$~\ang at $\sim 30$~days after explosion. As this feature, as for the \uti index, is strongly correlated with peak luminosity and lightcurve width, searching for a second order variation due to metallicity is challenging.}

\begin{figure}
  \centering
  \includegraphics[angle=0,width=0.5 \textwidth]{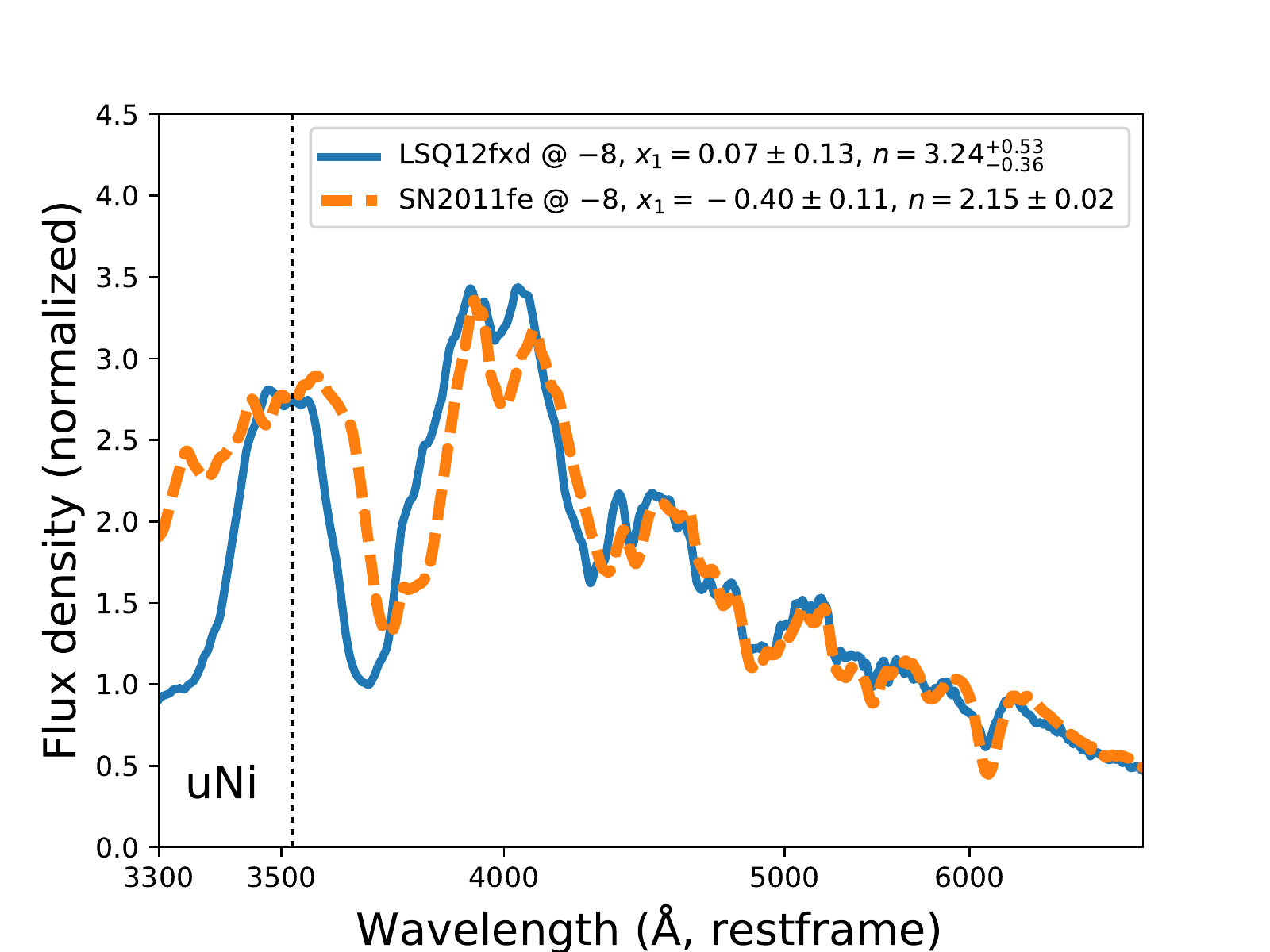}
  \caption{ Spectra of SN2011fe and LSQ12fxd at phase $\sim -8$~days. LSQ12fxd has, compared with SN2011fe, both a steeper early rise-time \citep{2015MNRAS.446.3895F} and \refcom{less flux in the \uniw region} (i.e. stronger \ion{Co}{ii} absorption at this phase).
  }
\label{fig:rise}
\end{figure}

A more direct way to probe Ni in the outermost layer is through observation of the very early lightcurve rise-time, parameterized as $ f \propto t^n$. A mixed ejecta (shallow \nifive) will cause an immediate, gradual flux increase while deeper \nifive with cool outer layers experience a few days of dark time before a sharply rising lightcurve \citep{2014ApJ...784...85P, 2016ApJ...826...96P}. In these models, the effects of outer ejecta mixing has largely disappeared $\sim$ one week after explosion.
\citet{2015MNRAS.446.3895F} examine a sample of SNe with very early observations, finding varying rise-time power-law indices $n$ between $1.48$ and $3.7$, suggesting that either the outermost \nifive layer and/or the shock structure varies significantly between events.
A sample of SNe with both early lightcurve data and U-band spectra would allow a direct comparison between pre-peak \uniw absorption and the amount of shallow \ion{Ni}{} predicted from the early lightcurve rise-time.

SN2011fe and LSQ12fxd in the \citet{2015MNRAS.446.3895F} study \rfctwo{are included in the sample studied here and have observations at a common phase of} $\sim 10$ days before peak  (shown in Fig.~\ref{fig:rise}). 
Compared with SN2011fe, LSQ12fxd has both a steeper rise-time immediately after explosion ($n=3.24$ vs. $n=2.15$) and more \ion{Co}{ii} absorption (less \uniw flux) in the line-forming region one week prior to lightcurve peak. 
\rfc{A scenario explaining both observations would involve \nifive mixed into the outermost ejecta regions in SN2011fe while being located slightly deeper and being denser in LSQ12fxd.}  
The comparison is non-trivial as these SNe  also vary significantly in lightcurve width and line velocities, but we note that neither $x_1$ nor \vsisix correlates strongly with early \uige and that \citet{2015MNRAS.446.3895F} find no correlation between $n$ and lightcurve width.

\section{Results: Impact on standardization}\label{sec:cosmo}

 Here we discuss the  strong correlation between post-peak \uti and peak luminosity, and the potential impact of \uca, for \sn standardization 
  (Sec.~\ref{sec:standard}). The residual magnitude correlation with host environment after standardization is explored in Sec.~\ref{sec:host}. \rfctwo{In Sec.~\ref{sec:reddening} we explore whether the systematic effects from reddening corrections could significantly affect these results.}

\subsection{\sn luminosity standardization with \uti and \uca}\label{sec:standard}

Many spectroscopic features show strong correlations with \sn luminosity.
These include \rsi \xspace and \rca, introduced by \citet{1995ApJ...455L.147N}, and the equivalent width (or ``strength'') of the \sifour feature \citep{2008A&A...492..535A,2011A&A...529L...4C,2011AA...526A.119N}.
As discussed in Sec.~\ref{sec:composite}, \uti displays a similarly strong sensitivity to peak luminosity, especially at post-peak phases where contamination by the blue edge of the \cahk feature is less likely.
For some SNe a direct connection to \rca, measured as the ratio between flux at the edges of the \cahk feature, could exist.
We now further explore the \uti-luminosity correlation.

\begin{figure*}
 \includegraphics[angle=0,width=0.32 \textwidth]{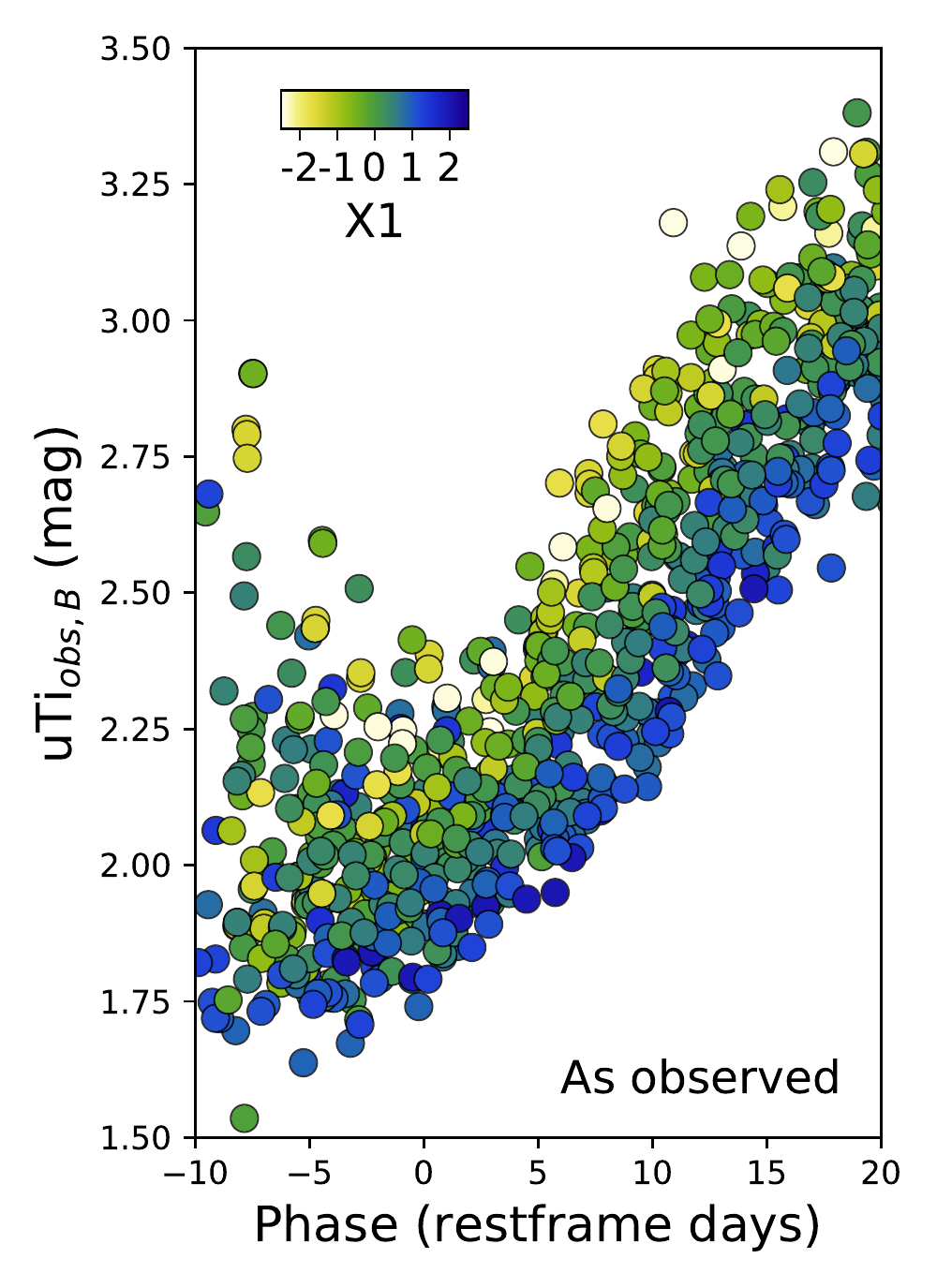}  
 \includegraphics[angle=0,width=0.32 \textwidth]{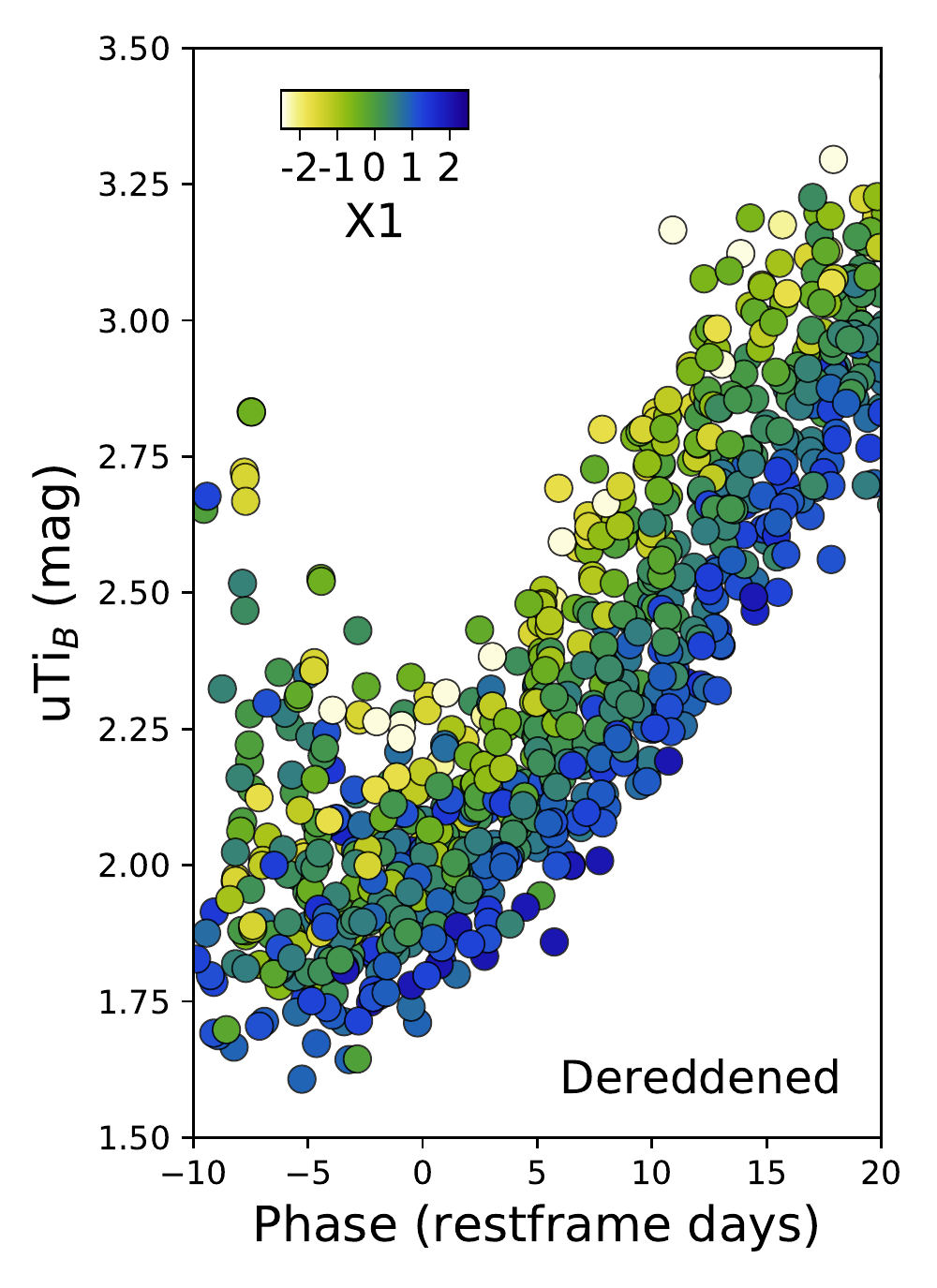}  
 \includegraphics[angle=0,width=0.32 \textwidth]{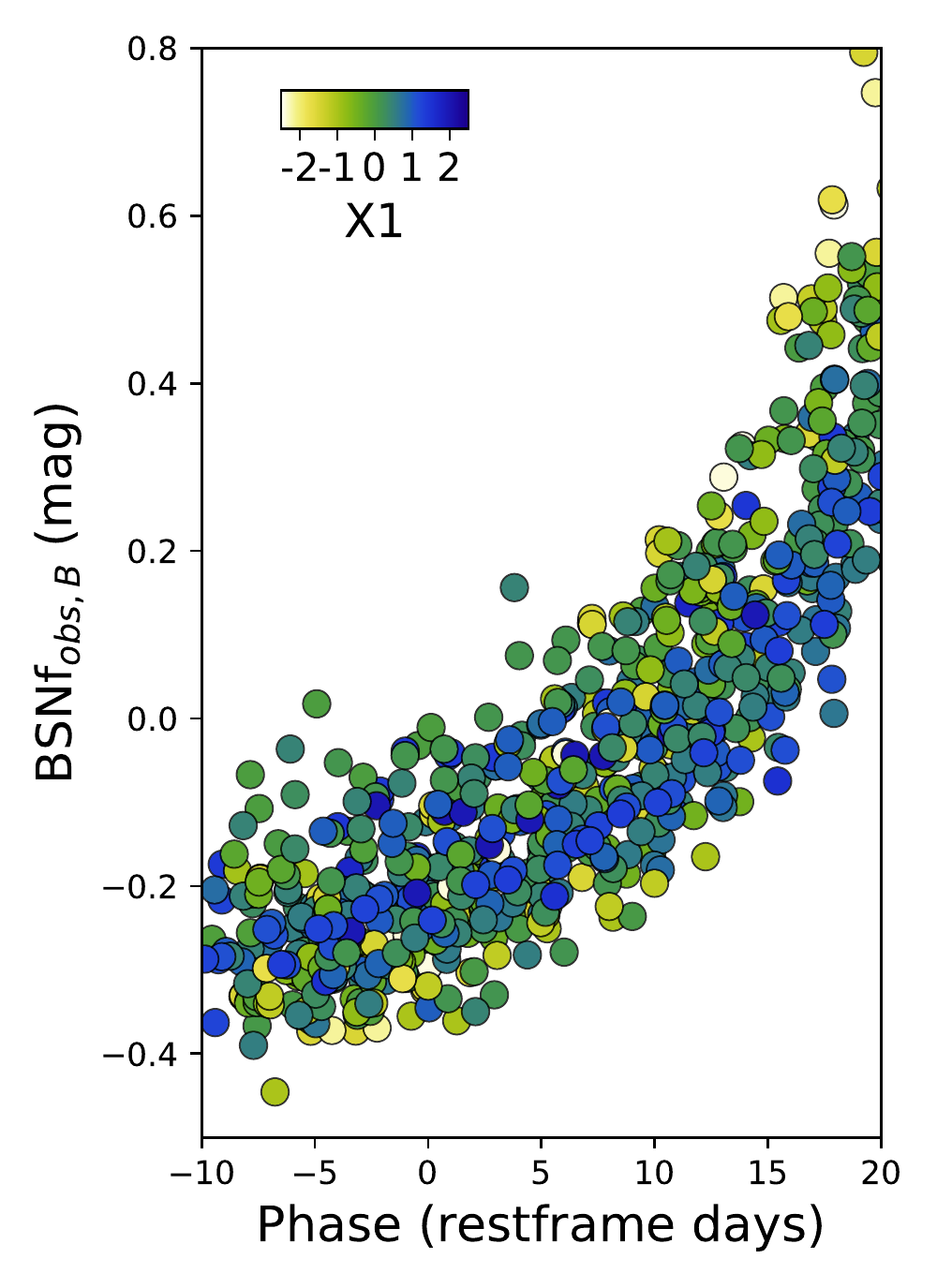}  
 \caption{\uti vs. phase with markers colored by \salt $x_1$.  Left panel shows the \uti color prior to F99 dereddening, mid panel after this correction. The right panel displays the $B_{\mathrm{SNf}}-V_{\mathrm{SNf}}$ color evolution for reference, calculated from dereddened restframe spectra.  
    Pre-peak observations show a scatter induced by the edge of the \cahk feature. \uti colors after peak show a strong stable correlation with lightcurve width.
  }
\label{fig:utilightcurve}
\end{figure*}

In Fig.~\ref{fig:utilightcurve} (mid panel) we show the \uti change with phase per \refcom{SN}, and with measurements color-coded by lightcurve width (\salt $x_1$).
From around lightcurve peak and later, we find a persistent and strong relation, where SNe with wider lightcurves have bluer \uti colors.
To confirm that this correlation is not driven by the dereddening correction or changes in the $B_{\mathrm{SNf}}$ band, Fig.~\ref{fig:utilightcurve} contains two modified color curves: The first (left panel) shows the observed color $\mathrm{uTi}_{\mathrm{obs,B}}$, i.e. \uti  recalculated according to Equation \ref{eq} but without dereddening spectra, and the second (right panel)  shows the $B_{\mathrm{SNf}}-V_{\mathrm{SNf}}$ color evolution (calculated based on restframe and dereddened spectra). We confirm that the strong $x_1$ trend is present also without dereddening but not visible for $B_{\mathrm{SNf}}-V_{\mathrm{SNf}}$.

We evaluate \sn standardization using combinations of post-peak \uti as a replacement for $x_1$, and pre-peak \uca as an additional standardization parameter (see Table~\ref{tab:hubb}). We assume a fixed $\Lambda CDM$ cosmology, continue to remove red SNe ($c>0.2$) and use SNe in the $0.03<z<0.1$ range to reduce scatter from peculiar velocities. We further fix the \salt $\beta$ parameter (the magnitude dependence of $c$) to the (blinded) value determined from the full SNfactory sample (derived without cuts based on color or first phase). The two base fits include either the magnitude dependence of $x_1$ (``$\alpha$'') or the magnitude dependence of post-peak \uti. Two permutations of these are made: one (``cut'') where the sample is limited to SNe with observations at all phases (early and late), and another where pre-peak \uca is added as a second standardization parameter. To allow comparisons of $\chi^2$ between runs we add a fixed  dispersion of $0.090$ mag, the value required to produce ${\chi^2/\mathrm{dof}=1}$ for the initial $x_1$ run (first row in table).
\rtrt{The uncertainty of each U-band index is composed of the sum of variance due to statistical uncertainties of the spectra and the propagated reddening variance, and is generally at the $\sim 0.03$ mag level (see Table~\ref{tab:snprops}). The measurement correlations between U-band indices and \salt fit parameters are negligible.}

\begin{table*}
  \centering
  \caption{Standardization fit results.}
  \resizebox{\textwidth}{!}{	
  \begin{tabular}{|l|c|c|c|c|c|c|}
    \hline
    Fit parameters & SNe & $\chi^2$ & $\chi^2/$dof & HR RMS (mag) & Host mass step (mag) & LsSFR step (mag)  \\
    \hline
\hline
\Tstrut $x_1$ & $73$ & $70.76$ & $1.00$ & $0.135 \pm 0.011$ & $0.098 \pm 0.031$ & $-0.151 \pm 0.028$ \\ [0.1em]
{\uti}@p6 & $57$ & $44.18$ & $0.80$ & $0.116 \pm 0.011$ & $0.042 \pm 0.031$ & $-0.075 \pm 0.031$ \\[0.1em]
$x_1$ (cut) & $43$ & $43.22$ & $1.05$ & $0.136 \pm 0.015$ & $0.094 \pm 0.037$ & $-0.156 \pm 0.035$ \\[0.1em]
{\uti}@p6 (cut) & $43$ & $27.21$ & $0.66$ & $0.105 \pm 0.012$ & $0.034 \pm 0.033$ & $-0.068 \pm 0.034$ \\[0.1em]
$x_1$ + {\uca}@m6 & $52$ & $40.90$ & $0.83$ & $0.122 \pm 0.012$ & $0.081 \pm 0.033$ & $-0.138 \pm 0.030$ \\[0.1em]
{\uti}@p6 + {\uca}@m6 & $43$ & $18.22$ & $0.46$ & $0.086 \pm 0.010$ & $0.022 \pm 0.030$ & $-0.065 \pm 0.030$ \\[0.1em]
\hline
  \end{tabular} }
  \label{tab:hubb}
  \tablefoot{ The first column shows which standardization parameters are included (in addition to \salt color), where \emph{cut} fits are restricted to SNe with measurements both at pre-peak and post-peak phases. The number of SNe included is given in the second column. The intrinsic dispersion was fixed to $0.090$~mag for all runs. The size of a step based on global host-galaxy mass or local age (LsSFR) were calculated as in R17 and are shown in the final two columns. }
\end{table*}

\rfctwo{Our primary conclusion is that post-peak \uti standardizes \sne very effectively, with an RMS of \rmsti~mag. A traditional $x_1$ standardization yields a higher RMS of \rmssalt~mag for these SNe.}
This is remarkable in many aspects: it is a single color measurement made in a fairly wide phase range and using a fixed wavelength range that produces a lower $\chi^2$ fit.
\rtrt{For the reduced sample of \nbrti SNe with both measurements this is an improvement over $x_1$ with $\Delta \chi^2 = $ \deltchitisalt, which is significant at greater than $3\, \sigma$ (see also Table~\ref{tab:aic}).}
  We further investigate the potential effects of sample selection by redoing the fits based on a ``cut'' sample, where only SNe with both pre-peak ($-8$ to $-4$ days) and post-peak ($4$ to $8$) data are included. We see no significant differences between the full and cut samples. The reduced dispersion for \uti fits is thus not due to sample selection.

\begin{figure*}
  \centering
  \includegraphics[angle=0,width=0.4 \textwidth]{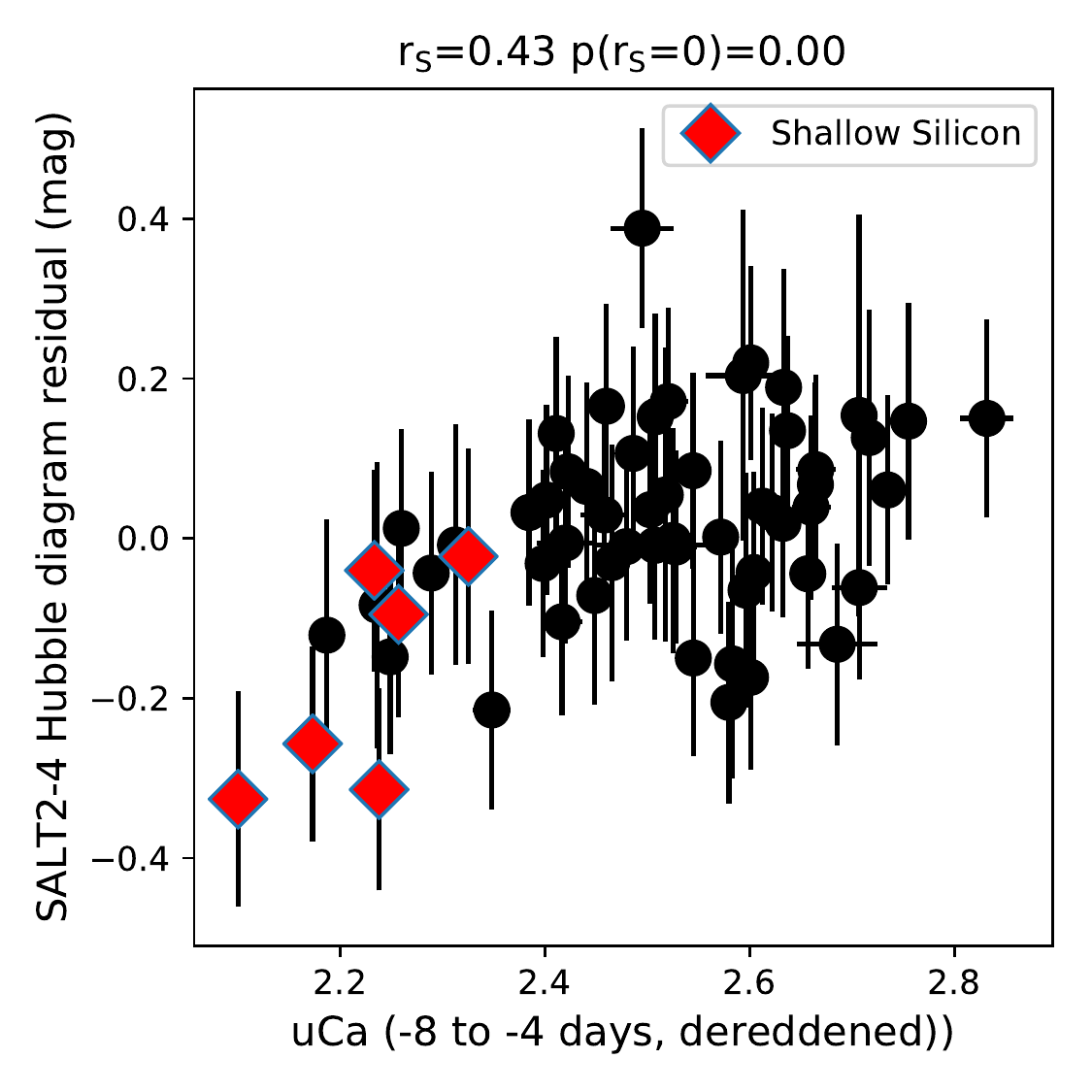} 
  \includegraphics[angle=0,width=0.38 \textwidth]{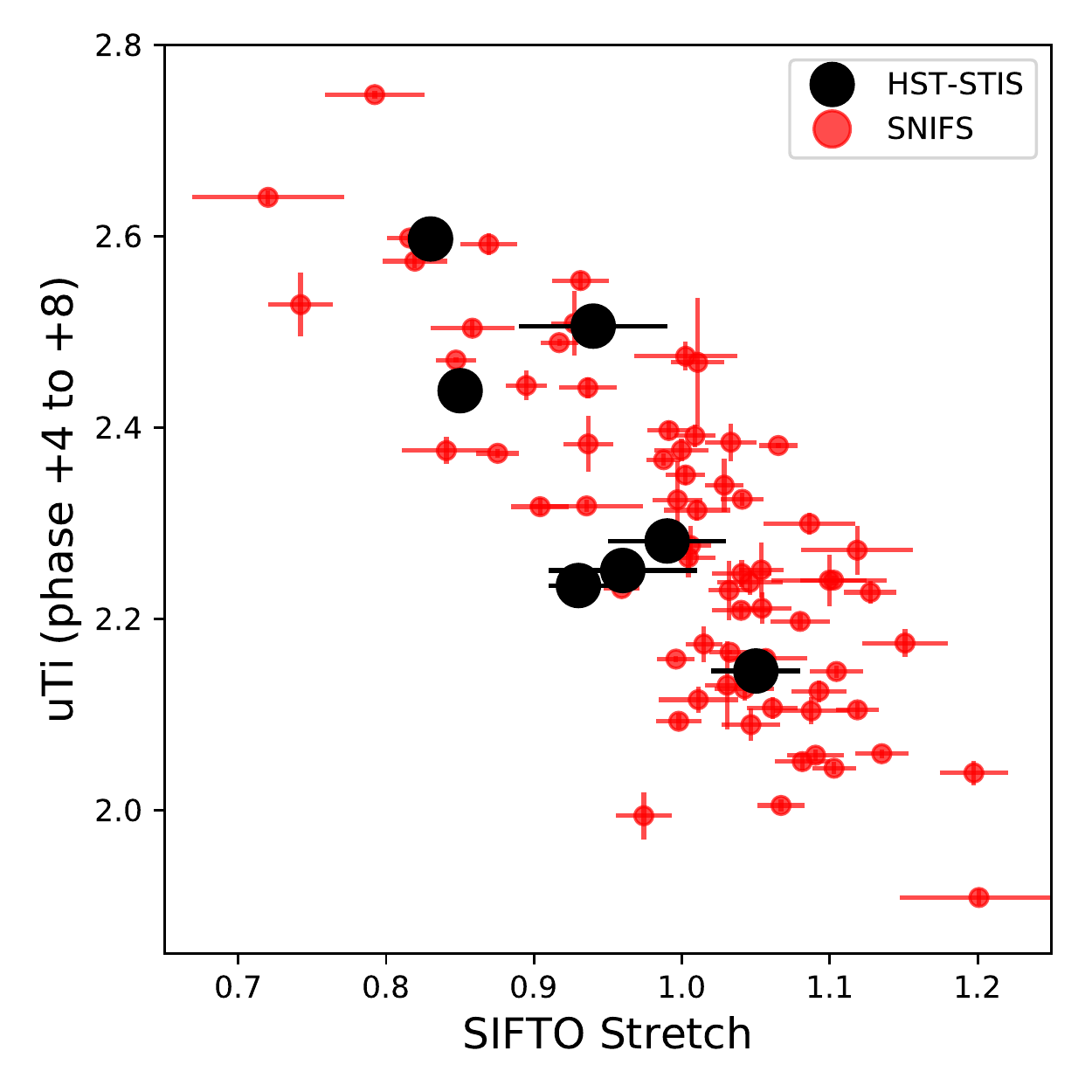}
  \caption{
    \emph{Left:}
    Origin of pre-peak \uca correlation with \salt Hubble residuals. The Spearman correlation coefficient of $r_s=$ \cahubblersi indicates a moderate correlation, with the hypothesis of no correlation at $>99\%$ confidence.
    \emph{Right:} Comparing the post-peak \uti color integrated from HST-STIS spectra with SIFTO stretch. Spectra were dereddened according to a similar procedure as the SNfactory sample and use lightcurve data from \citet{2012MNRAS.426.2359M}. \rfctwo{The SNIFS sample presented here is included for comparison, with \salt $x_1$ values converted to SIFTO stretch using the relation provided by \citet{2010A&A...523A...7G}. \refcom{The trend with \uti agrees betwen these two data sets.} }
    }
\label{fig:stis}
\end{figure*}

\rfc{ The Spearman rank correlation coefficient between \salt Hubble residuals and pre-peak \uca is $r_s=$ \cahubblersi, and the hypothesis of no correlation can be rejected at greater than $99\%$ confidence (Fig.~\ref{fig:stis}, left panel).} We therefore test adding pre-peak \uca measurements as a further standardization parameter. Combining post-peak \uti and pre-peak \uca produces a Hubble diagram RMS of \rmstica~mag, while using only \uca and $x_1$ reduced the RMS to \rmssaltca~mag. The driving trend of this improvement can also be seen in Fig.~\ref{fig:stis}: \sne with large \uca index are too bright after \salt standardization. Half of these are classified as Branch Shallow Silicon objects -- this connection is further explored in Sec.~\ref{sec:subset}.

We also note that the combined \uti + \uca fit produces a much reduced $\chi^2$ value, beyond what can be expected just through adding another fit parameter. When rerunning the standardization without a fixed intrinsic dispersion we obtain $\chi^2=38.6$ for $40$ degrees of freedom, thus there is no need to add any additional dispersion to reach $\chi^2/\mathrm{dof}=1$. As the internal \salt model error propagates an effective intrinsic dispersion of $0.055$ mag, other fit methods are required to investigate whether a fit without any added dispersion can be attained. For comparison, the \uti fit requires an intrinsic dispersion of $\sigma_{int} =$ \dispti mag and the $x_1$ fit requires $\sigma_{int} =$ \dispsalt mag.

\begin{table*}
  \centering
\caption{ \rtr{Each line shows Hubble residual fit quality for a given standardization method, measured relative to a reference fit based on post-peak \uti data without any host property corrections (first line). Each comparison is made using only the SNe in common (Nbr SNe) for a given measurement. The penultimate column shows the difference in $\chi ^2$ assuming Hubble residuals are described using one or two Gaussian distributions (the latter when a host property step is included).  The final column gives the likelihood according to the the sample-size corrected Akaike Information Criteria (AICc), again relative to the first line \uti model.}}
\begin{tabular}{|ll|c|c|c|}
  \hline
  Standardizing property & Host step & Nbr SNe & $\chi ^2- \chi ^2_{uTi}$  & P(AICc) Ratio \\ 
  \hline
  \hline
\Tstrut  {\uti}@p6 & None & $57$ & $0$ &  $1.0$ \\
  & Global mass & $47$ & $-1.9$ &  $0.19$ \\
  & LsSFR & $47$ & $-4.8$  & $0.84$ \\
  \hline
  $x_1$ & None     & $57$ & $20.5$ &  $3.6e-5$  \\
  & Global mass & $47$ & $5.6$ &  $0.0045$ \\
  & LsSFR  & $47$ & $1.9$ & $0.029$ \\
  \hline
  {\uti}@p6 + {\uca}@m6 & None & $43$ & $-17.2$ &  $1521.5$ \\
  & Global mass & $35$ & $-12.0$ & $5.0$\\
   & LsSFR & $35$ & $-15.9$ & $33.7$ \\
  \hline

\end{tabular}
  \label{tab:aic}
  
\end{table*}

As a further test we evaluate the fit quality using the sample-size corrected Akaike Information Criteria (AICc), which penalizes models with additional fit parameters. In Table~\ref{tab:aic} each line compares \uti standardization (without any host galaxy property correction) with one other combination of standardization property and host parameter. Each comparison includes all SNe available for that combination of data, and shows both the difference in $\chi^2$ and the AICc probability ratio.
  \rtrt{Models including both \uti and \uca are strongly preferred over only using \uti, with a P-value of $<0.001$, even though penalized for adding another fit parameter. Using only \uti is similarly favored compared with the $x_1$ fit.}

\rfc{The significance of these improvements can also be numerically investigated by re-fitting the standardization after randomly redistributing the \uca measurements among SNe. When coupled to \uti, \randomtica out of $10000$ random simulations yielded a similarly low RMS, equivalent to a $P$-value of $<10^{-5}$. When combined with $x_1$, \randomsaltca out of $10000$ did so.}

\rfctwo{Finally, the HST-STIS sample presented by \citet{2012MNRAS.426.2359M} also included a small sample of spectra covering the \utiw region and overlapping with the post-peak phase studied here (phase $+4$ to $+8$ days). Lightcurve width information (``stretch'' and $B-V$, determined by SIFTO) exists for seven of these SNe (PTF10wof, PTF10ndc, PTF10qyx, PTF10qjl, PTF10yux, PTF09dnp, PTF10nlg). With these data we can check an external dataset for a similar correlation. We deredden the spectra as was done previously with the SNf data and calculate the \uti color. Fig.~\ref{fig:stis} (right panel) shows a strong correlation for this small sample, compatible with the \uti trend discussed above. We find that this trend agrees well with the SNIFS measurement presented here, after converting the latter to SIFTO stretch values.}

\subsection{The SN progenitor environment}\label{sec:host}

The R17 analysis of the local host galaxy environment found that \sne in younger environments are \rbias\xspace~mag (\rsigma) fainter than \sne in older environments, after \salt standardization (based on a larger sample than used here). The corresponding analysis of global host galaxy mass found \sne in lower mass galaxies to be \rbiasm~mag fainter than those in more massive hosts. We recover these trends for the subset of SNe in this analysis with R17 measurements, finding a \lstepsalt~mag step for LsSFR and \mstepsalt~mag for global host galaxy mass.
 
When this step analysis is performed based on the standardization residuals where \uti replaced $x_1$ the step sizes are reduced to \mstepti~mag for mass and \lstepti~mag for LsSFR (given in the final two columns of Table~\ref{tab:hubb}). \rfctwo{\uca has less impact for environmental steps, producing modest step size reductions to \msteptica~mag and \lsteptica~mag.}
$\Delta \chi^2$ and AICc probabilities for these models, again relative to applying \uti but no host data, can also be found in Table~\ref{tab:aic}. We find that the full model including \uti, \uca and LsSFR provides the smallest $\chi^2/\mathrm{dof}$, but that the AICc find fits including \uti and \uca but \emph{no} host corrections to be preferred considering the number of parameters. The fit quality of the $x_1$ models rapidly increases as host information is included. Adding host information to the U-band parameter models is not justified as $\chi ^2$ is only modestly improved.

\begin{figure}
  \centering
  \includegraphics[angle=0,width=0.99 \columnwidth]{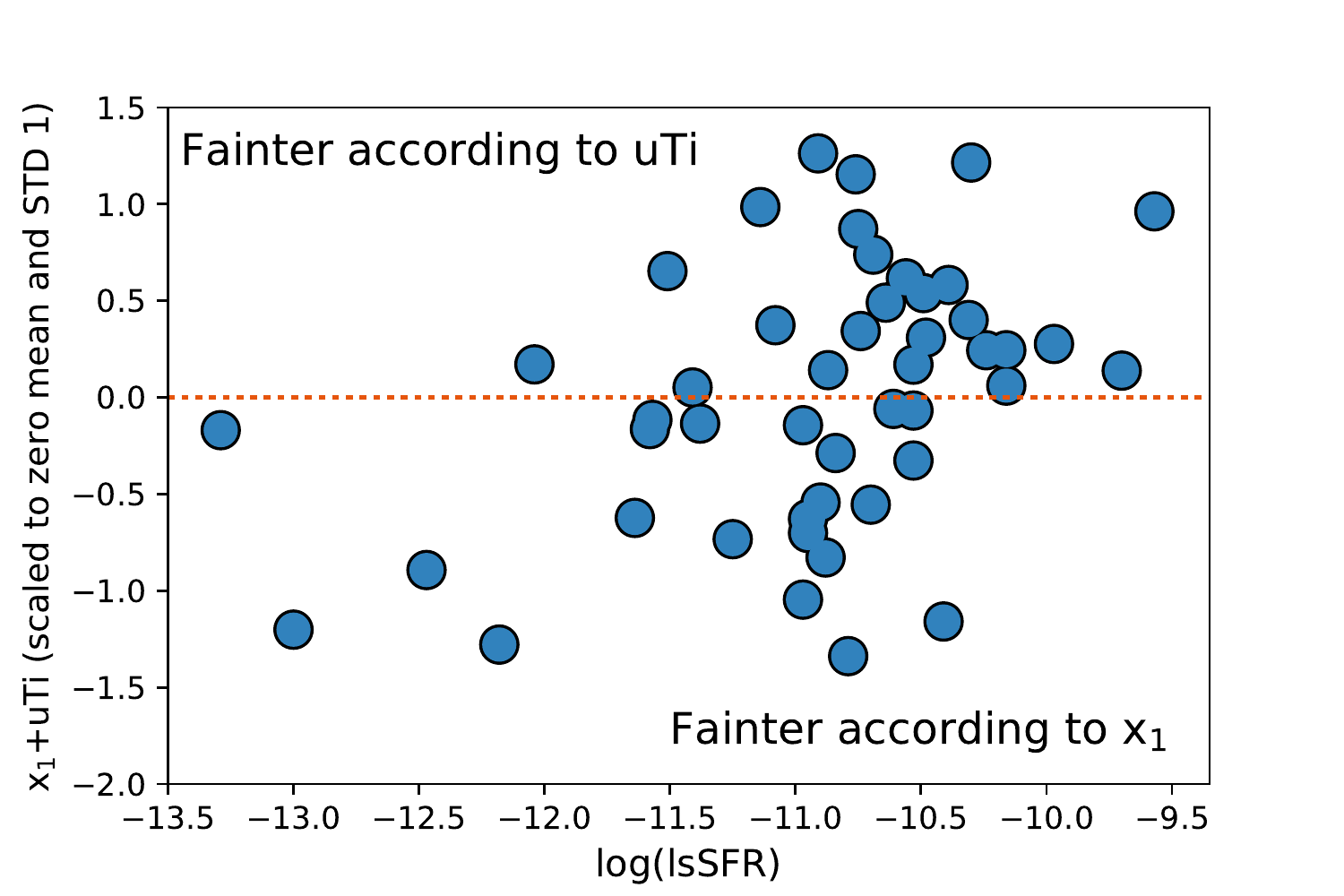} 
  \caption{
    The difference in standardization based on $x_1$ and \uti is examined by comparing their sum (as they are anti-correlated) against local specific star formation (LsSFR).
    \rtrt{A smaller $x_1+$\uti (y-axis) value corresponds to a SN considered fainter according to $x_1$ relative to what \uti predicts.}
    }
\label{fig:x1ti}
\end{figure}

In Fig.~\ref{fig:x1ti} we search for the SNe for which lightcurve width and \uti predict different magnitudes. As \uti and $x_1$ are anti-correlated, we can do this by normalizing both distributions to zero mean and unity RMS and then plotting the sum of the two transformed values for each SN.
\refcom{Current $x_1$ standardization produces SN magnitudes that are too bright in passive environments (low LsSFR), i.e. the $x_1$ parameter assumes these to be intrinsically fainter than they actually are and overcorrects their magnitudes. As is visualized in Fig. 11, \uti still predicts these SNe to be intrinsically faint, but not by as much as $x_1$, thus generating smaller magnitude corrections and avoiding overcorrection. Similarly, in actively star forming regions \uti does not predict SNe to be as intrinsically overluminous as predicted by $x_1$.}
\refcom{The variation in predicted peak magnitude as a function of host galaxy environment suggests an underlying property} 
that varies with age and affects the explosion duration (lightcurve width) without a corresponding change of the peak energy/temperature would cause a trend like the one observed. Further studies are needed to investigate whether, for example, progenitor size could act in this way.

\begin{figure*}
  \includegraphics[angle=0,width=0.49 \textwidth]{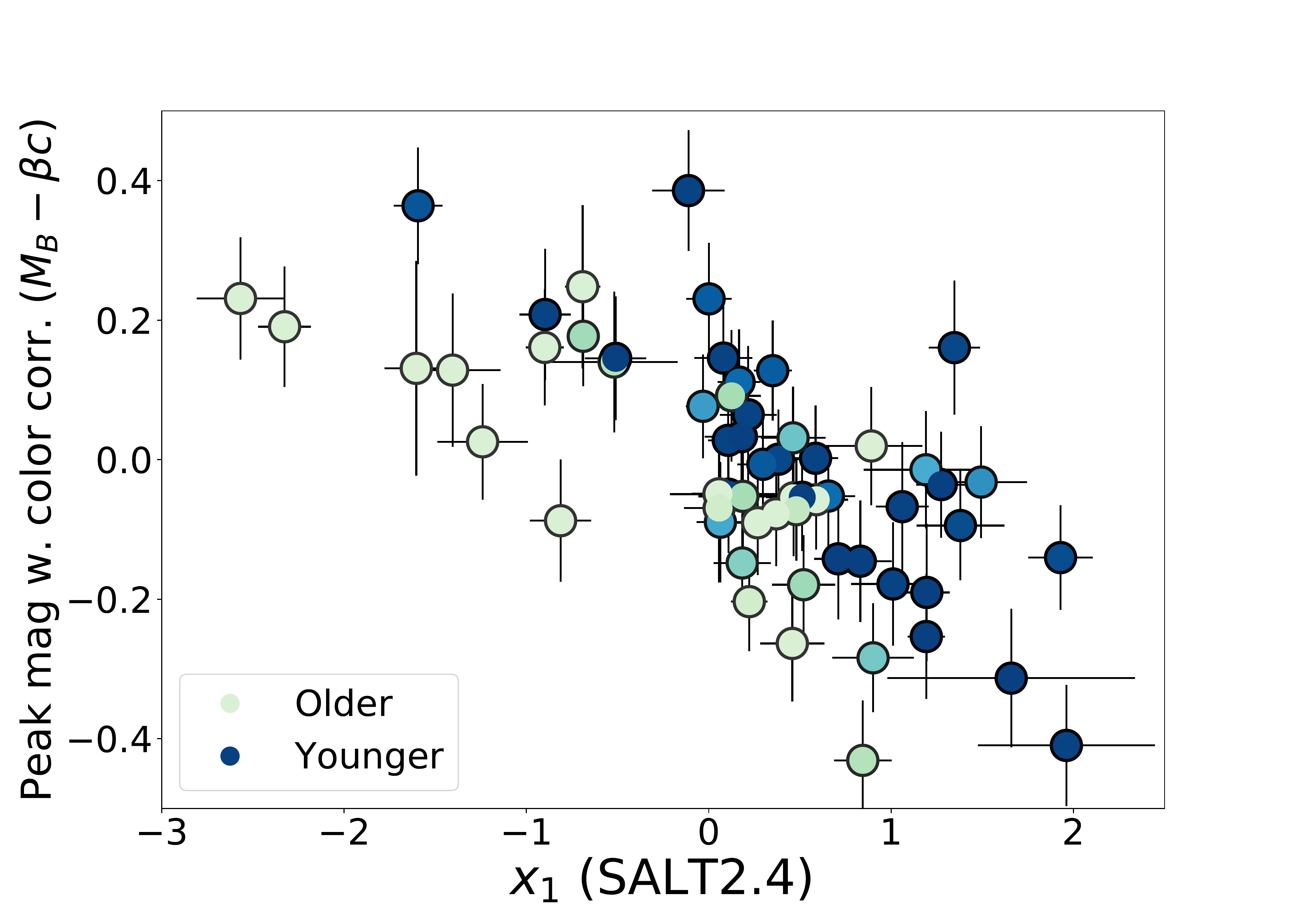} 
  \includegraphics[angle=0,width=0.49 \textwidth]{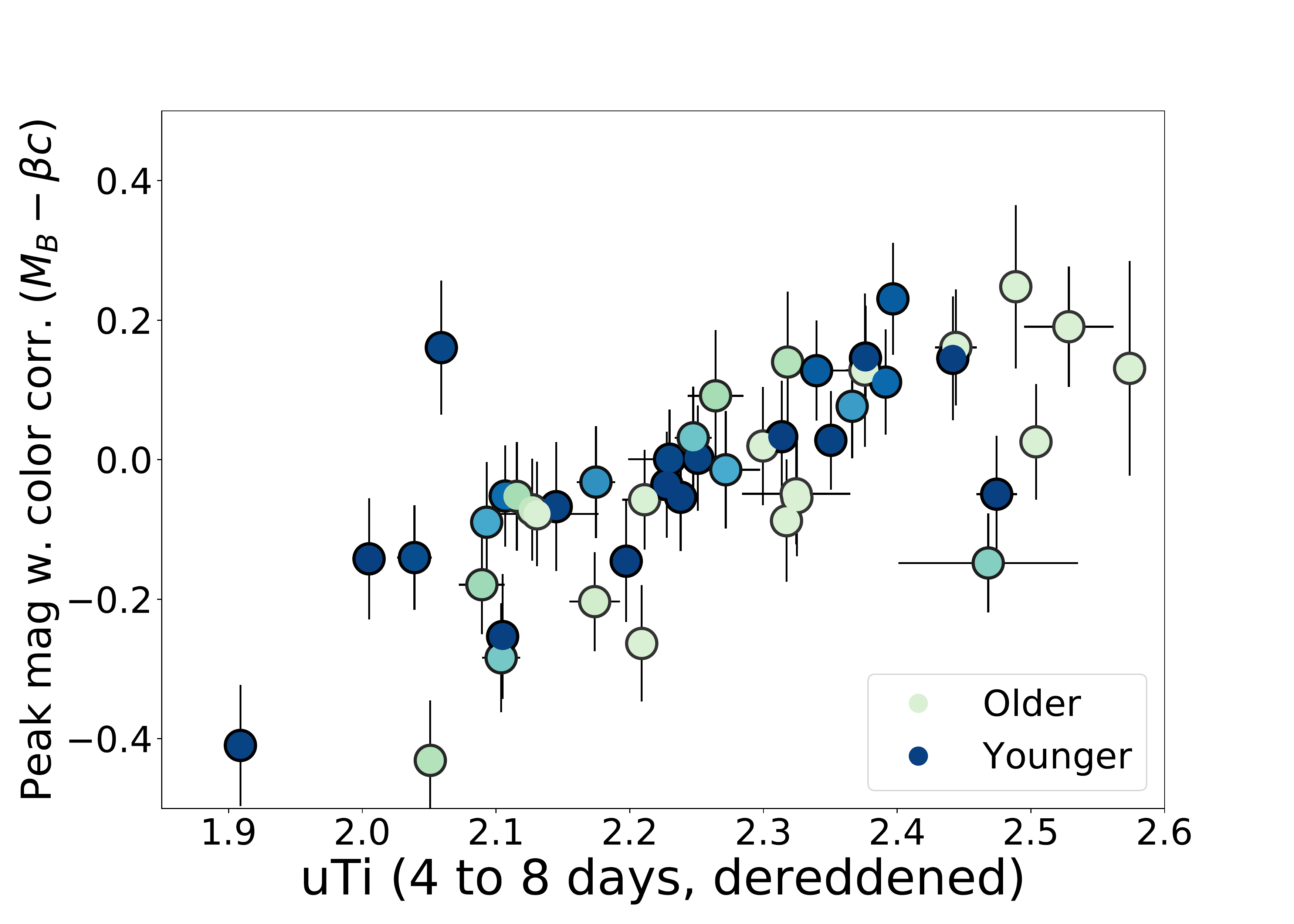} 
  \caption{ Comparing peak SN magnitude (including color correction) with \salt $x_1$ (left panel) and \uti (right panel). Lighter shaded points were found in R17 to originate in locally passive regions, and likely from old progenitors, while dark blue points were found in regions dominated by star formation. \refcom{These are separated in $x_1$ (left), but not in \uti (right).}
  }
\label{fig:hruti2}
\end{figure*}

The dependence on LsSFR can be visualized for this sample by comparing peak magnitude (for clarity, after color correction) versus \salt $x_1$ (left panel of Fig.~\ref{fig:hruti2}). For fixed $x_1$, SNe in passive (``delayed'') environments are brighter. Performing such a comparison for \uti shows the dependence on LsSFR  to be much reduced (right panel of Fig.~\ref{fig:hruti2}).

\subsection{U-band indices and the choice of color curve}\label{sec:reddening}

\begin{figure*}
  \includegraphics[angle=0,width=0.66 \textwidth]{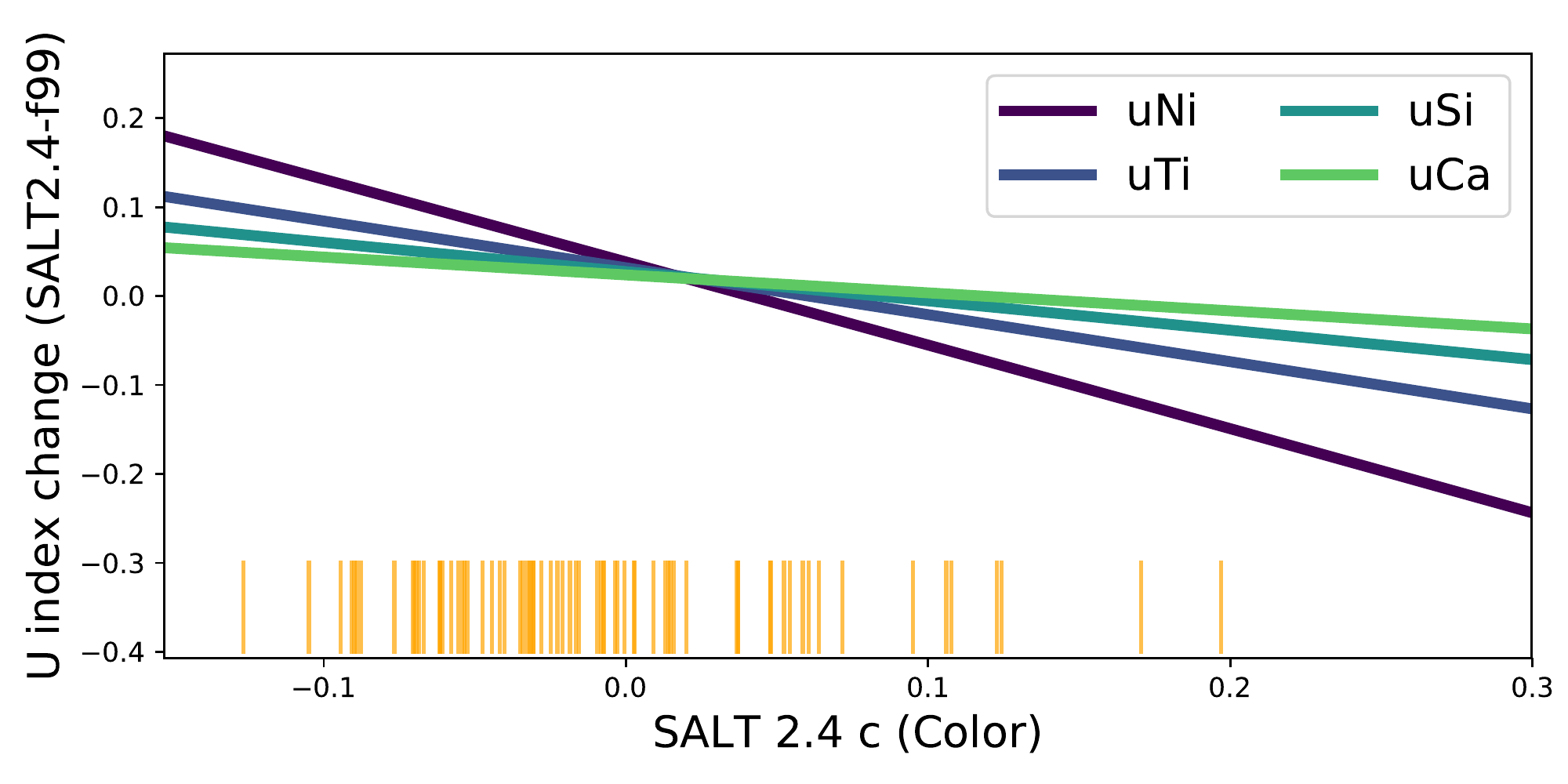}
  \includegraphics[angle=0,width=0.339 \textwidth]{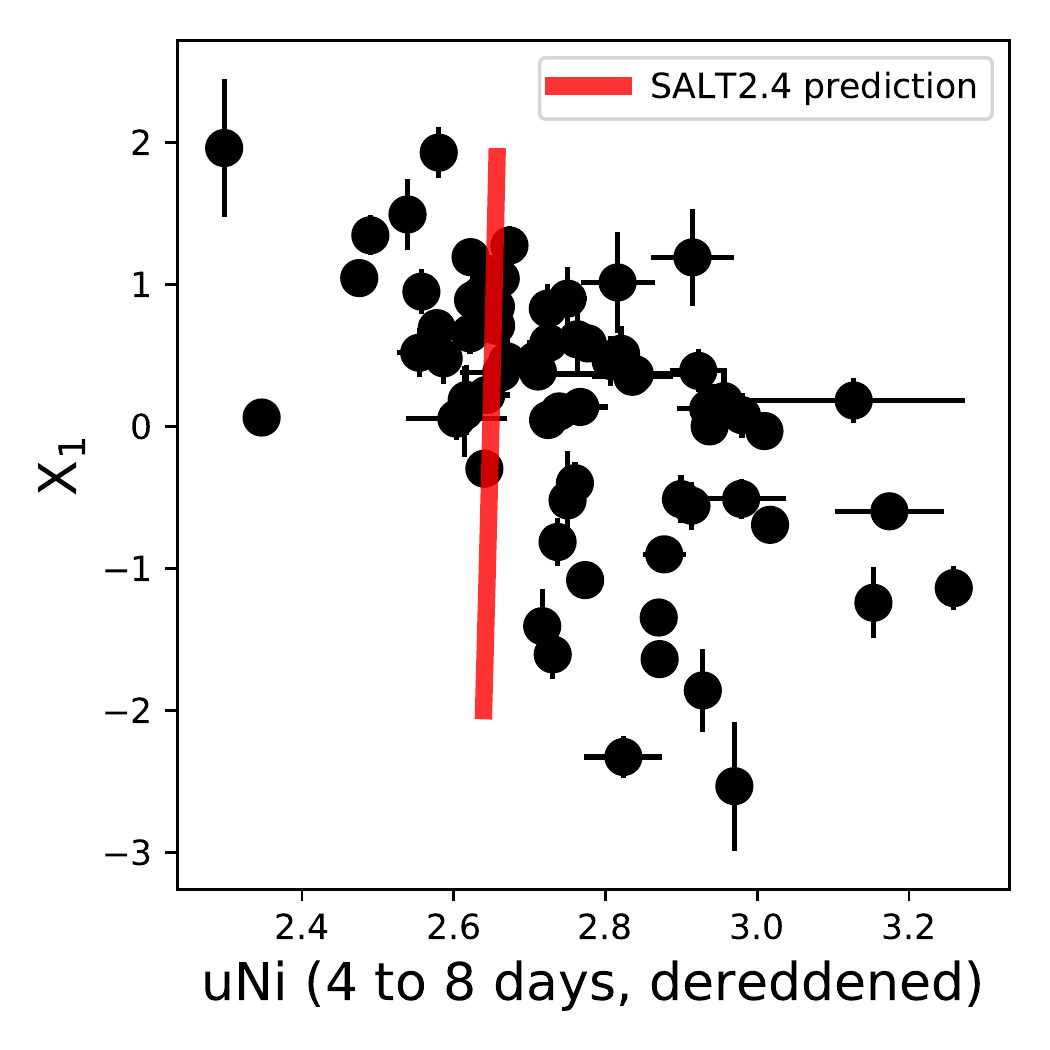}
  \caption{ \emph{Left:} \rfc{Directly comparing the difference in U-band color indices, \rfctwo{u(Ni|Ti|Si|Ca)$^{\mathrm{SALT}}$-u(Ni|Ti|Si|Ca)$^{\mathrm{F99}}$}.} While systematic differences grow large for highly reddened objects, these are limited for the current sample. Orange, short, vertical lines show the \salt $c$ values of SNe in this sample.
    \emph{Right:} SALT $x_1$ vs \uni at phase $4$ to $8$ days. The red line shows the corresponding SALT model predictions. 
  }
  \label{fig:deltadered}
\end{figure*}

Here we first study the potential systematic error caused by dereddening using the F99 color curve, if in fact all \sne actually followed the \salt color curve. The systematic (theoretical) change in the \refcom{\uni, \uti, \usi and \uca} color indices between F99 and \salt, as a function of the color parameter is shown in Fig.~\ref{fig:deltadered}. We use the same conversion between $E(B-V)$ and \salt $c$ as previously. 
More than $90\%$ of the sample has $|c|<0.15$, a range where the \emph{maximum} possible variation for the \uti, \usi and \uca colors are limited to less than $0.1$ mag -- small considering the U-band parameter value ranges found here. We therefore conclude that the standardization effects discussed above were not driven by systematic effects from the dereddening process.

An empirical \sn standardization model, like \salt, relies on the combination of a color curve and a spectral model to predict how the intrinsic spectrum varies (for SALT the latter is parameterized by $x_1$). 
\rfctwo{Comparing the \salt spectral model with observations in the \uniw region, where empirical and dust color curves start to strongly deviate, we note a clear functional difference -- the \uni color decreases with wider lightcurve width in a way that is not captured by the the \salt model (Fig.~\ref{fig:deltadered}, right panel). Such a mismatch between the \salt template and the observed SED could, if correlated with broad-band colors, modify the derived effective color curve and potentially bias cosmological parameter constraints if the \sn sample distributions vary with redshift/look-back time.
}

\section{Discussion}\label{sec:disc}

\subsection{Literature subclasses}\label{sec:subset}

\begin{table*}[!htbp]
  \centering
  \caption{Mean and its uncertainty of Hubble diagram residual for Branch Shallow Silicon and SN1991T-like SNe (the latter a subset of the former), for three different standardization methods, as discussed in Sec.\ref{sec:cosmo}. Results shown are from the ``cut'' sample, for which the three standardization methods can be compared. Note that the uncertainties are highly correlated.}
  \begin{tabular}{|l|c|cc|cc|cc|}
    \hline
    &  & \multicolumn{2}{c|}{$x_1$} & \multicolumn{2}{c|}{\uti} & \multicolumn{2}{c|}{\uti + \uca}  \\
    Sample & Nbr & HR mean (mag) & $\chi ^2$ & HR mean (mag) & $\chi ^2$ & HR mean (mag) & $\chi ^2$ \\
    \hline
    \hline
    \Tstrut SS & 5 & $-0.194 \pm 0.052$ & $14.4$ & $-0.100 \pm 0.058$ & $8.0$ & $-0.032 \pm 0.051$ & $4.4$\\[0.1em]
    91T & 2 & $-0.279 \pm 0.024$ & $8.6$ &  $-0.190 \pm 0.023$ & $4.3$ & $-0.095 \pm 0.011$ & $1.1$ \\[0.1em]
    \hline
  \end{tabular} 
  \label{tab:ss}
\end{table*}

\begin{figure*}
  \includegraphics[angle=0,width=0.48 \textwidth,clip, trim=0 0 0 0.3cm]{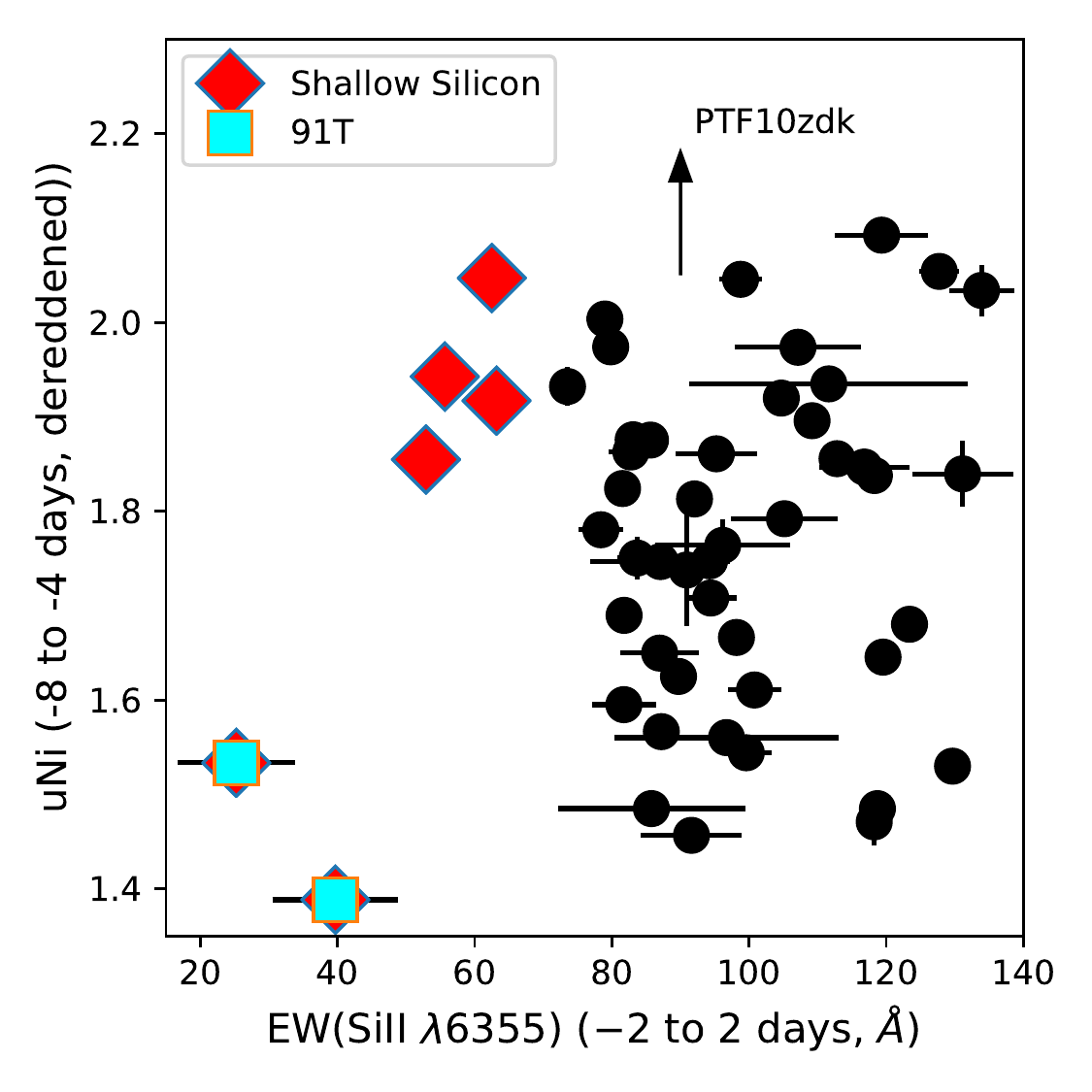}
  \includegraphics[angle=0,width=0.5 \textwidth,clip, trim=0.0 0.15cm 0.0cm 0.2cm]{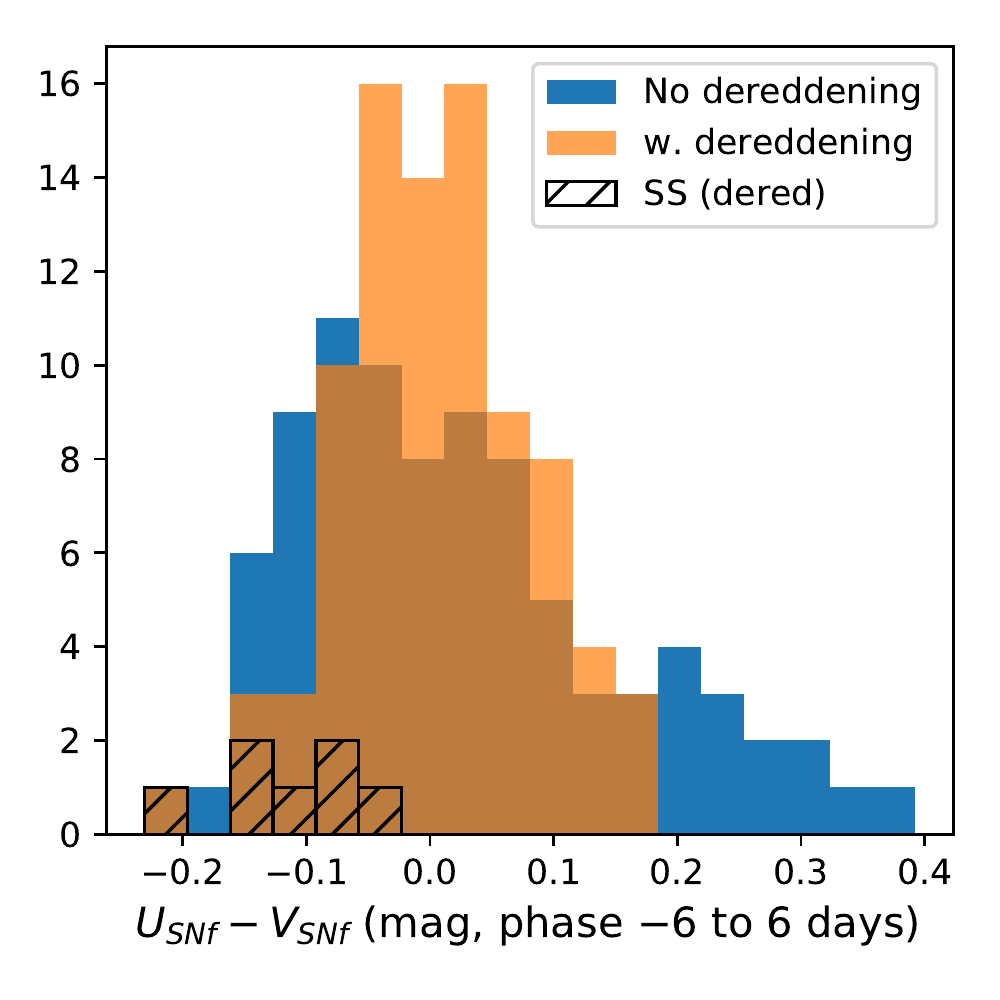}
  \caption{\emph{Left:} \uni index measurements of SN1991T-like and Shallow Silicon-like \sne compared with \ewsisix. The SN1991T-like SNe are disconnected from the remaining sample.
    \emph{Right:}
    Distribution of $U_{\mathrm{SNf}}-V_{\mathrm{SNf}}$ for spectra within $6$ days of lightcurve peak. The figure shows data with (orange) and without (blue) reddening correction, where the linear evolution with phase was removed for each (returning distributions with zero mean).  The Shallow Silicon (SS) subset of the dereddened $U_{\mathrm{SNf}}-V_{\mathrm{SNf}}$ distribution is shown by the cross-hatched histogram.
}
\label{fig:subset3}
\end{figure*}

A potential link with Shallow Silicon SNe \citep{2006PASP..118..560B} was highlighted in connection with \uca and SN standardization (Fig.~\ref{fig:stis}).
In particular, SN1991T-like objects, a core group among SS SNe,  have flat early U-band spectra as one of their defining features \citep{1992ApJ...384L..15F,2012ApJ...757...12S}. There are eight SS SNe in this sample, out of which two are SN1991T-like. Sub-classification of the SNfactory sample will be further discussed in Chotard et al. (in prep). Here we note that the mean Hubble diagram residual bias for SS and 91T-like objects gets progressively smaller when standardizing using U-band indices, as shown in Table~\ref{tab:ss}. 
\rtrt{The two 91T-like SNe have very similar \uni index, \ewsisix and Hubble residuals, while they are separated from both other SS SNe and other \sn subtypes (see Fig.~\ref{fig:subset3}, left panel).} We find no evidence of a continuous distribution connecting these to the main SN population, further suggesting that these SNe are more closely related to Super Chandrasekhar-mass \sne \citep[see discussion in][]{2012ApJ...757...12S,2014MNRAS.440.1498S}.

\citet{2013ApJ...779...23M,2015ApJ...803...20M} have suggested that \sne can be divided into  two subsets based on the \refcom{SWIFT} u$-$v color at $\pm 6$ days from peak. In Fig.~\ref{fig:subset3} we show the $U_{\mathrm{SNf}}-V_{\mathrm{SNf}}$ color as determined from a similar wide phase range.  We do not find strong signs of distinct populations in the $U_{\mathrm{SNf}}-V_{\mathrm{SNf}}$ color, especially not after reddening corrections are applied. \refcom{The SWIFT UVOT u band is bluer compared with $U_{\mathrm{SNf}}$ while $V_{\mathrm{SNf}}$ extends slightly redder than the UVOT v band, thus filter sensitivity differences possibly cause different intrinsic \sn features to be probed.
\citet{2017MNRAS.466..884C} did not find significant signs of subsets based on modeling ground-based photometry of more distant SDSS and SNLS SNe.}   
\citet{2013ApJ...779...23M} also highlighted high-velocity SNe and/or Branch Shallow Silicon SNe as having blue u$-$v colors, trends compatible with our data and highlighted in  Fig.~\ref{fig:subset3}.
\refcom{Finally, we note that \citet{2013ApJ...779...23M} showed both SN2011fe and SNF20080514-002 to have similar blue UVOT u$-$v colors, while our analysis began with the spectroscopic differences found between these two SNe (Fig.~\ref{fig:syn_comp}). This further points to the difficulty in constraining spectroscopic variations using broadband photometry, and vice versa.}

\begin{figure}
  \centering
  \includegraphics[angle=0,width=0.99 \columnwidth]{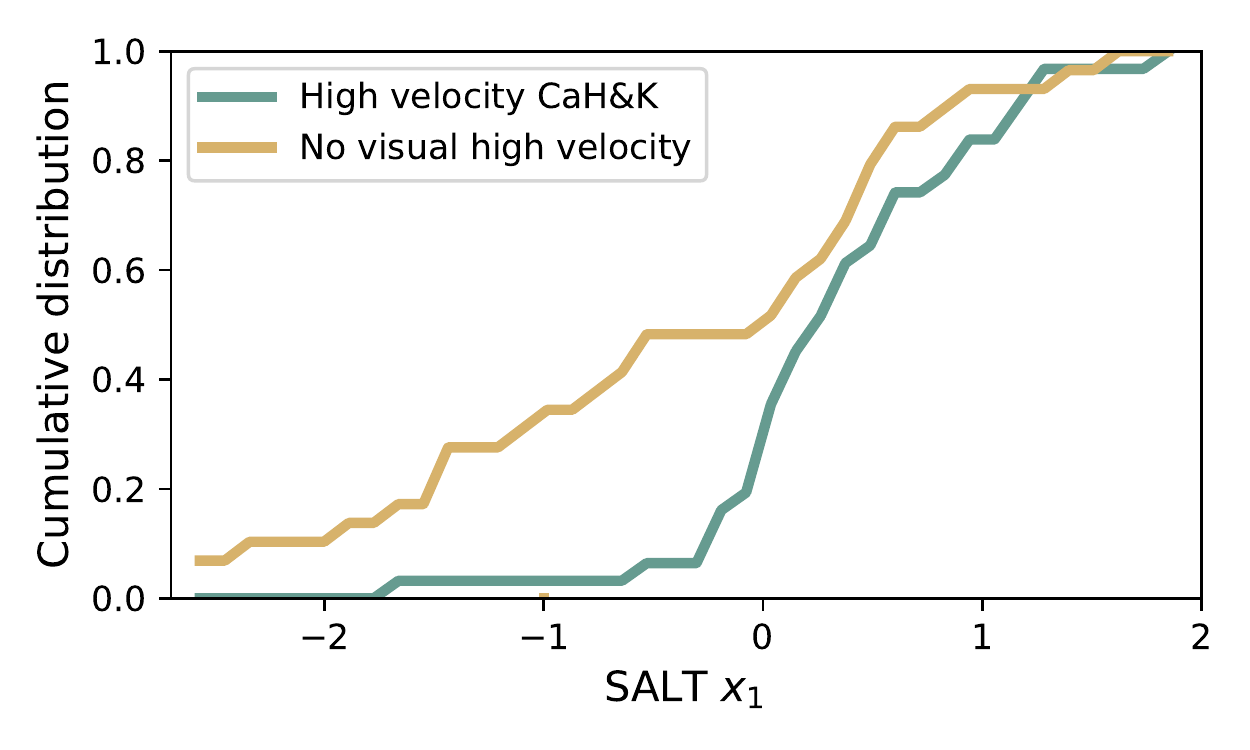}
  \caption{
  Cumulative $x_1$ distributions for SNe with and without visually identified High Velocity Features (based on spectra with phase $<-2$ day).}
\label{fig:subset4}
\end{figure}

The existence of High Velocity Features has been another approach to subclassification involving U-band data. A blinded visual search for \ion{Ca}{} HVFs was made with the purpose of dividing the sample into SNe with and without HVFs at phase $\sim-2$ days. All SNfactory spectra were randomly reordered and the \cahk and \cair region displayed, together with the expected photospheric line position based on the \sisix velocity at that phase. A scanner (J.N.) classified each spectrum as either having clear HVF, having good data but no HVF or being too noisy for an accurate classification. For the small SN subset with multiple high S/N spectra prior to $-2$ days, we found the visual classification to agree, showing that we consistently classified these.
\rfc{We show the \salt $x_1$ cumulative distributions of SNe with and without the HVF classification in Fig.~\ref{fig:subset4} (listed in Table.~\ref{tab:snprops}). As have previous studies \citep{2014MNRAS.437..338C, 2015MNRAS.451.1973S}, we find narrower lightcurve SNe ($x_1 \lesssim -0.5$) to only very rarely show HVFs at a phase whereas a majority of the $x_1 > -0.5$ SNe show HVFs. \rfctwo{A KS-test rejects a common parent population at greater than $99\%$ confidence.} If the presence of HVFs were  solely caused by, for example, the effective radius at a certain phase, to which $x_1$ could be correlated through the ejecta mass, a continously increasing probability of detecting HVFs would be expected. The data presented here is more suggestive of a rapid probability change with $x_1$, such as would be caused if $x_1 \lesssim -0.5$ originate from a distinct explosion channel.
}
In terms of the U-band spectral indices, this difference can be seen as a consistently small RMS for \usi and \uca at early phases among the low $x_1$-subset in Fig.~\ref{fig:urms_x1bin}.

\subsection{Outlook}\label{sec:wfirst}

The wavelength limits of the U-band features used here were set based on a comparison of only two SNe, examined at two phases. These wavelength regions were then analyzed at three phase intervals chosen to roughly capture typical \sn spectral variations on weekly scales, but chosen prior to any knowledge regarding when certain features come to dominate.
An improved future analysis will include phase-dependent wavelength regions, where observations at different phases are combined with appropriate weights.
Integration of flux within fixed wavelength limits carry less stringent signal-to-noise requirements compared with measurements of individual spectroscopic features. \rtrt{Thus, low-resolution spectroscopy (or possibly narrow-band filters) could measure the U-band indices directly at high redshifts.}

A sample of nearby \sne with cadenced U-band spectroscopy starting at early phases ($<-10$ days) would allow novel studies of the explosion process through the temporal development of the \uniw region.
Simultaneously, \cahk features map IME variations and \uti provides accurate temperature estimates after peak light. Such early data is also needed to finally determine whether the lack of HVFs among narrower lightcurve SNe is a consequence of a less energetic explosion, or if this signals a physically different explosion mechanism and/or circumstellar material.

\rfctwo{The limit at which HVFs disappear ($x_1 \sim -0.5$) agrees with where \citet{2014MNRAS.440.1498S}, using SNfactory SNe with coverage extending to late phases, find that the derived ejecta mass (\mej) points to sub-Chandrasekhar mass explosions. The seven SNe in the sample analyzed here with derived \mej and \mni masses from \citet{2014MNRAS.440.1498S} \rtr{are insufficient for further} tests of \mej and \mni correlations, again motivating larger samples which combine several types of observations. These could include also ejecta/\ion{Ni}/IGE constraints derived through  nebular \ion{Co}{ii} lines \citep{2015MNRAS.454.3816C}, second maximum position \citep{2015MNRAS.448.1345D} and early lightcurve rise-time \citep{2015MNRAS.446.3895F}, as well as measurements of ejecta asymmetry from nebular lines \citep{2011MNRAS.413.3075M}.}

We see signs that the currently most widely used SN model, \salt, can be improved. Updates to the SN template in this region could propagate into a change in the effective color law. Creating an unbiased template for the intrinsic U-band spectrum is a requirement for current and future ground-based \sn analyzes like DES and LSST, where the rest-frame U-band is redshifted to the wavelengths most efficiently observed from the ground.

\section{Summary and conclusions}\label{sec:conc}

\rfc{Comparing spectra of two supernovae at early and peak phases led to the subdivision of the U-band into four regions whose flux ratio relative to B-band were used to construct the \uni, \uti, \usi and \uca indices. \refcom{We use a sample of \samplesize \sne to analyze these indices} at three representative phase ranges: $-8$ to $-4$ days (pre-peak), $-2$ to $2$ days (peak) and $4$ to $8$ days (post-peak). Based on these measurements we find: }

\begin{itemize}
\item SYNAPPS comparisons show that Ni/Co absorption can explain differences in the bluest feature (``uNi'').  \rfc{ \uige at pre-peak phases can be used as a probe of Ni abundance, to complement constraints from  the early lightcurve rise-time, bolometric lightcurves and nebular emission.} 
\item In similar tests, SYNAPPS fits also show that Ti absorption differences dominate the \utiw wavelength range. 
  The \uti index is an extremely sensitive luminosity indicator. \rfctwo{When used instead of the \salt $x_1$ parameter the RMS scatter around the Hubble diagram falls to \rmsti~mag.} Here we use measurements made in the post-peak phase region, but note that the \uti index at all phases after peak is strongly correlated with lightcurve width.
\item \rtrt{Adding pre-peak \uca as a third standardization parameter further reduces the RMS around the Hubble diagram. A fit with \uti and \uca  yields a \rmstica mag scatter with $\chi^2/\mathrm{dof} \sim 1$ without the need to introduce intrinsic dispersion.}
\item \rtrt{Including U-band indices in \sn standardization reduces the  dependence on host-galaxy environment. The difference between lightcurve width and \uti can be used to study \sn progenitor scenarios.}
\item \rtrt{Shallow Silicon and 91T-like SNe have biased \salt Hubble residuals, which are substantially reduced when U-band parameters are included in the fit. The two 91T-like SNe in this sample have similar U-band properties and magnitudes, which are offset from the remaining sample.}
\item  We confirm previous results that narrower-lightcurve \sne  very rarely show HVFs. We find a sharp transition in their incidence at $x_1 = -0.5$.
\end{itemize}

\begin{acknowledgements}
We thank the technical staff of the University of Hawaii 2.2 m telescope, and Dan Birchall for observing assistance. We recognize the significant cultural role of Mauna Kea within the indigenous Hawaiian community, and we appreciate the opportunity to conduct observations from this revered site. This work was supported in part by the Director, Office of Science, Office of High Energy Physics of the U.S. Department of Energy under Contract No. DE-AC02-05CH11231.
Support in France was provided by CNRS/IN2P3, CNRS/INSU, and PNC; LPNHE acknowledges support from LABEX ILP, supported by French state funds managed by the ANR within the Investissements d’Avenir programme under reference ANR-11-IDEX-0004-02. 
Support in Germany was provided by DFG through TRR33 ``The Dark Universe'' and by DLR through grants FKZ 50OR1503 and FKZ 50OR1602.
In China the support was provided from Tsinghua University 985 grant and NSFC grant No 11173017. Some results were obtained using resources and support from the National Energy Research Scientific Computing Center, supported by the Director, Office of Science, Office of Advanced Scientific Computing Research of the U.S. Department of Energy under Contract No. DE-AC02- 05CH11231. We thank the Gordon \& Betty Moore Foundation for their continuing support.
Additional support was provided by NASA under the Astrophysics Data Analysis Program grant 15-ADAP15-0256 (PI: Aldering). We also thank the High Performance Research and Education Network (HPWREN), supported by National Science Foundation Grant Nos. 0087344 \& 0426879.
\end{acknowledgements}

\bibliographystyle{aa}
\bibliography{jnref}

\appendix

\section{Element contributions to SYNAPPS fits.}

\begin{figure*}
  \includegraphics[angle=0,width=0.8 \textwidth]{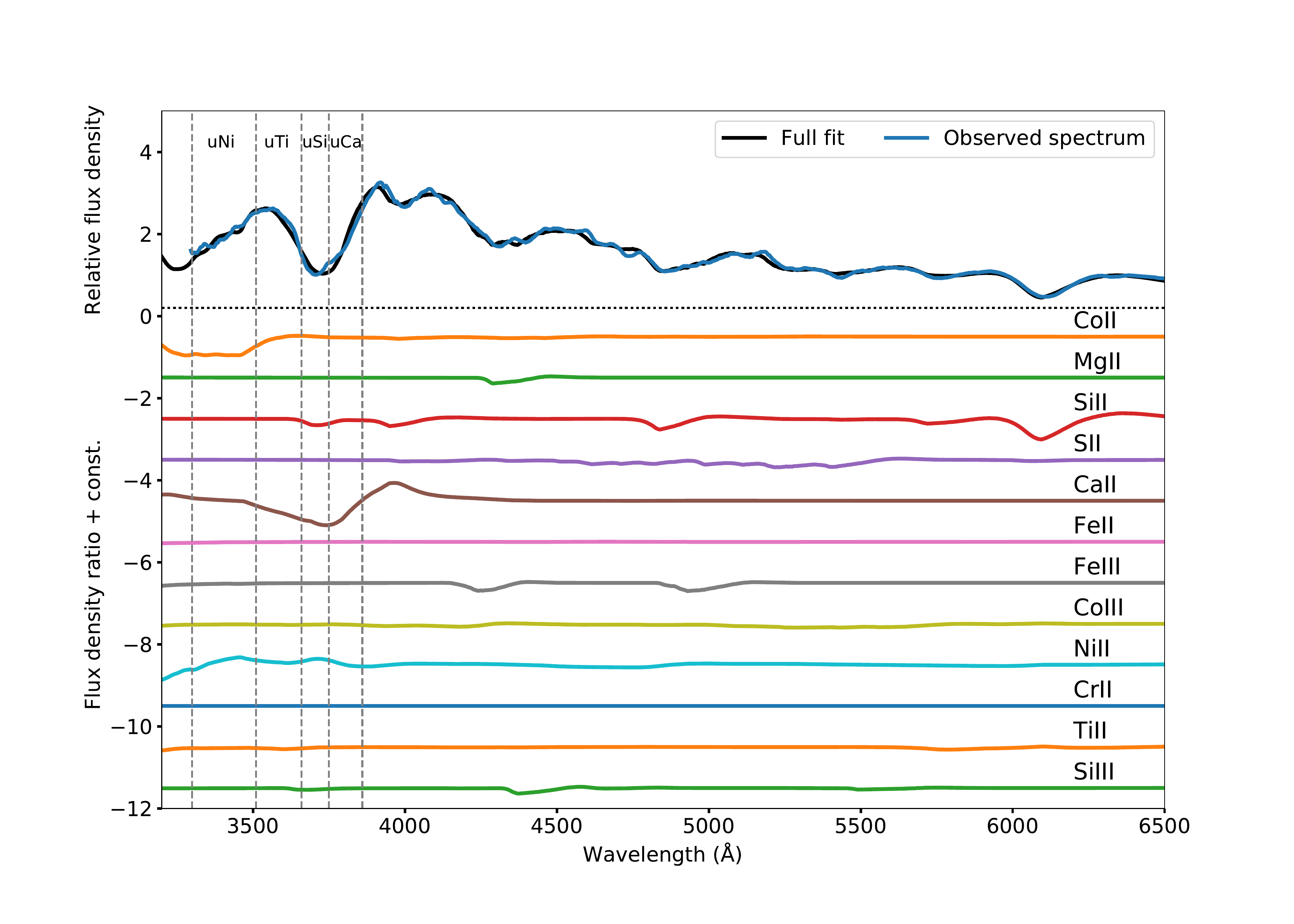}
  \caption{SYNAPPS fit to SN2011fe $10$ days prior to peak. The top panel compares the observed spectrum to the full SYNAPPS fit. \rfc{The lower panels show the contribution of each ion to the final spectra. These are plotted as the ratio between the spectrum with and without each ion activated.}
    \rfctwo{Dashed, vertical, grey lines indicate the U-band index region boundaries, with designations given at the top.}
} 
\label{fig:fesyn}
\end{figure*}

\begin{figure*}
  \includegraphics[angle=0,width=0.8 \textwidth]{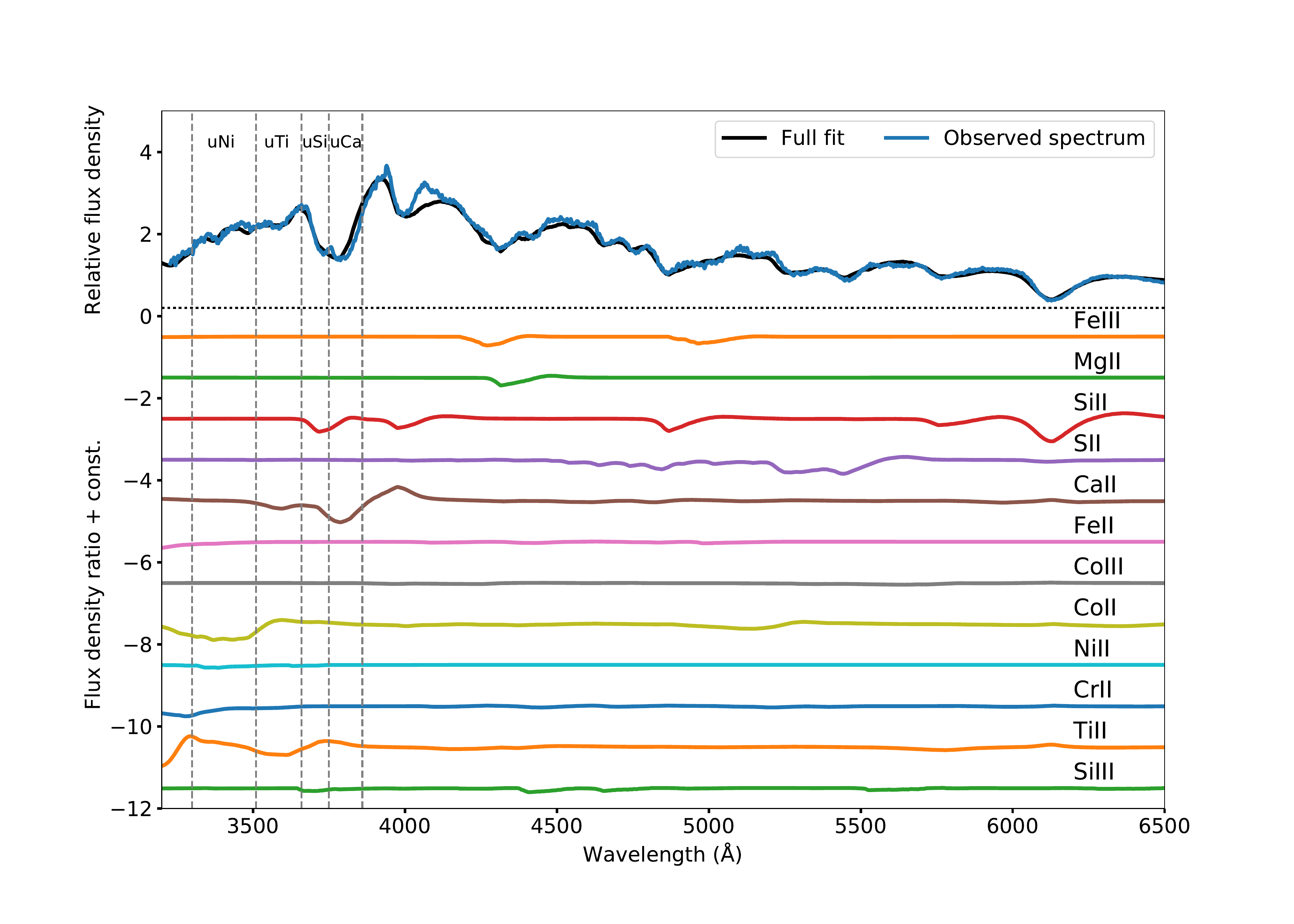}
  \caption{SYNAPPS fit to SNF20080514-002 at peak light. The top panel compares the observed spectrum to the full SYNAPPS fit.  \rfc{The lower panels show the contribution of each ion to the final spectra. These are plotted as the ratio between the spectrum with and without each ion activated.}
        \rfctwo{Dashed, vertical, grey lines indicate the U-band index region boundaries, with designations given at the top.}
}
\label{fig:snfsyn}
\end{figure*}

\rfctwo{We discuss in Sec.~\ref{sec:lines} how the observed spectral differences between SN2011fe and SNF20080514-002 can be explained by changing the early Ni/Co abundance and Ti abundance at peak. Here we show the complete set of ion contributions to the best fit SYNAPPS spectrum of SN2011fe $10$ days prior to peak (Fig.~\ref{fig:fesyn}) and for SN200805014-002 close to peak (Fig.~\ref{fig:snfsyn}). These fits match the observed data well. }

\end{document}